\documentclass[12pt]{iopart}
\usepackage{graphicx}
\usepackage{dcolumn}
\usepackage{bm}

\usepackage{rotating}

\newcommand{\Pm}{\mathrm{Pm}}

\providecommand\br{\bm{\rm r}}

\begin{document} 
\bibliographystyle{unsrt}
\title[A non local shell model of turbulent dynamo]{A non local shell model of turbulent dynamo}
\author{F. Plunian$^{1,2}$\footnote{Franck.Plunian@ujf-grenoble.fr}, R. Stepanov$^3$
\footnote{rodion@icmm.ru}}
\address{
$^1$Laboratoire de G\'eophysique Interne et Tectonophysique, CNRS, Universit\'e Joseph Fourier,
Maison des G\'eosciences, B.P. 53, 38041 Grenoble Cedex 9, France}
\address{$^2$Laboratoire des Ecoulements G\'{e}ophysiques et
Industriels, CNRS, Universit\'e Joseph Fourier, INPG, B.P. 53, 38041 Grenoble Cedex 9, France}
\address{$^3$Institute of Continuous Media Mechanics, Korolyov 1, 614013 Perm, Russia}
\begin{abstract}
We derive a new shell model of magnetohydrodynamic (MHD) turbulence in which the energy transfers are not necessary local.
Like the original MHD equations, the model conserves
the total energy, magnetic helicity, cross-helicity and volume in phase space (Liouville's theorem) apart from the effects of external forcing, viscous dissipation and magnetic diffusion. In the absence of magnetic field the model exhibits a statistically stationary kinetic energy solution with a Kolmogorov spectrum. The dynamo action from a seed magnetic field by the turbulent flow and the non linear interactions are 
studied for a wide range of magnetic Prandtl numbers in both kinematic and dynamic cases.
The non locality of the energy transfers are clearly identified.
\end{abstract}
\submitto{\NJP}
\maketitle
\section{Introduction} \label{intro}
Pioneering shell models of hydrodynamic turbulence were developed in the seventies \cite{Obukhov71,Lorenz72,Gledzer73,Desnyansky74,Siggia77}, aiming at reproducing the main turbulence features, as intermittency, with a low order model of equations. Such shell models were also called wave packet representation \cite{Nakano88} for the Fourier space is logarithmically divided into shells of logarithmic width $\lambda$ such that each wave packet (or shell) $k_n$ is defined by $k_0 \lambda^{n-1} < k \le k_0 \lambda^{n}$. Those models are local in the sense that each shell interacts with only the first neighbours like the DN model (named after Desnyansky and Novikov \cite{Desnyansky74}), or the two first neighbours like the GOY model (named after Gledzer, Ohkitani and Yamada \cite{Gledzer73,Yamada87}). The latter has been intensively studied (for a review, see \cite{Frisch95,Bohr98,Pope00,Biferale03} and references therein) and subjected to improvements leading to the so-called Sabra model \cite{Lvov98,Lvov99,Ditlevsen00} or extensions using the wavelet decomposition \cite{Benzi96}. The GOY and subsequently Sabra models have been used in different contexts like convection \cite{Mingshun97}, rotation \cite{Hattori04} or intermittency \cite{Jensen91,Pierotti97}. It has been shown \cite{Bowman04} that the DN model is a spectral reduction of the GOY model, showing in some sense the consistency of one model against the other.

To our knowledge only one non-local shell model of turbulence has been developed so far, by Zimin and Hussain \cite{Zimin95}, projecting the Navier-Stokes equations onto a wavelet basis, and reducing the number of variables from statistical assumptions. 
In such non-local model each shell may interact with any other shell. 
Since then, the original model has been improved by Zimin in order to include left- or right-handed polarity of the solenoidal basis functions as in the complex helical wave decomposition, and used by Melander and Fabijonas \cite{Melander97,Melander02,Melander03}.

The extension of shell models to MHD turbulence has been done including either first neighbours interactions \cite{Grappin86,Gloaguen85,Carbone94,Biskamp94} or two first
neighbours interactions \cite{Frick83,Frick84,Brandenburg96a,Frick98,Stepanov06} eventually including Hall effect \cite{Hori05, Frick03}.
However there is a number of situations in MHD turbulence in which assuming the locality of energy transfers may become somewhat spurious even if the turbulence is considered as isotropic \cite{Stepanov06}. This is true for example when a large scale external magnetic field is imposed leading to Alfven waves \cite{Kraichnan65,Kraichnan67,Iroshnikov63}. This problem has been tackled using a MHD shell model implementing non local interactions with the externally imposed magnetic field scale \cite{Biskamp94}. However it has been shown recently using an other method \cite{Verma99} that the other non local interactions are also important and may rule out the predicted Iroshnikov-Kraichnan $k^{-3/2}$ spectrum.
Non local interactions are also at the heart of the dynamo problem i.e. when the magnetic field is produced by the turbulent motion instead of being externally applied. For example at large value of the magnetic Prandtl number, the magnetic spectrum is observed to peak at scales much smaller than the viscous scale \cite{Haugen03,Schekochihin02a,Schekochihin02b} showing a direct evidence of the importance of non local energy transfers.
In presence of helicity, 
two possible mechanisms may generate a large scale magnetic field as observed in planets and stars: either a local inverse cascade \cite{Pouquet76}  or a non local direct transfer from small to large scales as predicted by the mean-field theory \cite{Krause80}. Which mechanism prevails is still not clear. Recently, the importance of non-local interactions has been shown in both hydrodynamic \cite{Alexakis05a}
 and MHD \cite{Mininni05,Alexakis05b,Carati06} turbulence. For recent reviews on MHD turbulence and the dynamo problem, see e.g. \cite{Verma04,Brandenburg05}

In the present paper our aim is to introduce a new non local shell model of turbulence which can be used either in its hydrodynamic or MHD form. As we are interested by MHD applications, we shall introduce the MHD model only, the hydrodynamic one being easily deduced from the former, setting the magnetic field to zero.
Our model can be understood as a non local version of the Sabra model. This involves similar rules for complex conjugations and imposes the value of shell spacing equal to the golden number. 
Our first attempt to derive a non local shell model of MHD turbulence was based on the Zimin model \cite{Zimin95}. However we realized that this model was unable to describe the non local interaction between a large scale velocity and two small neighboring scales of the magnetic field. We believe that the one which is described here is more relevant to actual isotropic MHD turbulence.  
\section{General concept}
\subsection{The model}
The model is defined by the following set of equations
\begin{eqnarray}
\dot{U}_n &=& i k_n \left[Q_n(U,U,a)+Q_n(B,B,-a)\right]
- \nu k_n^2 U_n + F_n, \label{eq_u} \\
\dot{B}_n &=& i k_n \left[Q_n(U,B,b)+Q_n(B,U,-b)\right] 
- \eta k_n^2 B_n, \label{eq_b}
\end{eqnarray}
where
\begin{eqnarray}
\fl Q_n(X,Y,c)= \sum^{N}_{m=1}T_m [ c_m^1 X_{n+m}^*Y_{n+m+1} 
 +  c_m^2 X_{n-m}^* Y_{n+1}  
+ c_m^3 X_{n-m-1} Y_{n-1} ] \label{Qn}
\end{eqnarray}
represents the non linear transfer rates.
The parameters $\nu$ and $\eta$ are respectively the kinematic viscosity and the magnetic diffusivity,
$F_n$ is the forcing of turbulence, and $k_n=\lambda^n$ with $\lambda^2 = \lambda +1$ \cite{Lvov98}.
For $N=1$ in (\ref{Qn}), we recognize the local Sabra model. The additional non local interactions for $N \ge 2$ correspond to all other possible triad interactions except the ones involving two identical scales.
Expressions for the kinetic energy $E_U$ and helicity $H_U$, magnetic energy $E_B$ and helicity $H_B$, and cross helicity $H_C$ are given by
\begin{eqnarray}
\fl	E_U = \sum_{n}E_U(n), \;\;\; E_U(n)=\frac{1}{2} |U_n|^2, &&  \quad
	H_U = \sum_{n}H_U(n), \;\;\; H_U(n)=\frac{1}{2} (-1)^n k_n |U_n|^2 \label{kinetic}\\
\fl	E_B = \sum_{n}E_B(n), \;\;\; E_B(n)=\frac{1}{2} |B_n|^2,&&  \quad
	H_B = \sum_{n}H_B(n), \;\;\; H_B(n)=\frac{1}{2} (-1)^n k_n^{-1} |B_n|^2 \label{magnetic}\\&&  \quad
	H_C = \sum_{n}H_C(n), \;\;\; H_C(n)=\frac{1}{2}  (U_n B_n^* + B_n U_n^*).\label{cross helicity}
\end{eqnarray}

In the inviscid and non-resistive limit ($\nu=\eta=0$), the total energy
$E=E_U+E_B$, magnetic helicity and cross helicity must be conserved ($\dot{E}=\dot{H}_B=\dot{H}_C=0$).
This leads to the following expression for the coefficients $a_m^i$ and $b_m^i$:
\begin{eqnarray}
	a_m^1 = k_{m} + k_{m+1}&
	a_m^2 = \frac{-k_{m+1} - (-1)^m}{k_{m}}&
	a_m^3 =   \frac{k_{m} - (-1)^m}{k_{m+1}}\nonumber\\
	b_m^1 = (-1)^{m+1}&b_m^2 = 1&b_m^3 =  -1.
	\label{coefficients}
\end{eqnarray}
\\
In the case of pure hydrodynamic turbulence (without magnetic field), the coefficients $a_m^i$ are 
derived from the kinetic energy and helicity conservations ($\dot{E_U}=\dot{H}_U=0$), leading again to the same expression as (\ref{coefficients}). 
The coefficients $T_m$ are free parameters depending on $m$ only, that we
choose of the form $T_m= k_{m-1}^{\alpha}/\lambda(\lambda +1)$.
The coefficient $1/\lambda(\lambda +1)$ is chosen such that the terms for $m=1$ in $Q_n$ correspond to the local Sabra model. The local Sabra model corresponds to $\alpha = - \infty$.
\subsection{All possible interactions}
\label{interactions}
\begin{figure}[ht]
\begin{center}
\begin{tabular}{@{}c@{\hspace{0em}}c@{}}
   \raisebox{4cm}{$q$} & \includegraphics[width=0.3\textwidth]{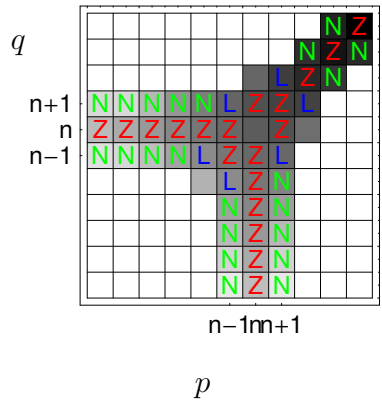}\\*[0.0cm]
       & $p$
   \end{tabular}
  \end{center}
\caption{Probability of interactions between three modes belonging to shells $n,p$ and $q$.
The white squares correspond to a null probability.} 
\label{toto}
\end{figure}

In our shell model we see from (\ref{Qn}) that only a discrete number of triads are allowed.
For example, $Q_n$ does not contain any term involving interactions between the modes $n+m$ and $n+m+2$.
The reason why there is only a discrete number of allowed triads comes from the fact that the shells are logarithmically spaced  and that the geometrical factor $\lambda$ satisfies
\begin{equation}
	\lambda^2 \ge \lambda +1.
	\label{lambda}
\end{equation}
To identify all the allowed triads in a shell model, let us consider three vectors
$(\textbf{k}_1,\textbf{k}_2,\textbf{k}_3)$ satisfying
\begin{equation}
	\textbf{k}_1 + \textbf{k}_2 + \textbf{k}_3 = 0.
	\label{triad formula}
\end{equation}
Assuming that $\textbf{k}_1$ and $\textbf{k}_2$ belong to shell $n$ and $p$,
we have
\begin{equation}
	k_0 \lambda^{n-1} < |\textbf{k}_1| \le k_0 \lambda^{n}, \quad
	k_0 \lambda^{p-1} < |\textbf{k}_2| \le k_0 \lambda^{p}.
\end{equation}
From (\ref{triad formula}), this implies
\begin{equation}
k_0 |\lambda^{n-1} - \lambda^{p-1}| \le	|\textbf{k}_3| \le k_0 (\lambda^n + \lambda^p).
\label{kp}
\end{equation}
Now using the inequalities (\ref{lambda}) and (\ref{kp}) we can show that
$\textbf{k}_3$ belongs to shell $q$ which depends on $p$ in the following way:
\begin{equation}
\begin{tabular}{@{\hspace{-2cm}}l@{\hspace{1cm}}l@{}}
	$p \le n-3 \quad        \Rightarrow \quad n-1 \le q \le n+1$  & $p = n+1   \quad\Rightarrow\quad\quad\quad\quad\;\, q \le n+2$ \\
	$p =   n-2 \quad        \Rightarrow \quad n-2 \le q \le n+1$  & $p = n+2   \quad\Rightarrow\quad\quad\;\, n \le q \le n+3$ \\
	$p = n-1 \quad          \Rightarrow \quad \quad \quad \quad \;\;\, q \le n+1$            & $p \ge n+3 \quad\Rightarrow\quad p-1 \le q \le p+1$ \\
	$p =n \quad\quad \;\;\, \Rightarrow \quad \quad \quad \quad \;\;\, q \le n+2$. &
	\end{tabular}
	\label{possible}
	\end{equation}
This is illustrated in Figure \ref{toto} in the plane $(p,q)$ where the grey (resp. white) squares indicate allowed (resp. not allowed) triads ($n,p,q$). 
 The demonstration of (\ref{possible}) is given in Appendix \ref{appendixpossible}.
In Figure \ref{toto}, the terms corresponding to the original Sabra local model are denoted by ``L'', the additional non
local terms by ``N'' and the terms of Zimin's model by ``Z''. In every case the possible shells $(p,q)$  are symmetric with respect to the diagonal in the $(p,q)$ plane .
\subsection{Evaluation of $\alpha$}
\label{sec:alpha}
There is one free parameter left, $\alpha$, which corresponds to the non locality
strength. It is not an easy task (if possible at all) to figure out what $\alpha$ should be in the general case.
However we tried to estimate it in the case of homogeneous isotropic turbulence (without
magnetic field). For that,
we take two random
vectors $\textbf{k}_1$ from shell $n$, and $\textbf{k}_2$
uniformly distributed in whole space, and
we calculate the probability that $\textbf{k}_3 = - (\textbf{k}_1 + \textbf{k}_2)$ and $\textbf{k}_2$ belong respectively to shells $p$  and $q$. It is the simplest estimate of the probability that the three modes $\textbf{k}_1$, $\textbf{k}_2$ and $\textbf{k}_3$ interact together. A high (resp. low) value of this probability 
is given in Figure \ref{toto} by the dark (resp. light) colour of the grey squares. In this representation a white square
correspond to a null probability.
The probability spectra along the $p$
and $q$ directions 
are found to scale as $k^{-7/2}$
for the "L" and "N" terms
and  $k^{-5/2}$
for the "Z" terms (which is consistent
with the Zimin's model \cite{Zimin95}).
In order to have terms $T_m c_m^2$ and $T_m c_m^3$ in (\ref{Qn}) scaling in $k_m^{-7/2}$,
we have to take $\alpha = -7/2$. We note that this derivation of $\alpha$ does not imply the terms $T_m c_m^1$ (and neither the diagonal terms of Figure \ref{toto}). The latters are determined a posteriori from the conservation laws (not included in the probability diagram of Figure \ref{toto}).\\
In section \ref{sec:freedecay} we shall find that the value $\alpha=-5/2$ is the most appropriate to describe accurately the large scale slope of the kinetic energy spectrum. However, 
in the rest of the paper we shall vary the value of $\alpha$ in order to investigate the
role of $\alpha$ in the non local interactions.
\subsection{Energy transfers}
\begin{figure}[ht]
\begin{center}
   \includegraphics[width=0.7\textwidth]{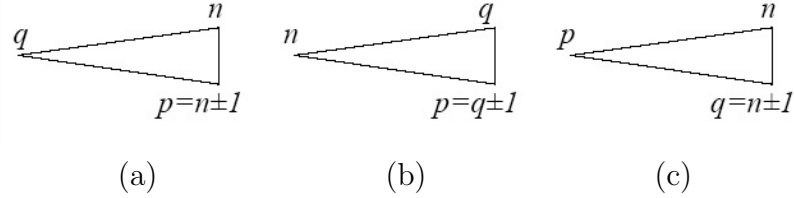}\\
   (a)   \quad \quad \quad \quad \quad \quad \quad (b)  \quad \quad \quad \quad \quad \quad \quad (c) 
   \end{center}
\caption{The three types of possible triad $(X_q,Z_p,Y_n)$ : (a) $q \le n-1$ and $p=n \pm 1$, (b) $q \ge n+1$ and $p=q \pm 1$,
(c) $q = n\pm 1$ and $p \le n - 1$. Among these cases those with repeated subscripts ($q=p$, $p=n$ or $q=n$) are forbidden in our model.} 
\label{triads}
\end{figure}

To study the non local interactions, we introduce the transfer rate ${\cal T}_{XY}(q,n)$ from
$X$-energy lying in shell $q$ to $Y$-energy lying in shell $n$. It can operate only within triads,
implying an additional energy $Z$ lying in all other shells $p$ different from $q$ and $n$. 
Denoting $T(X_q|Z_p|Y_n)$
the transfer rate from $X_q$ to $Y_n$ and which involves $Z_p$ as a mediator,
the transfer rate from $X_q$ to $Y_n$ can then be written as
\begin{equation}
	{\cal T}_{XY}(q,n) = \sum_{p \ne q,n} T(X_q|Z_p|Y_n).
	\label{TXYqn}
\end{equation}
From our model (\ref{Qn}) we see that for each couple $(X_q,Y_n)$, only a discrete number of $Z_p$  can act as mediators.
The possible triads are represented schematically in Figure \ref{triads}.
Therefore ${\cal T}_{XY}(q,n)$ takes the following form
\begin{equation}
\fl	{\cal T}_{XY}(q,n) = \left\{ \begin{tabular}{@{}l@{\hspace{1em}}l@{\hspace{3em}}l@{}}
	                                    $T(X_q|Z_{n+1}|Y_n)+ T(X_q|Z_{n-1}|Y_n)$,& for $q \le n-2$\\
	                                    $T(X_{q}|Z_{n+1}|Y_n) + \sum_{p\le n-2}T(X_{q}|Z_p|Y_n) $,& for $q = n-1$\\
	                                    $T(X_{q}|Z_{n+2}|Y_n) + \sum_{p\le n-1}T(X_{q}|Z_p|Y_n) $,& for $q = n+1$\\
	                                    $T(X_q|Z_{q+1}|Y_n)+T(X_q|Z_{q-1}|Y_n)$,& for $q \ge n+2$
	                                 \end{tabular}
	                                 \right..
\end{equation}
Now, we have to define $T(X_q|Z_p|Y_n)$ for any $X_q$, $Z_p$ and $Y_n$. For that,
we can re-write the model (\ref{eq_u}) (\ref{eq_b}) in the following generic form
\begin{equation}
	\dot{Y}_n = \sum_{p,q}i k_n M_n(X_q,Z_p) + \cdots
\end{equation}
where $M_n(X_q,Z_p)$ is a symmetric bilinear form (given in Appendix \ref{Apptransfers}) representing the quadratic terms. The dots represent the dissipation and forcing terms appropriate to $Y_n$.

Now let us denote by $S(X_q,Z_p|Y_n)$ the rate of energy within the triads $(X_q,Z_p,Y_n)$ which is transferred from the couple $(X_q,Z_p)$ to $Y_n$. It is naturally defined by
\begin{equation}
	S(X_q,Z_p|Y_n) = \Re\left\{i k_n Y_n^* M_n(X_q,Z_p)\right\}.
	\label{S}
\end{equation}
As $M_n$ is symmetric, we have
\begin{equation}
S(X_q,Z_p|Y_n) = S(Z_p,X_q|Y_n).
\label{symmetryS}
\end{equation}
In addition,and with the help of (\ref{coefficients}), we can show that for any triad $(X_q,Z_p,Y_n)$ the following relation is satisfied
\begin{equation}
	S(X_q,Z_p|Y_n) + S(Z_p,Y_n|X_q) + S(Y_n,X_q|Z_p) = 0,
	\label{SSS}
\end{equation}
meaning that the energy is conserved within each triad.

Then we look for $T(X_q|Z_p|Y_n)$ as a linear combination of $S(X_q,Z_p|Y_n)$, $S(Y_n,X_q|Z_p)$ and $S(Z_p,Y_n|X_q)$.
As in real MHD turbulence, $T(X_q|Z_p|Y_n)$ must satisfy the following conditions\\
\begin{eqnarray}
	T(X_q|Z_p|Y_n) &=& - T(Y_n|Z_p|X_q) \label{cond1}\\
	S(X_q,Z_p|Y_n) &=& T(X_q|Z_p|Y_n) + T(Z_p|X_q|Y_n).
	\label{cond2}
\end{eqnarray} 
The first condition (\ref{cond1})
means that the transfer rate from $X_q$ to $Y_n$ and from $Y_n$ to $X_q$ with the same mediator $Z_p$ are of equal intensity but opposite signs.
We note that this implies
\begin{equation}
	{\cal T}_{XY}(q,n) = - {\cal T}_{YX}(n,q).
	\label{antisymmetry}
\end{equation}
The second condition (\ref{cond2})
means that in any triad $(X_q,Z_p,Y_n)$ the transfer rate from $(X_q,Z_p)$ to $Y_n$ is equal to the sum of the transfer rates
from $X_q$ to $Y_n$ via $Z_p$ and from $Z_p$ to $Y_n$ via $X_q$.
Then combining (\ref{symmetryS}),(\ref{SSS}), (\ref{cond1}) and (\ref{cond2}) we end up with the following expression for $T$
\begin{equation}
	T(X_q|Z_p|Y_n) = \frac{1}{3}\left[S(X_q,Z_p|Y_n) - S(Y_n,Z_p|X_q)\right]
	\label{transferT}
\end{equation}
Furthermore we can show that the following energy balances at scale $n$ are satisfied
\begin{eqnarray}
	\dot{E}_U(n) +{\cal D}_{U}(n) - {\cal F}(n) &=& \sum_q[{\cal T}_{UU}(q,n) + {\cal T}_{BU}(q,n) ]  \\
	\dot{E}_B(n) +{\cal D}_{B}(n)&=& \sum_q[{\cal T}_{UB}(q,n) + {\cal T}_{BB}(q,n) ] 
	\label{balance}
\end{eqnarray}
with the kinetic and magnetic dissipation rates in shell $n$ defined by
\begin{equation}
	{\cal D}_{U}(n) = \nu k_n^2 |U_n|^2, \quad {\cal D}_{B}(n) = \eta k_n^2 |B_n|^2
\end{equation}
and 
\begin{equation}
	{\cal F}(n) = \frac{1}{2}(F_n U_n^* + F_n^* U_n).
	\label{forcing}
\end{equation}
\subsection{Energy fluxes}
We define the energy flux $\Pi_{XY}(n)$ as the rate of loss of $X$-energy lying in the shells $j<n$ to the $Y$-energy lying in the shell $j\ge n$.
Therefore, we have
\begin{eqnarray}
	\Pi_{XY}(n) &=& - \sum_{j=0}^{n-1} \sum_q \sum_p T_{XY}(q|p|j) \label{fluxesdef}\\
	            &=& - \sum_{j=0}^{n-1} \sum_q {\cal T}_{XY}(q,j).
\end{eqnarray}
The fluxes $\Pi_{UU}(n)$ and $\Pi_{BU}(n)$ coincide respectively with $\sum_{j=0}^{n-1} \Im \{k_j U_j^* Q_j(U,U,a)\}$ and $\sum_{j=0}^{n-1} \Im \{-k_j U_j^* Q_j(B,B,a)\}$. There is no such coincidence for $\Pi_{UB}(n)$ and $\Pi_{BB}(n)$.\\

The following flux balances are satisfied
\begin{eqnarray}
	\sum_{j=0}^{n-1} \left( \dot{E}_U(j) +{\cal D}_{U}(j) - {\cal F}(j)  \right) + \Pi_{UU}(n) + \Pi_{BU}(n) = 0\\
		\sum_{j=0}^{n-1} \left( \dot{E}_B(j) +{\cal D}_{B}(j) \right) + \Pi_{UB}(n) + \Pi_{BB}(n) = 0.
\end{eqnarray}
In a statistically stationary case they imply that
\begin{equation}
\fl	\Pi_{UU}(n) + \Pi_{BU}(n) + \Pi_{UB}(n) + \Pi_{BB}(n) = - \sum_{j=0}^{n-1} \left( {\cal D}_{U}(j) +{\cal D}_{B}(j) + {\cal F}(j) \right).
\end{equation}

In order to investigate the importance of non local versus local interactions,
we define the local part of the fluxes given by (\ref{fluxesdef}) in which only the local energy transfer rates $T_{XY}(n\pm 2|n\pm 1|n)$, $T_{XY}(n\pm 1|n\mp 1|n)$ and $T_{XY}(n\pm 1|n\pm 2|n)$
are involved. These fluxes correspond to those of the MHD version of the original (local) Sabra model, taking $N=1$ in (\ref{Qn}). The non local parts of the fluxes are defined as the total fluxes minus their local parts.

\subsection{Hydrodynamic forcing}
The hydrodynamic forcing is generally applied at scale $n_f$ and $n_{f+1}$ with $n_f=4$. It is of the form $F_n=A_n \exp^{i\phi_n(t)}$
where $\phi_n(t)\in \left[ 0, 2\pi \right]$ is constant during time intervals $t_c$, the constant value changing randomly from one time interval to the next one. In this way we obtain a statistically constant injection rate equal to $\epsilon = A_n^2 t_c$. We chose $t_c=10^{-2}$ for it is smaller than the turn-over time at the injection scale and larger than the viscous characteristic time.

For some arbitrary initial conditions on $U_n$ of small intensity we let the hydrodynamic
evolve until it reaches some statistically stationary state. Then introducing at a given time some small intensity of $B_n$ we solve the full problem until a statistically stationary MHD state is reached.
The time of integration needed to obtain good statistics depends on $\nu$ and $\eta$ but is typically of several hundreds of time unit (of the largest scale $n=0$). 
We define the magnetic Prandtl number as the ratio $\Pm= \nu / \eta$.

To study the free decaying turbulence, after having reached a statistically stationary state,
the forcing is set to zero.

\section{Results}
\label{results}
\subsection{Energy spectra}
\subsubsection{Free decaying turbulence}
\label{sec:freedecay}
\begin{figure}
\begin{center} \begin{tabular}{@{}c@{\hspace{0em}}c@{\hspace{0em}}c@{\hspace{0em}}c@{\hspace{0em}}c@{\hspace{0em}}c@{}}
        &
    \begin{turn}{0} $R_e=10^{-6}$ \end{turn}
        &
    \begin{turn}{0}$P_m=10^{-3}$\end{turn}
        &
   \begin{turn}{0} $P_m=1$\end{turn}
        &
       \begin{turn}{0} $P_m=10^{3}$\end{turn}
        &
       \\*[0.cm]
  \begin{turn}{90}$\quad \quad \quad \quad \alpha=-0.5$ \end{turn}&
    \includegraphics[width=0.28\textwidth,angle=90]{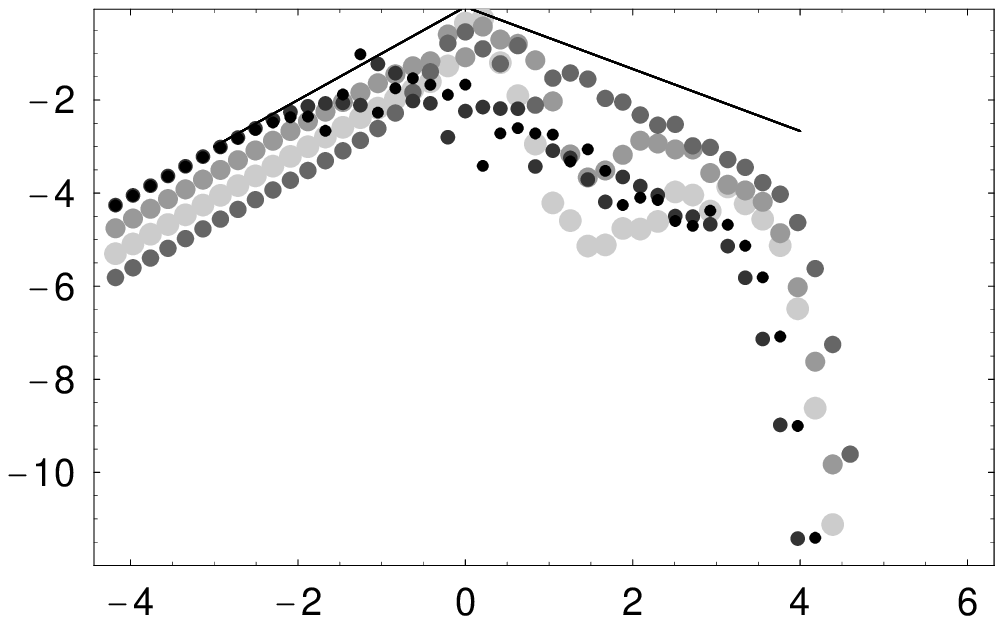}
        &
    \includegraphics[width=0.28\textwidth,angle=90]{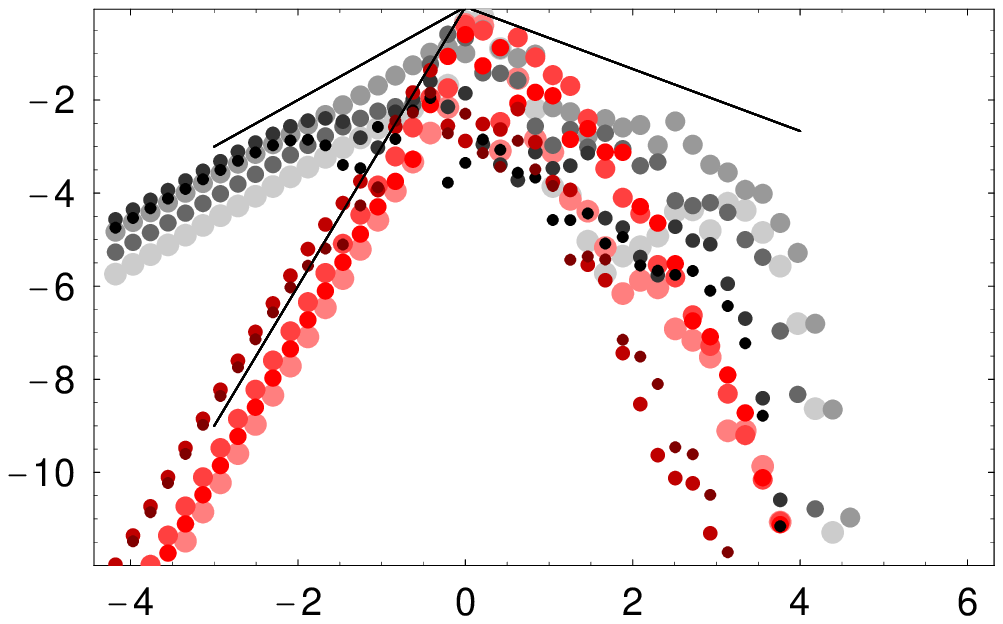}
        &
    \includegraphics[width=0.28\textwidth,angle=90]{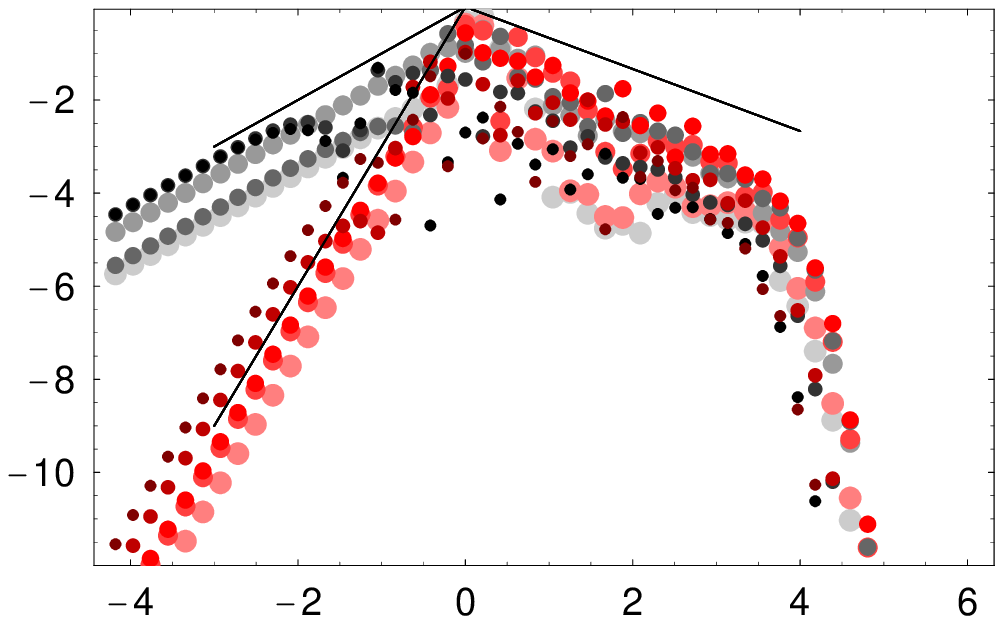}
        &
    \includegraphics[width=0.28\textwidth,angle=90]{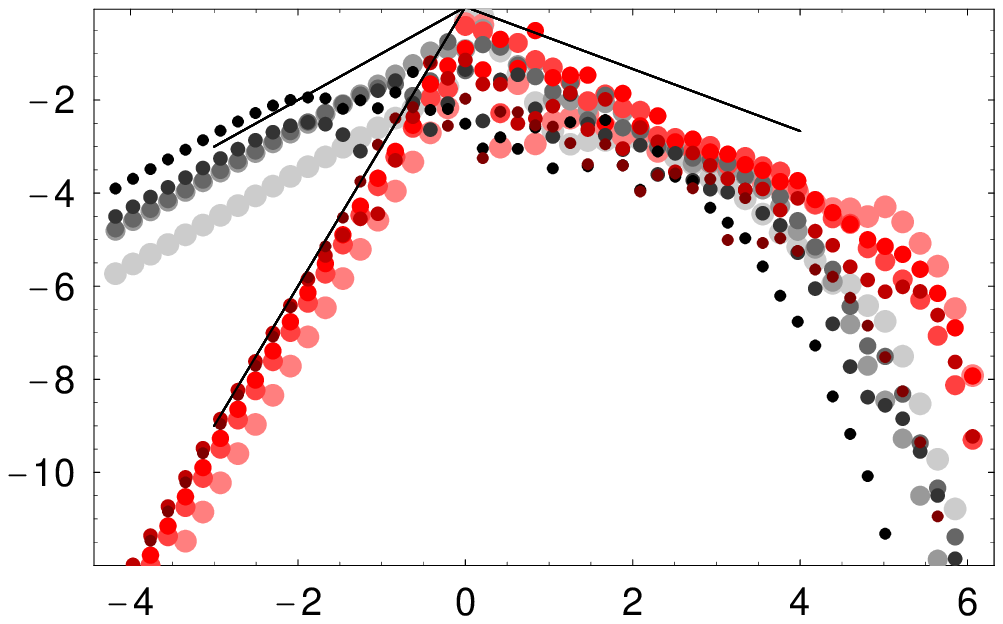}
        &
   \begin{turn}{90}$\quad \quad \quad \quad \log_{10}k$ \end{turn}
    \\*[0.cm]
    \begin{turn}{90}$\quad \quad \quad \quad  \alpha=-1$ \end{turn}&
    \includegraphics[width=0.28\textwidth,angle=90]{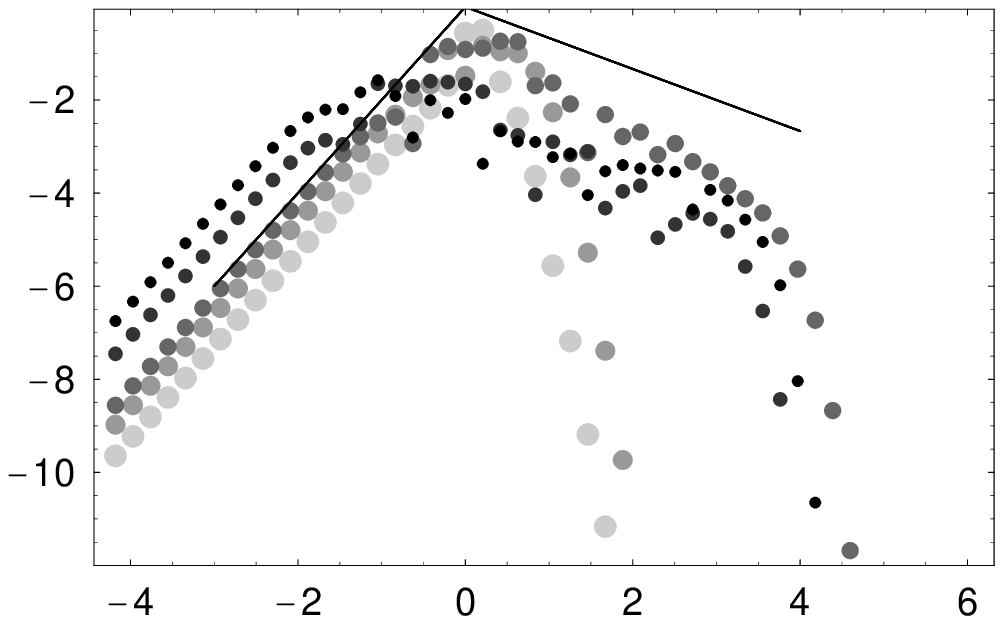}
        &
    \includegraphics[width=0.28\textwidth,angle=90]{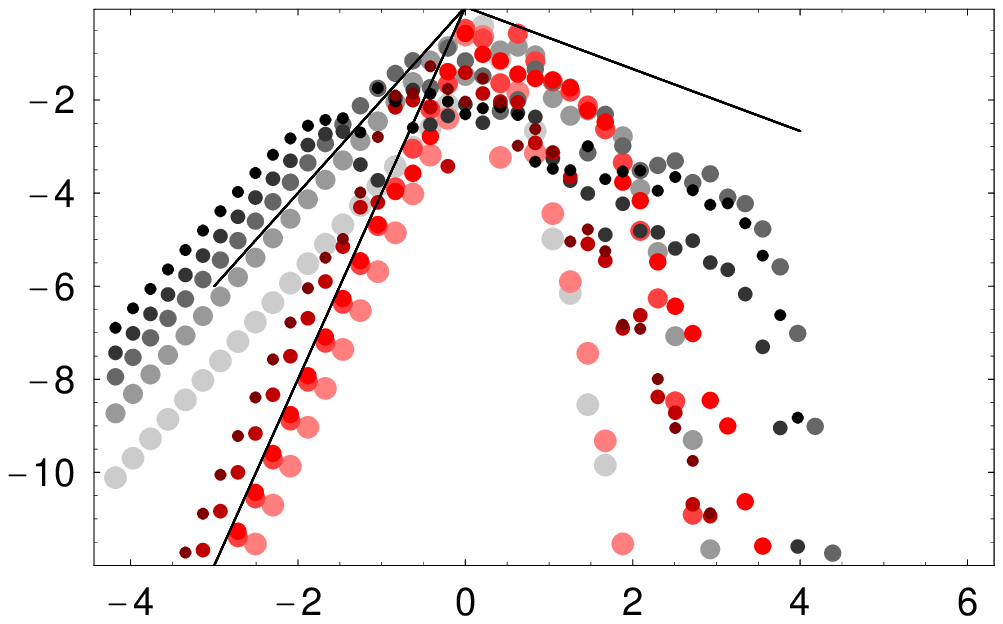}
        &
    \includegraphics[width=0.28\textwidth,angle=90]{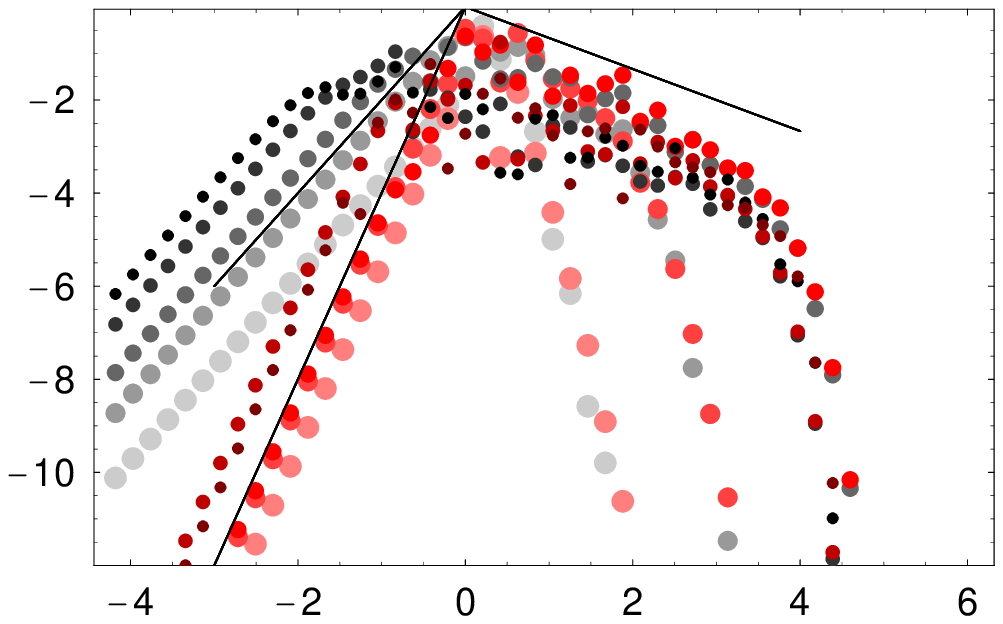}
        &
    \includegraphics[width=0.28\textwidth,angle=90]{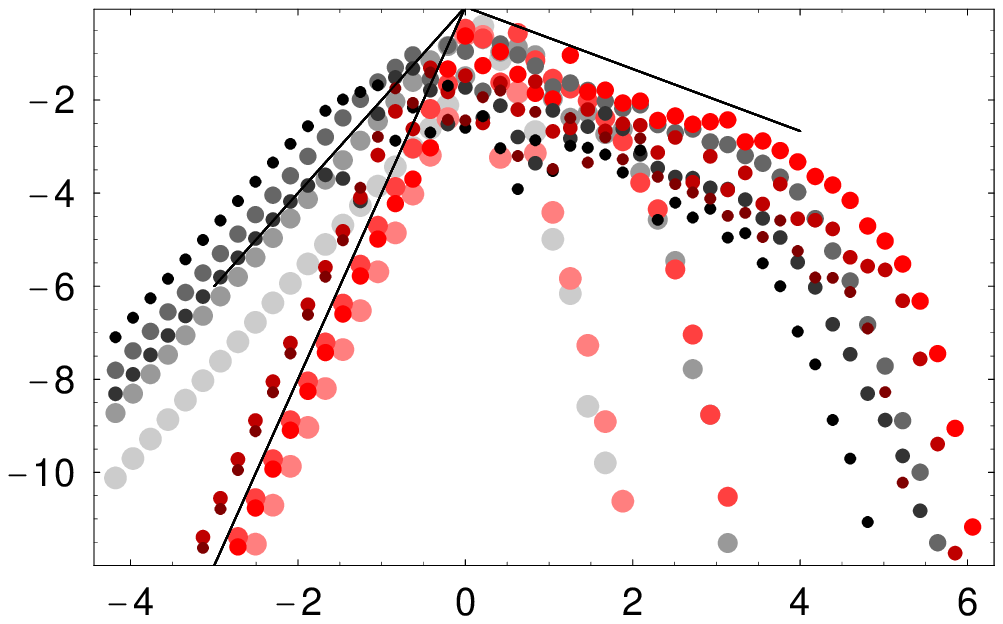}
            &
   \begin{turn}{90}$\quad \quad \quad \quad \log_{10}k$ \end{turn}
        \\*[0.cm]
        \begin{turn}{90}$\quad \quad \quad \quad  \alpha=-1.5$ \end{turn}&
    \includegraphics[width=0.28\textwidth,angle=90]{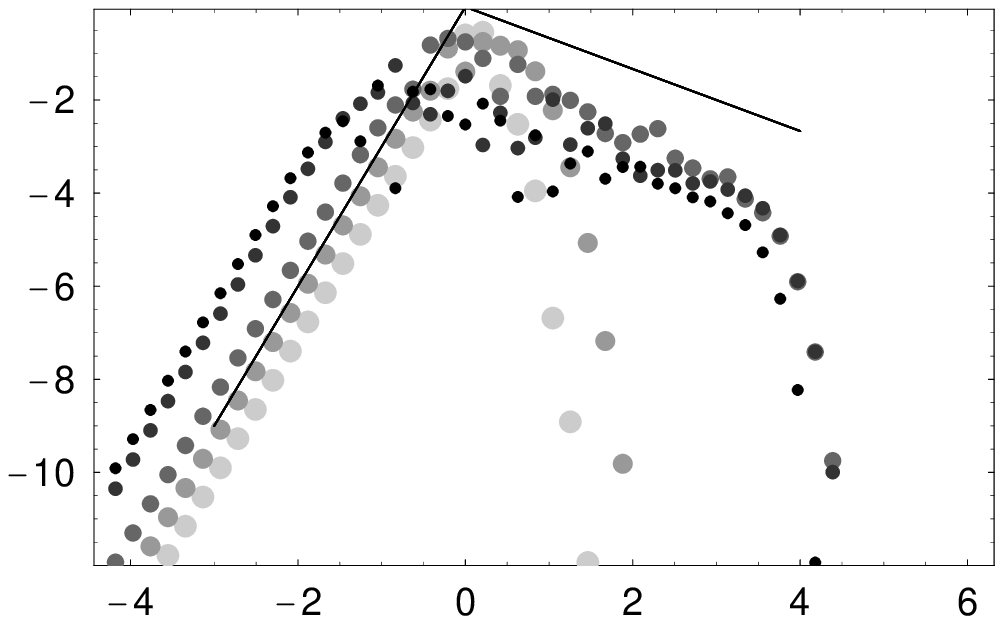}
        &
    \includegraphics[width=0.28\textwidth,angle=90]{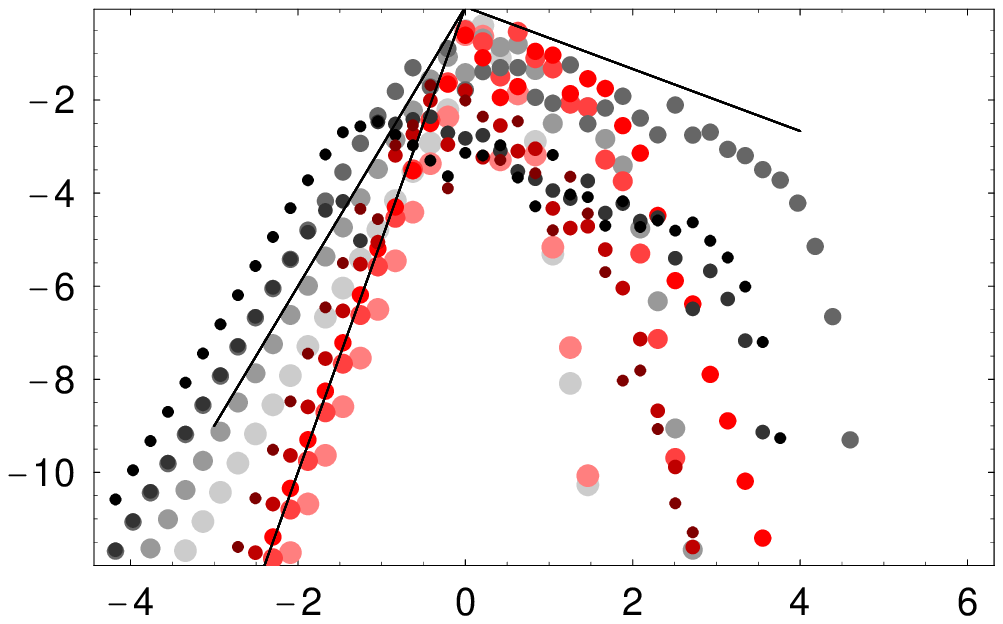}
        &
    \includegraphics[width=0.28\textwidth,angle=90]{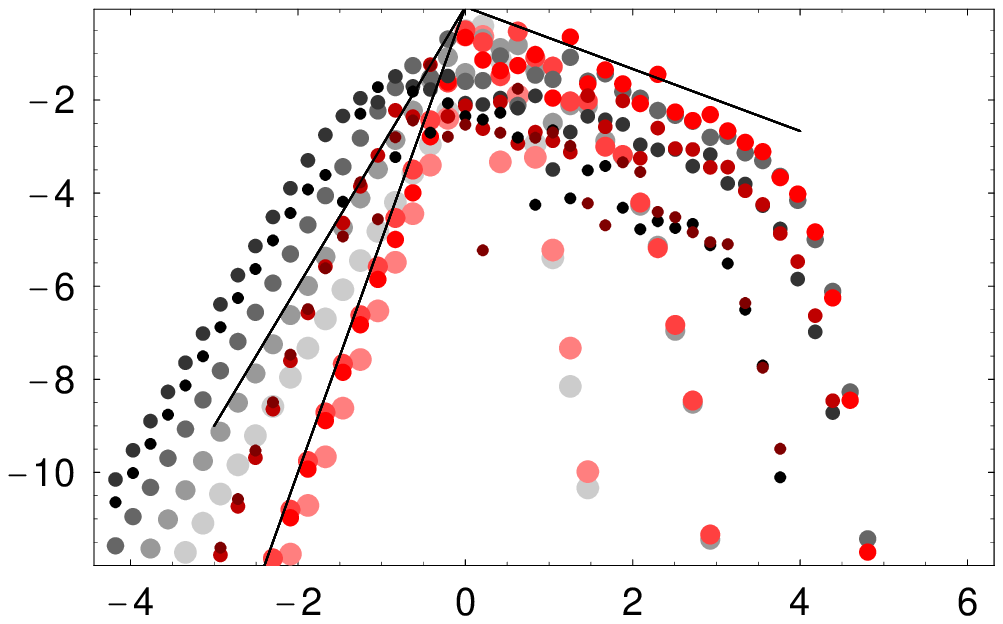}
        &
    \includegraphics[width=0.28\textwidth,angle=90]{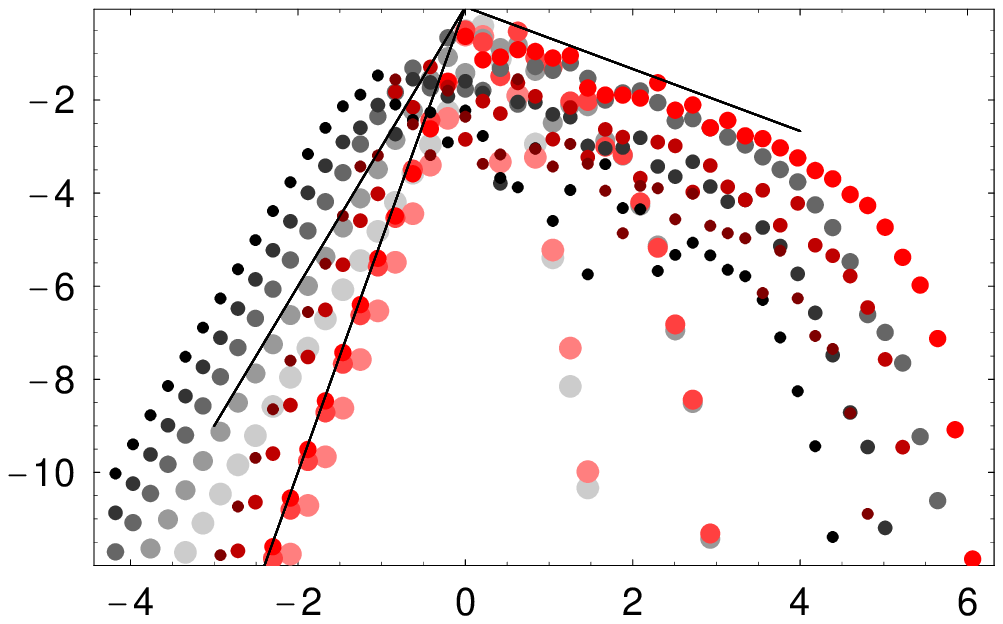}
            &
   \begin{turn}{90}$\quad \quad \quad \quad \log_{10}k$ \end{turn}
        \\*[0.cm]
        \begin{turn}{90}$\quad \quad \quad \quad \alpha=-2.5$ \end{turn}&
    \includegraphics[width=0.28\textwidth,angle=90]{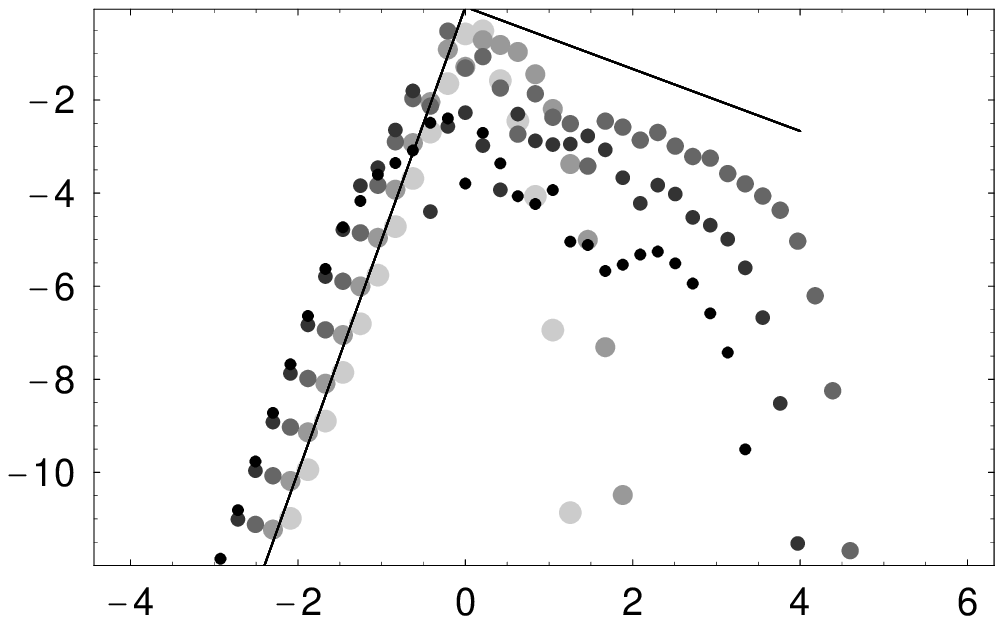}
        &
    \includegraphics[width=0.28\textwidth,angle=90]{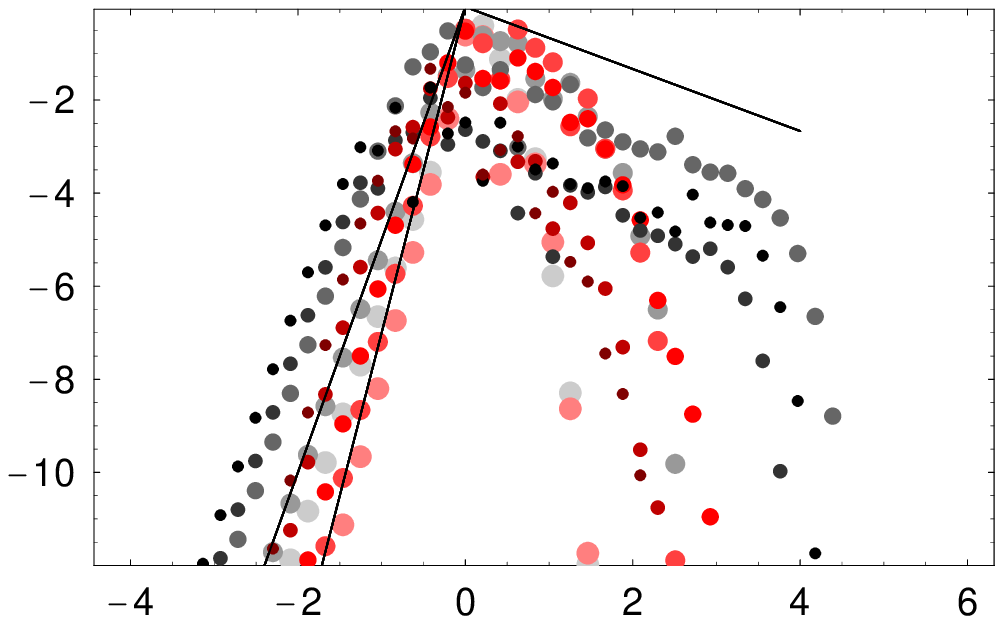}
        &
    \includegraphics[width=0.28\textwidth,angle=90]{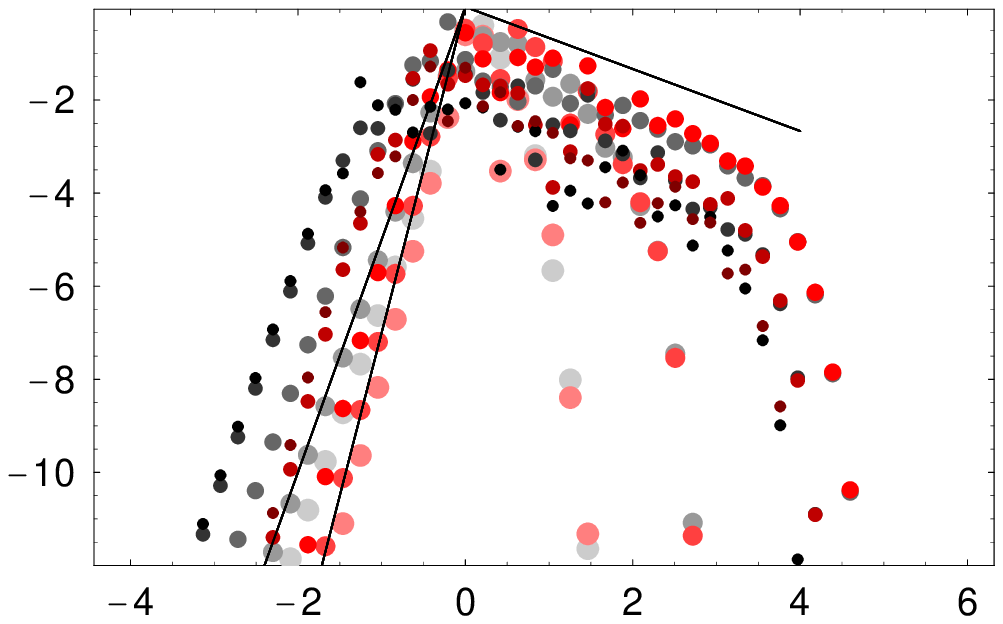}
        &
    \includegraphics[width=0.28\textwidth,angle=90]{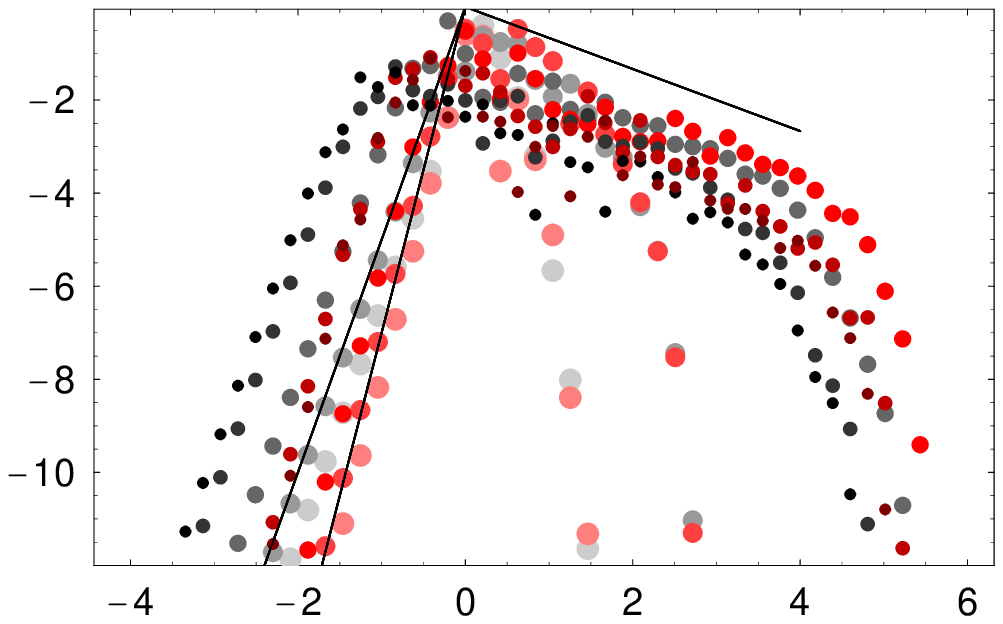}
            &
   \begin{turn}{90}$\quad \quad \quad \quad \log_{10}k$ \end{turn}
        \\*[0.cm]
        \begin{turn}{90}$\quad \quad \quad \quad \alpha=-\infty$ \end{turn}&
    \includegraphics[width=0.28\textwidth,angle=90]{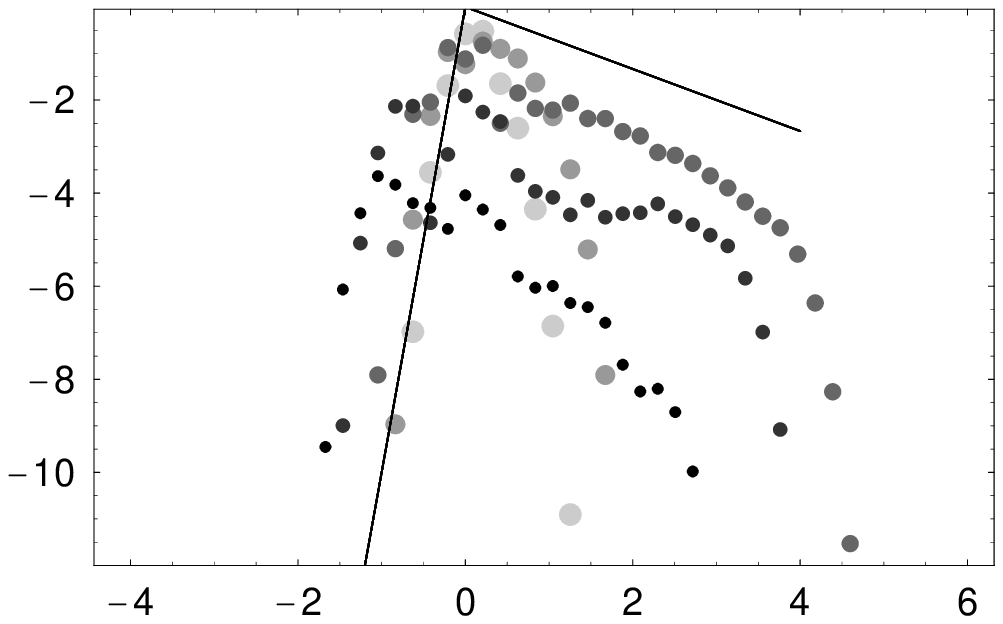}
        &
    \includegraphics[width=0.28\textwidth,angle=90]{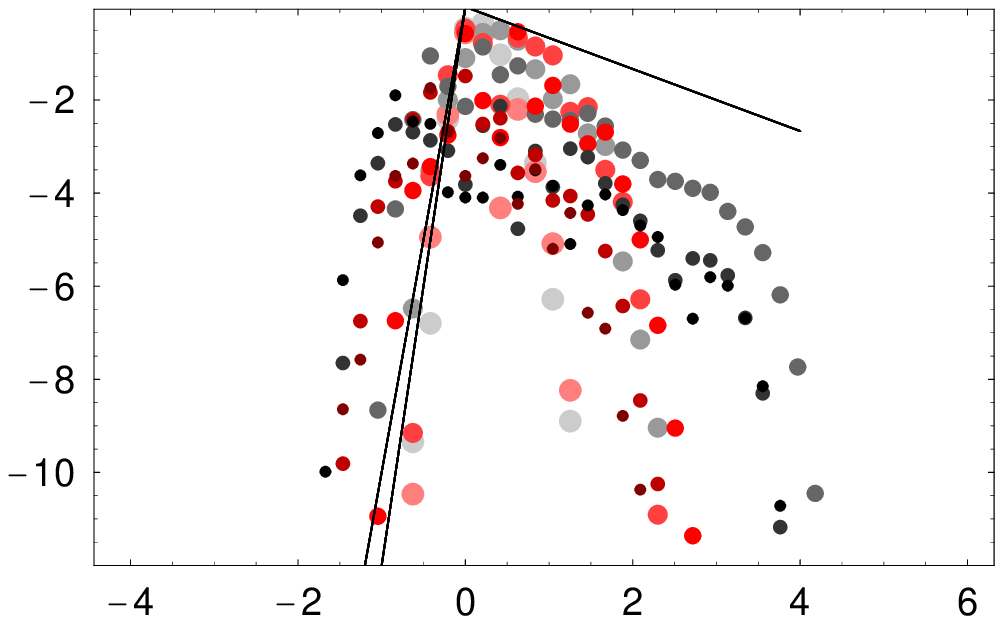}
        &
    \includegraphics[width=0.28\textwidth,angle=90]{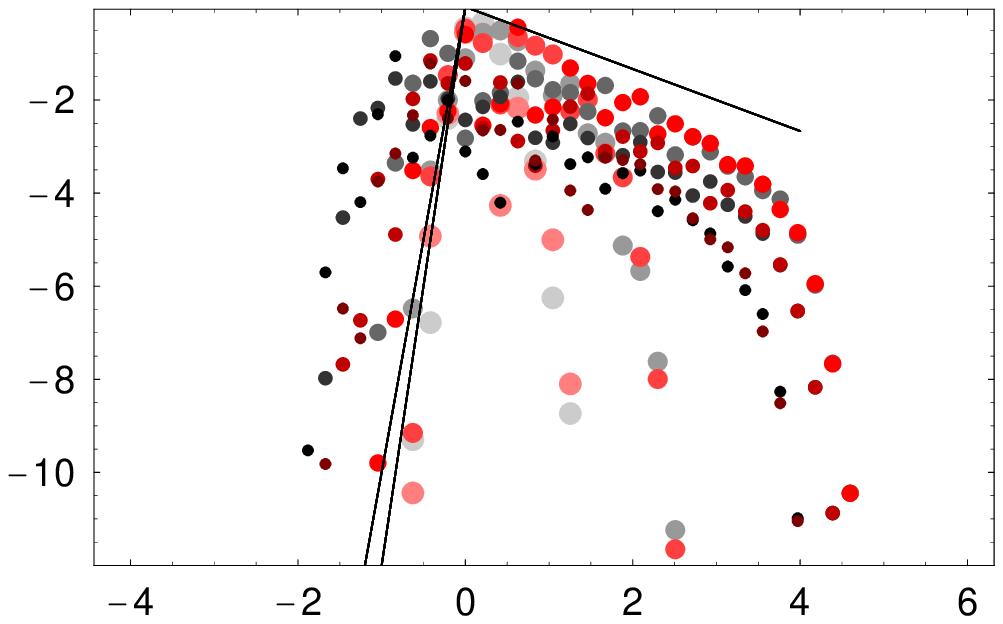}
        &
    \includegraphics[width=0.28\textwidth,angle=90]{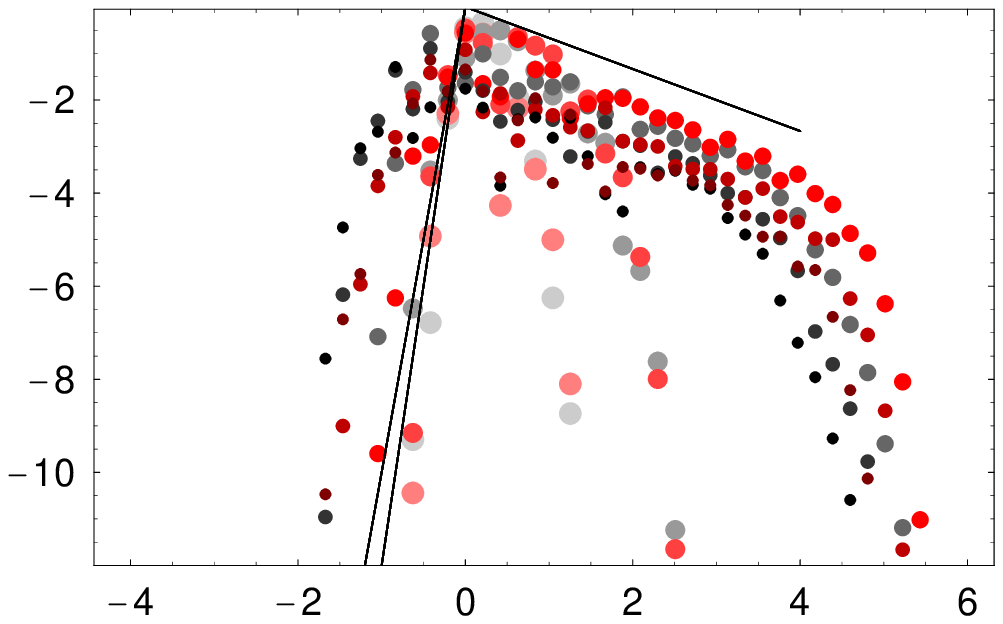}
            &
   \begin{turn}{90}$\quad \quad \quad \quad \log_{10}k$ \end{turn}
  \end{tabular}
  \end{center}
\caption{On top: free decaying turbulence for $R_e=10^{-6}$.
From second to fourth row: free decaying MHD turbulence for three values of $P_m$ ($R_e=10^{-6}$).
Each column corresponds to a given value of $\alpha$.} 
\label{freedecay}
\end{figure}

In this section we study free decaying turbulence without and with magnetic field.
The results are presented in Figure \ref{freedecay} for $\nu = 10^{-6}$, different values of $\alpha$ and $P_m$. 
The kinetic (magnetic) spectra are plotted with grey (red) dots at different times.
The initial conditions are such that the cross-helicity is close to zero at any scale. Trying the same simulation but with an initial condition with a cross-helicity equal to 1 leads to inertial ranges poorly defined.
The time sample at which the spectra are plotted is
$t=1, 2, 10, 100, 200$. The dots corresponding to $t=200$ are the darkest and the smallest. 

We observe that changing $\alpha$ does not change the slope of the inertial range
for both kinetic and magnetic spectra. This slope compares well with the Kolmogorov slope $k_n^{-2/3}$
($k^{-5/3}$ in spectral space) which is represented by the straight line with negative slope in each plot.
In the other hand, changing $\alpha$ changes drastically the slope at large scale. The kinetic and magnetic energy slopes are indicated by the two straight lines with positive slopes in each plot.

In table \ref{tab:slopes} these slopes are indicated for both spectra and for the different values of $\alpha$ that are considered.
\begin{table}[ht]
\begin{center}
\begin{tabular}{cc@{\hspace{2em}}c@{\hspace{2em}}c@{\hspace{2em}}c@{\hspace{2em}}c} \br
	$\alpha =$&$-\infty$&-2.5&-1.5&-1&-0.5\\ \mr
	Kinetic slope&10&5&3&2&1\\
	Magnetic slope&12&7&5&4&3\\ \br
\end{tabular}
\end{center}
\label{tab:slopes}
\caption{Slopes of large scales kinetic and magnetic spectra for several values of $\alpha$.}
\end{table}
From hydrodynamic turbulence theory, a slope in $k_n^5$ ($k^6$ in spectral space) is expected at large scales.
The only value of $\alpha$ which gives such slope is $\alpha = -5/2$.
The slope for the local Sabra model, corresponding here to $\alpha = -\infty$, leads to a slope 
in $k_n^{10}$ ($k^9$ in spectral space) much larger than the one predicted by the theory.
This is a first drastic difference between the local and non local models which emphasizes the
importance of including the non local interactions.

We note that there is always a difference of 2 between the magnetic and kinetic slopes,
leading to a magnetic spectrum slope in $k_n^7$ ($k^8$ in spectral space) for $\alpha = -5/2$.
Though $\alpha = -5/2$ seems to be the most appropriate for hydrodynamic and MHD turbulence, in the rest of the paper we shall investigate other values of $\alpha$ as well, in order to investigate the role of $\alpha$ in non local transfers.

\subsubsection{Forced turbulence}
In forced turbulence we consider shells larger than 0 ($k_n \ge 1$).
We could also have considered a cut-off at any other arbitrary negative shells. However, in order to reach a stationary state it is important to have a cut off scale above which the system is not solved (like some integral scale or the scale of the box in an experiment). Considering negative shells without limit would imply energy to fill all shells on a time scale of the order $k_n^{-1}$ and without reaching a stationary state.
 
\begin{figure}[ht]
\begin{center}
  \begin{tabular}{@{}c@{\hspace{0em}}c@{\hspace{0em}}c@{\hspace{0em}}c@{}}
$P_m=10^{-3}$ &$P_m=1$ &$P_m=10^5$ &\\*[0.cm]  
    \includegraphics[width=0.3\textwidth]{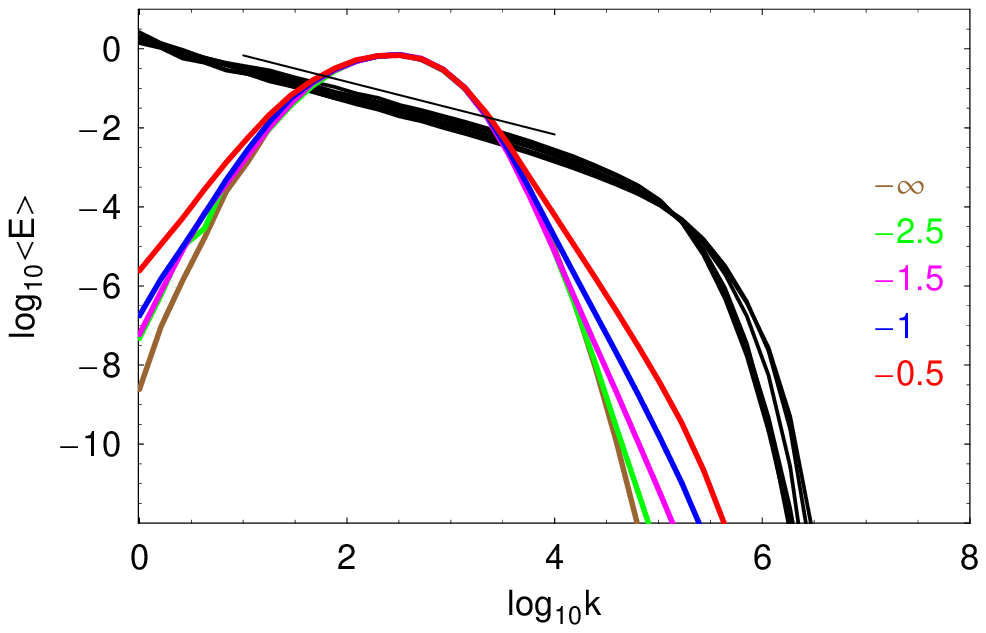}
    &
    \includegraphics[width=0.3\textwidth]{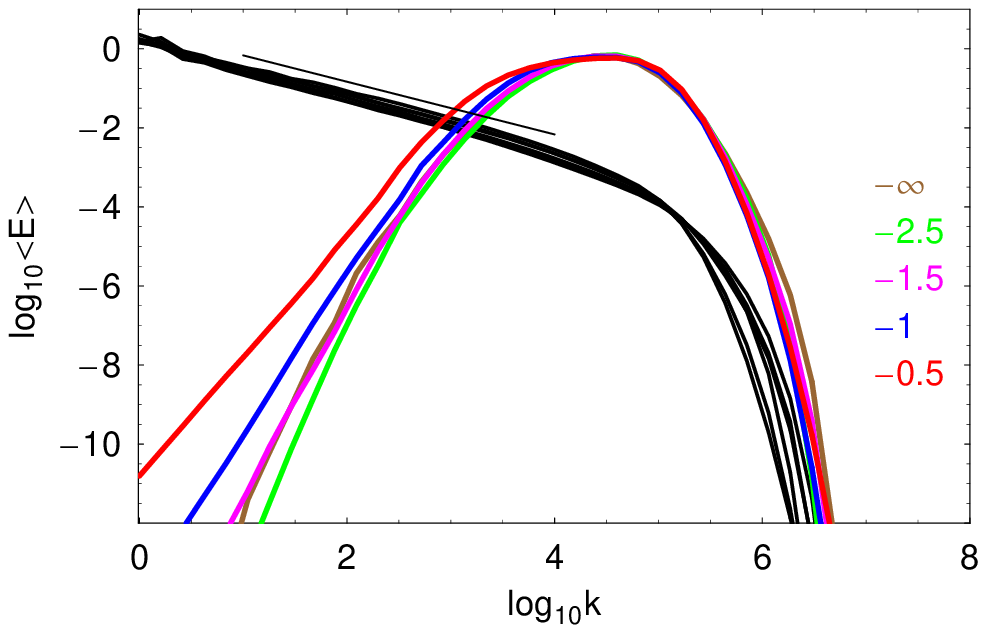}
    &
    \includegraphics[width=0.3\textwidth]{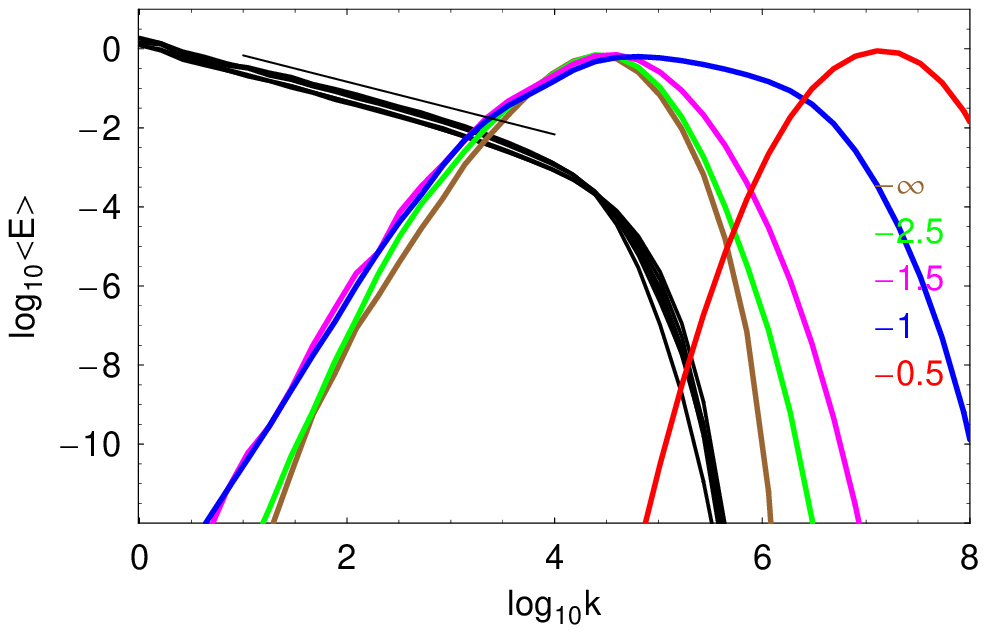}
    & \raisebox{2.5cm}{\begin{turn}{270} Kinematic \end{turn}}
    \\*[0.cm]
    \includegraphics[width=0.3\textwidth]{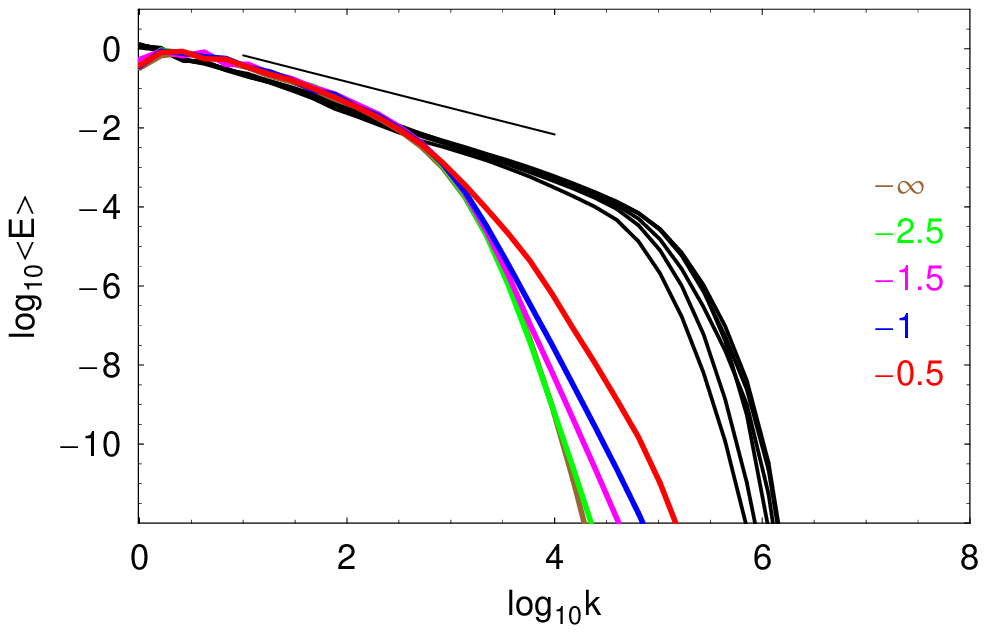}
    &
    \includegraphics[width=0.3\textwidth]{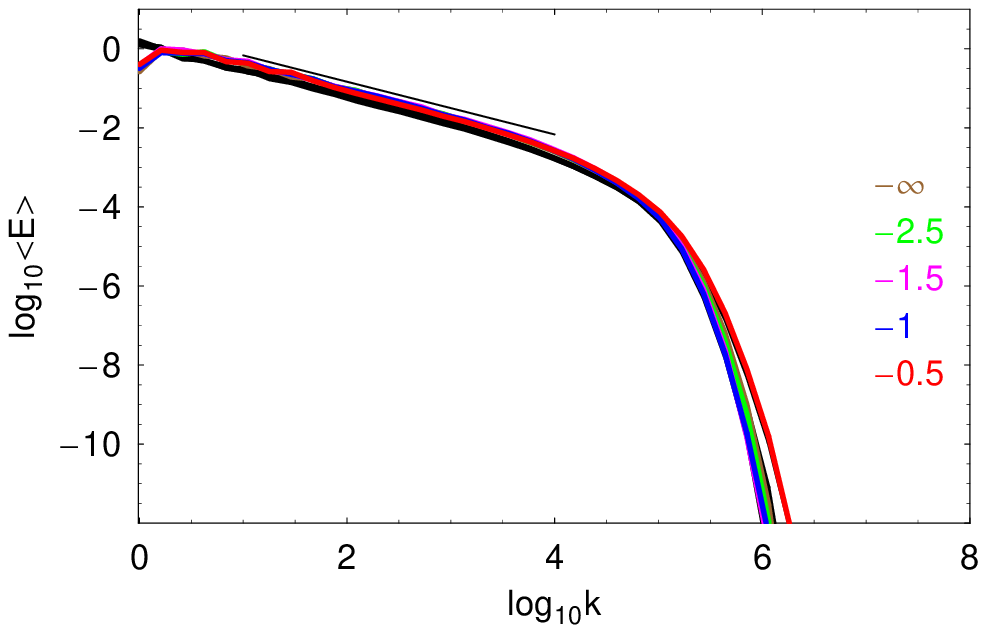}
    &
    \includegraphics[width=0.3\textwidth]{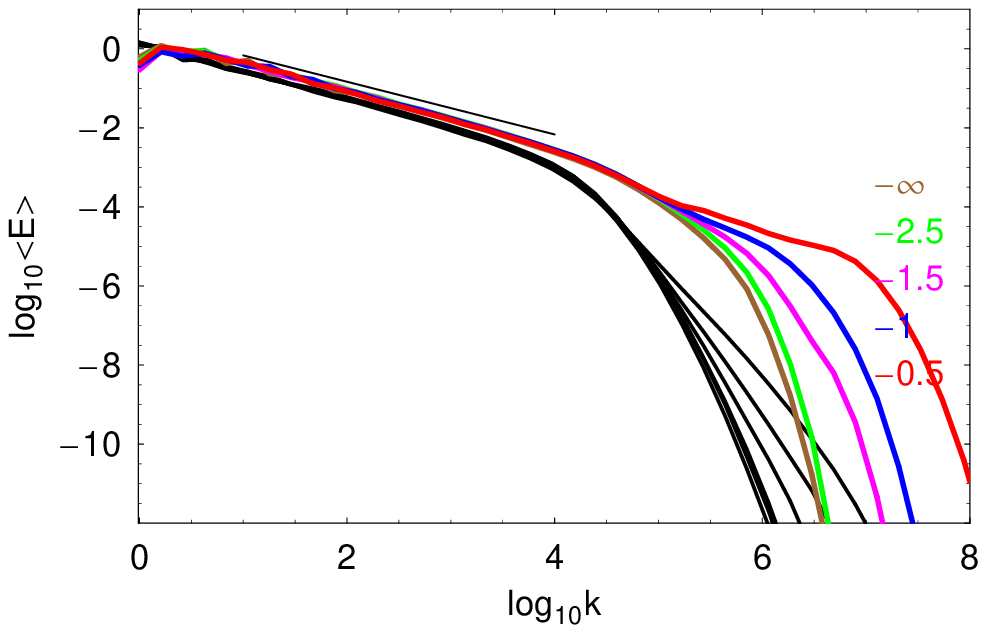}
    & \raisebox{2.5cm}{\begin{turn}{270} Dynamic \end{turn}}
    \\*[0.cm]
    \end{tabular}
    \end{center}
\caption{Spectra of kinetic (black curves) and magnetic (colour curves) energy 
for three values of $P_m$ (from left to right $P_m = 10^{-3}; 1; 10^{5}$) and several values of $\alpha$
(indicated by the labels).
The spectra on top (resp. bottom) correspond to the kinematic (resp. dynamic) regime.} 
\label{SpectraKD}
\end{figure}

In Figure \ref{SpectraKD} both kinetic and magnetic energy spectra are plotted for three values of $P_m $
and both kinematic and saturated dynamo regimes. In each plot, the curves correspond to different values of $\alpha$. In the kinematic regime (top plots) the Lorentz forces corresponding to the term $Q_n(B,B,-a)$ are small and can be neglected in the energy balance. Then the magnetic energy grows exponentially at any scale. In Figure \ref{SpectraKD} the magnetic spectra are normalized by the maximum value of $E_B(n)$.
In the saturated regime (bottom plots), the Lorentz forces act back onto the flow, leading to statistically stationary kinetic and magnetic energies.\\

In both kinematic and dynamic regimes and for $P_m\le 1$ (left and middle plots of Figure \ref{SpectraKD}),
the effect of $\alpha$ is not really significant. 
The kinetic spectra (black curves) are almost not sensitive to non local interactions showing that hydrodynamic interactions are mostly local as predicted by the Kolmogorov cascade. Besides the kinetic spectra always scale as $k_n^{-2/3}$ ($k^{-5/3}$ in the spectral space) in the inertial range.
We see some non local effects onto the magnetic spectra
(colour curves) which spread towards large scales or even small scales for $P_m=10^{-3}$.
However, the scale for which
the spectrum is maximum does not change and is roughly equal to $k_{\eta} \approx k_{\nu} P_m^{3/4}$ as again predicted by Kolmogorov arguments (see e.g. \cite{Stepanov06}). \\

In the other hand, the non local effect are much more significant for $P_m > 1$ (plots on the right), mainly for
the magnetic spectra at scales smaller than the viscous scales ($k_n \ge k_{\nu}$) and for $\alpha > -1$. 
In the kinematic regime the maximum of the magnetic energy spectrum occurs at scales smaller than $k_{\nu} P_m^{3/4}$ when  $\alpha > -1$. In the dynamic regime some magnetic bottle neck appears.
To understand why it is so, let us first recall that the flow scale which produces magnetic field in the most efficient way is the one for which the shear is the largest \cite{Stepanov06}. In the inertial range the flow shear scales as $k_n u_n \propto k_n^{2/3}$, and it is then maximum for $k_n \approx k_{\nu}$. Therefore the non local interactions relevant for the magnetic spectrum  extension towards smaller scales
are mostly those involving $U_{\nu}$. The non local terms involving $U_{\nu}$ and generating magnetic energy $E_B(n)$ with $n \gg \nu$ involves also $B_{n\pm 1}$. The corresponding non local term in (\ref{eq_b}) 
are in the form $k_n(T_{n-\nu} b^2_{n-\nu}U_{\nu}^*B_{n+1} + T_{n-\nu-1} b^3_{n-\nu-1}U_{\nu}^*B_{n-1})$
which scale as $k_n^{1+\alpha}$. Therefore we understand that for $\alpha +1 > 0$ the non local effect may be strong at small scales.  
\subsection{Energy fluxes}
\begin{figure}[ht]
\begin{center}
  \begin{tabular}{@{}c@{\hspace{0em}}c@{\hspace{0em}}c@{\hspace{0em}}c@{}}
$P_m=10^{-3}$ &$P_m=1$ &$P_m=10^4$ & \\*[0.cm]  
    
    \includegraphics[width=0.3\textwidth]{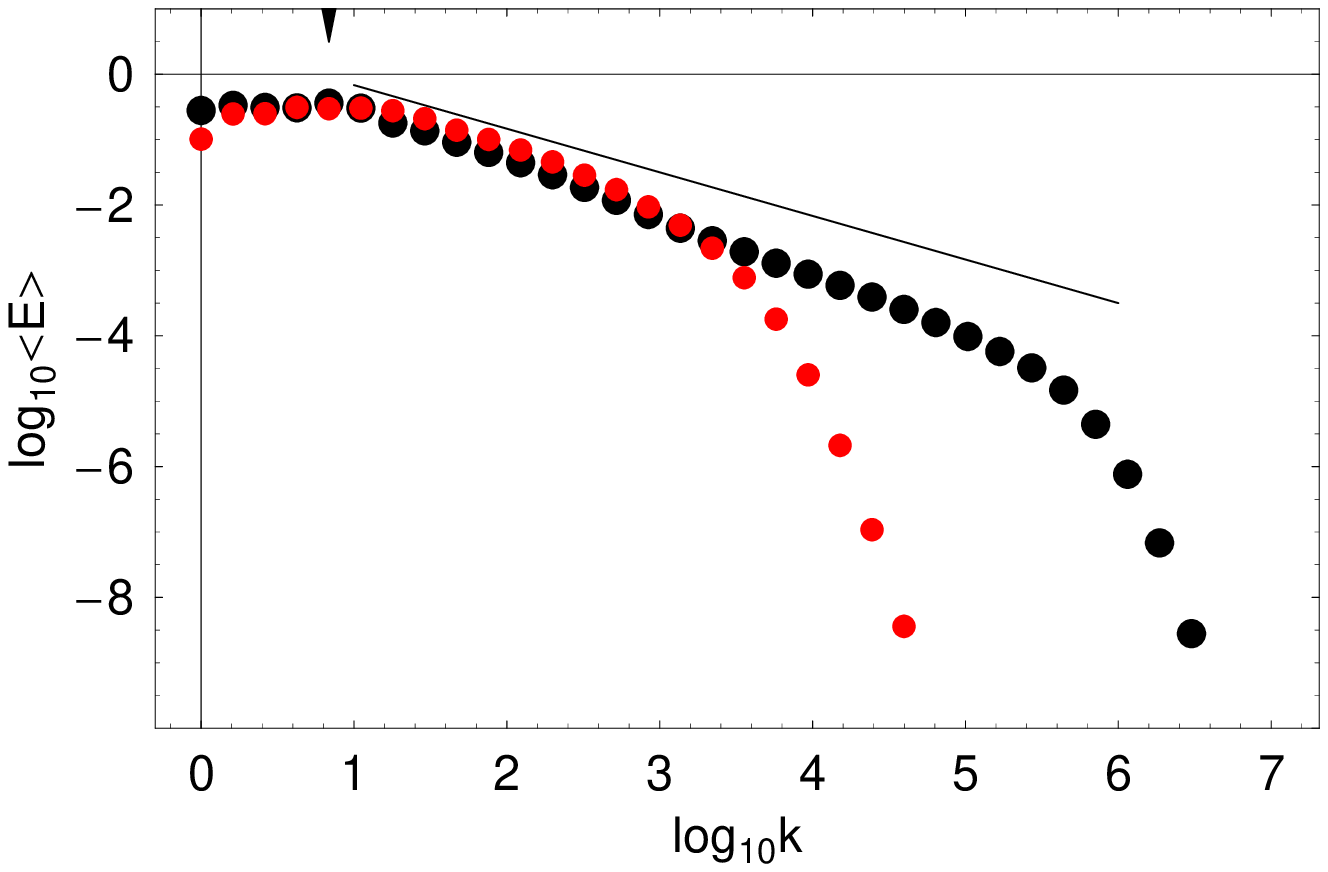}
    &
    \includegraphics[width=0.3\textwidth]{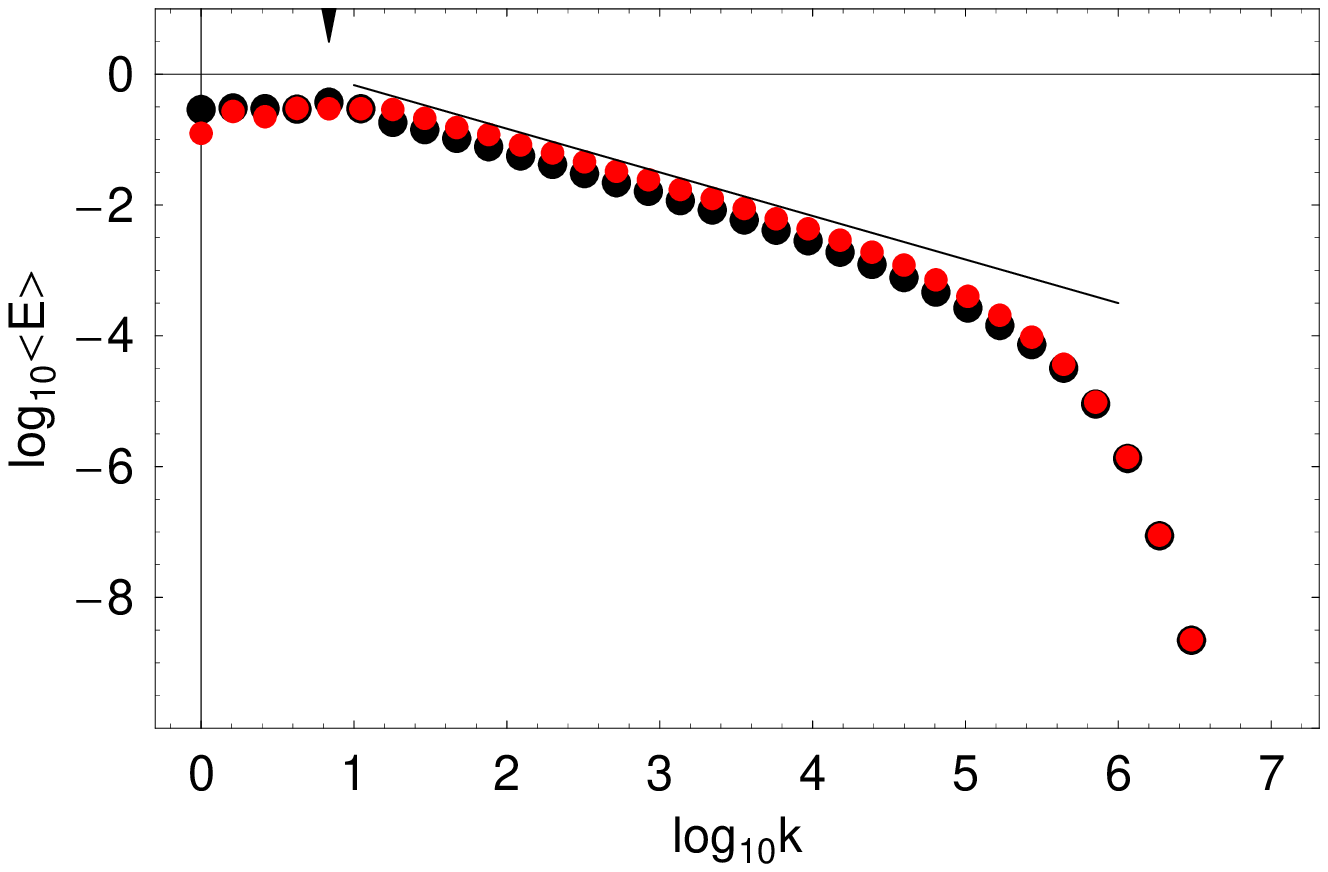}
    &
    \includegraphics[width=0.3\textwidth]{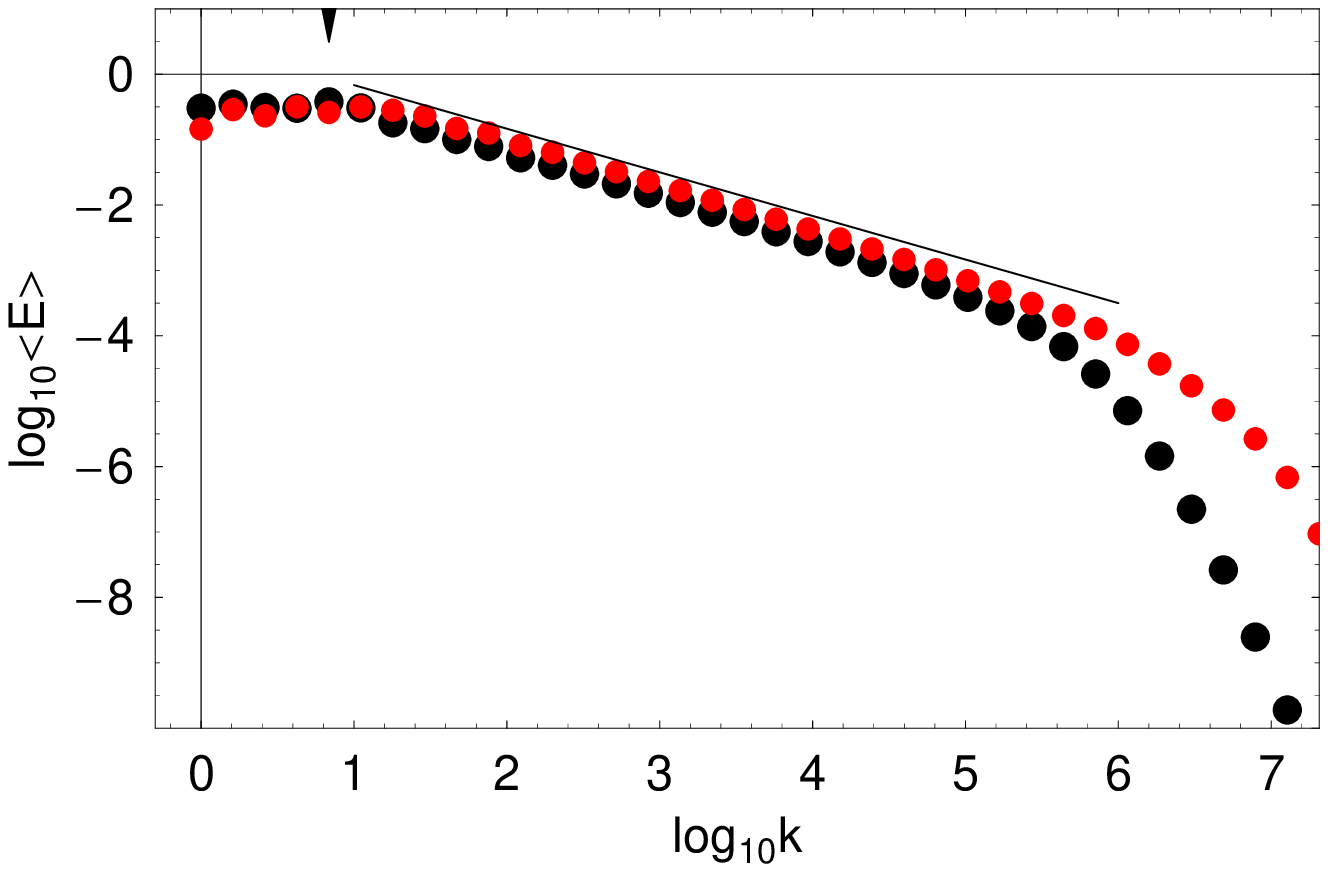}
    & \raisebox{2.5cm}{\begin{turn}{270} Spectra \end{turn}} 
    \\*[0.cm]
    \includegraphics[width=0.3\textwidth]{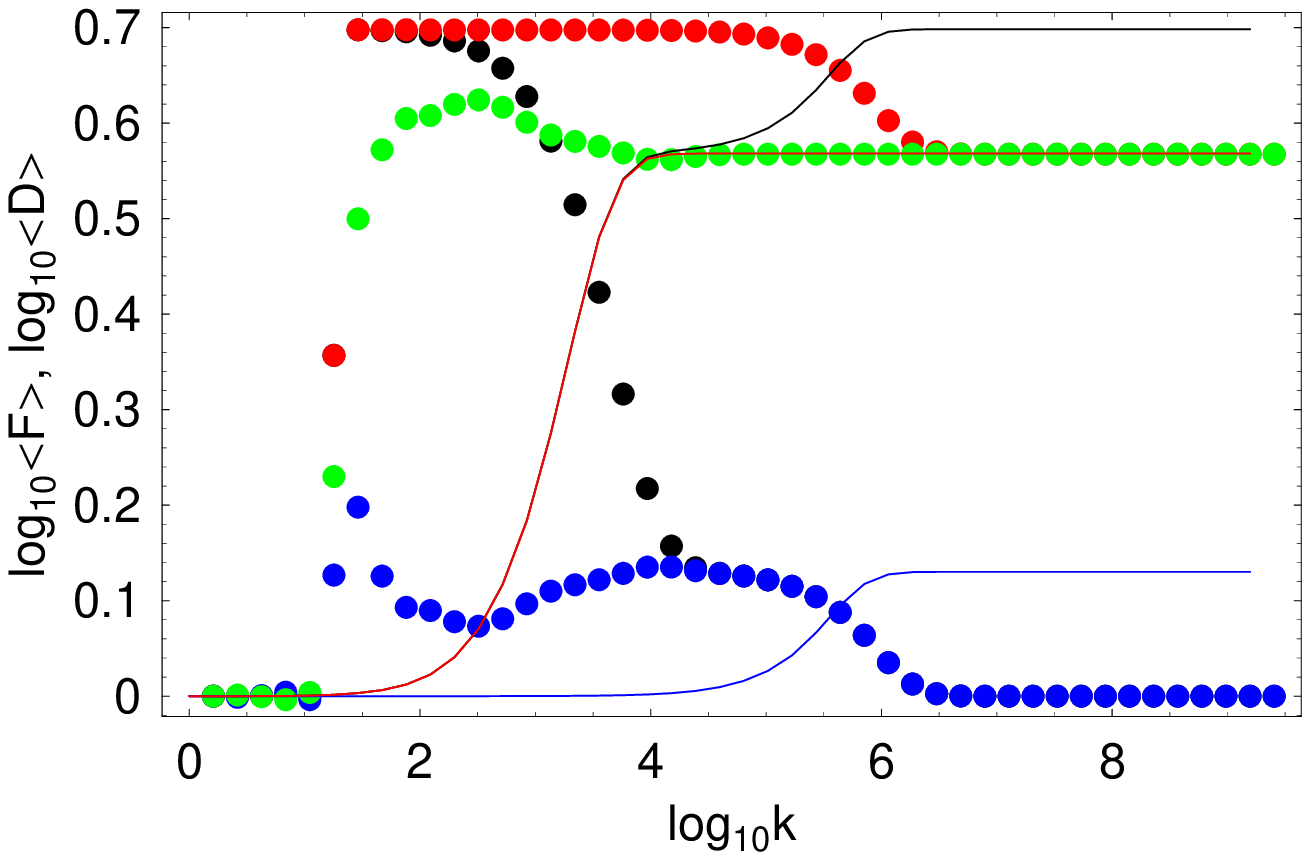}
    &
    \includegraphics[width=0.3\textwidth]{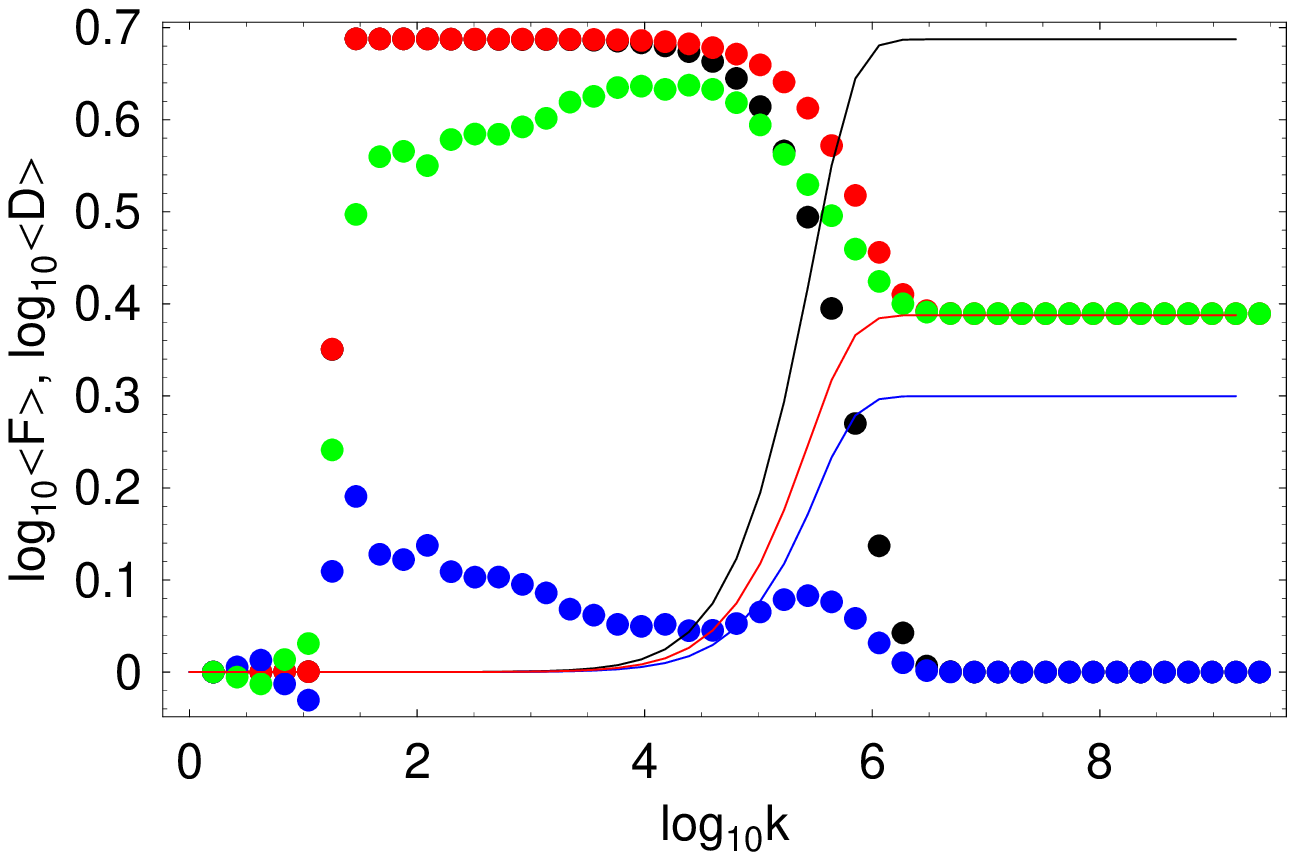}
    &
    \includegraphics[width=0.3\textwidth]{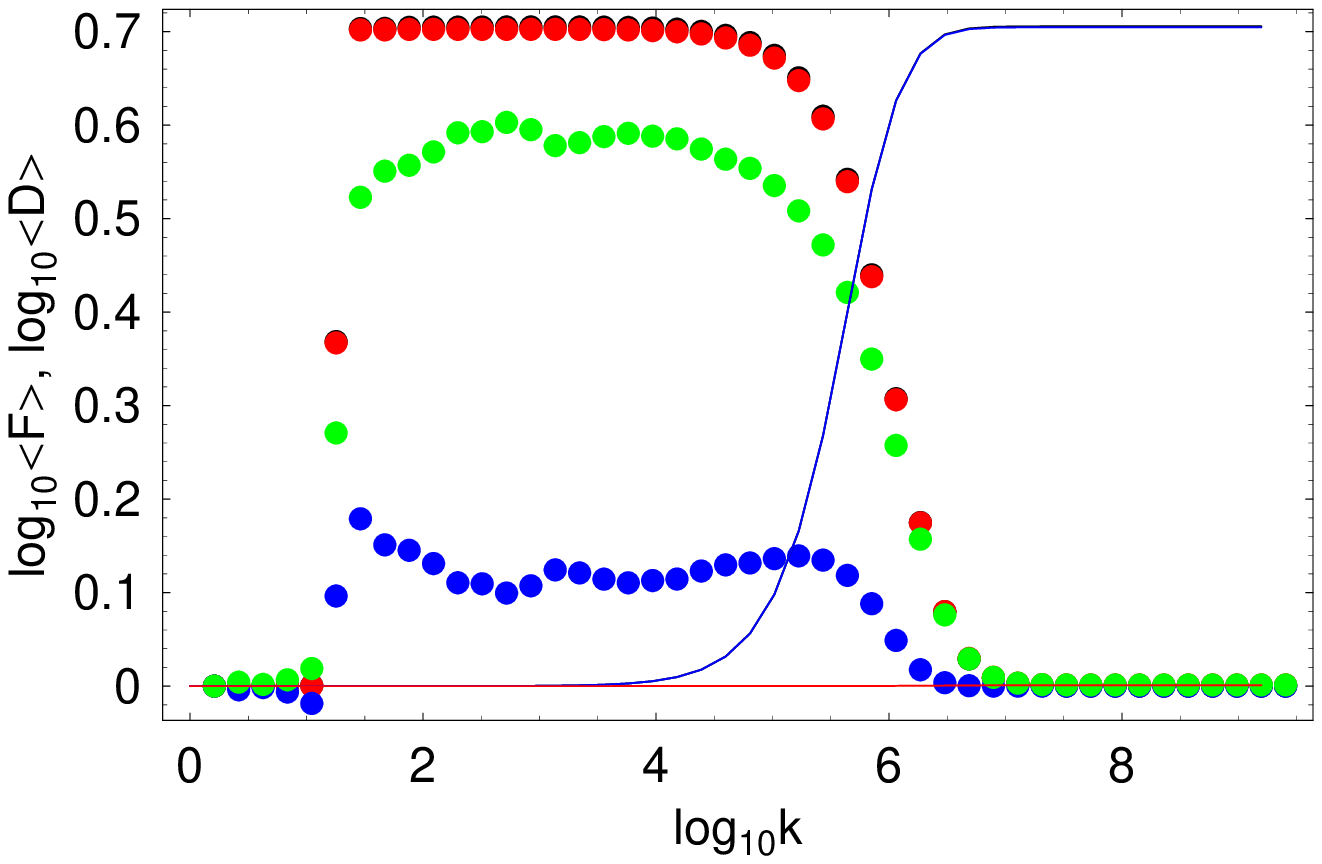}
    & \raisebox{2.5cm}{\begin{turn}{270} Fluxes \end{turn}} 
    \\*[0.cm]
    \includegraphics[width=0.28\textwidth]{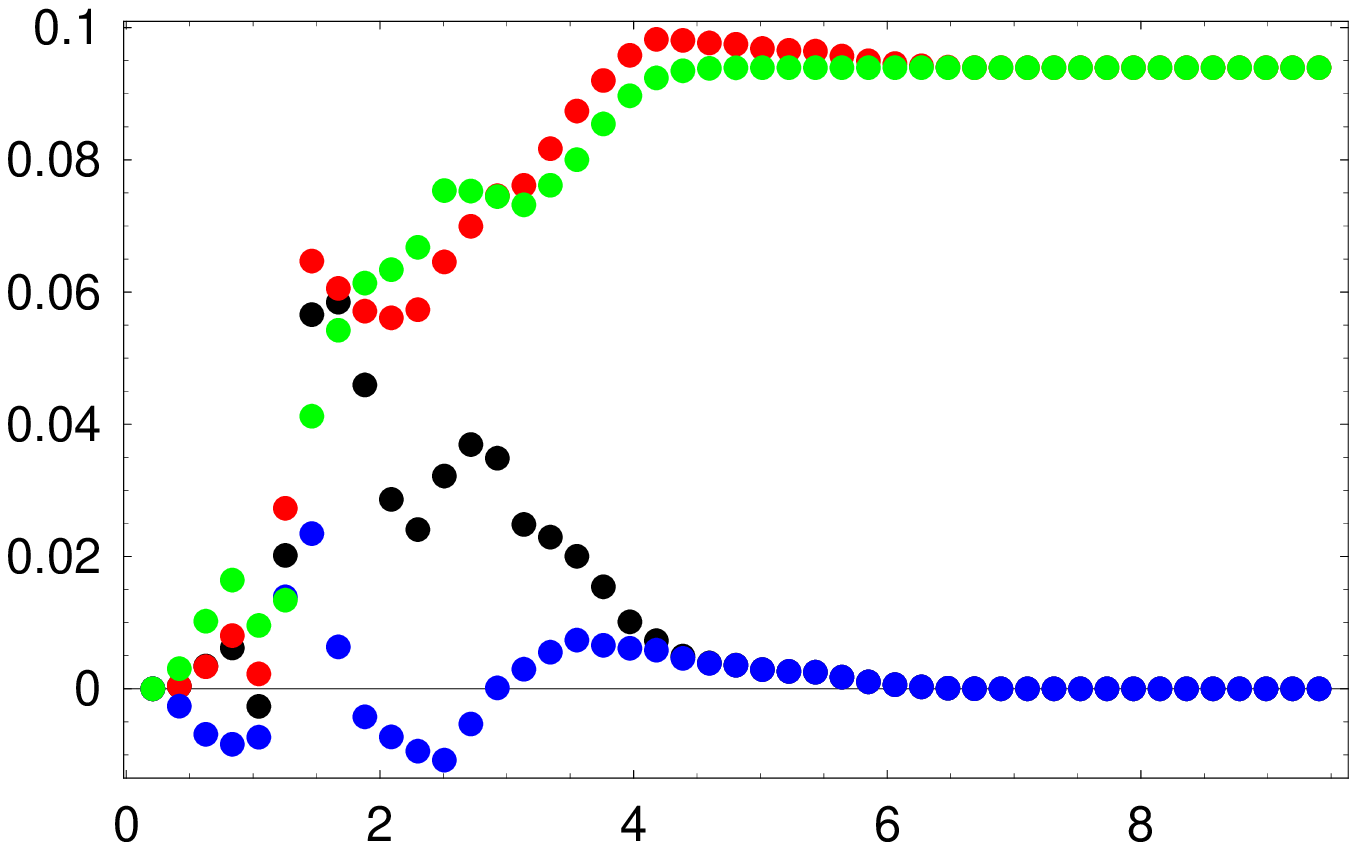}
    &
    \includegraphics[width=0.28\textwidth]{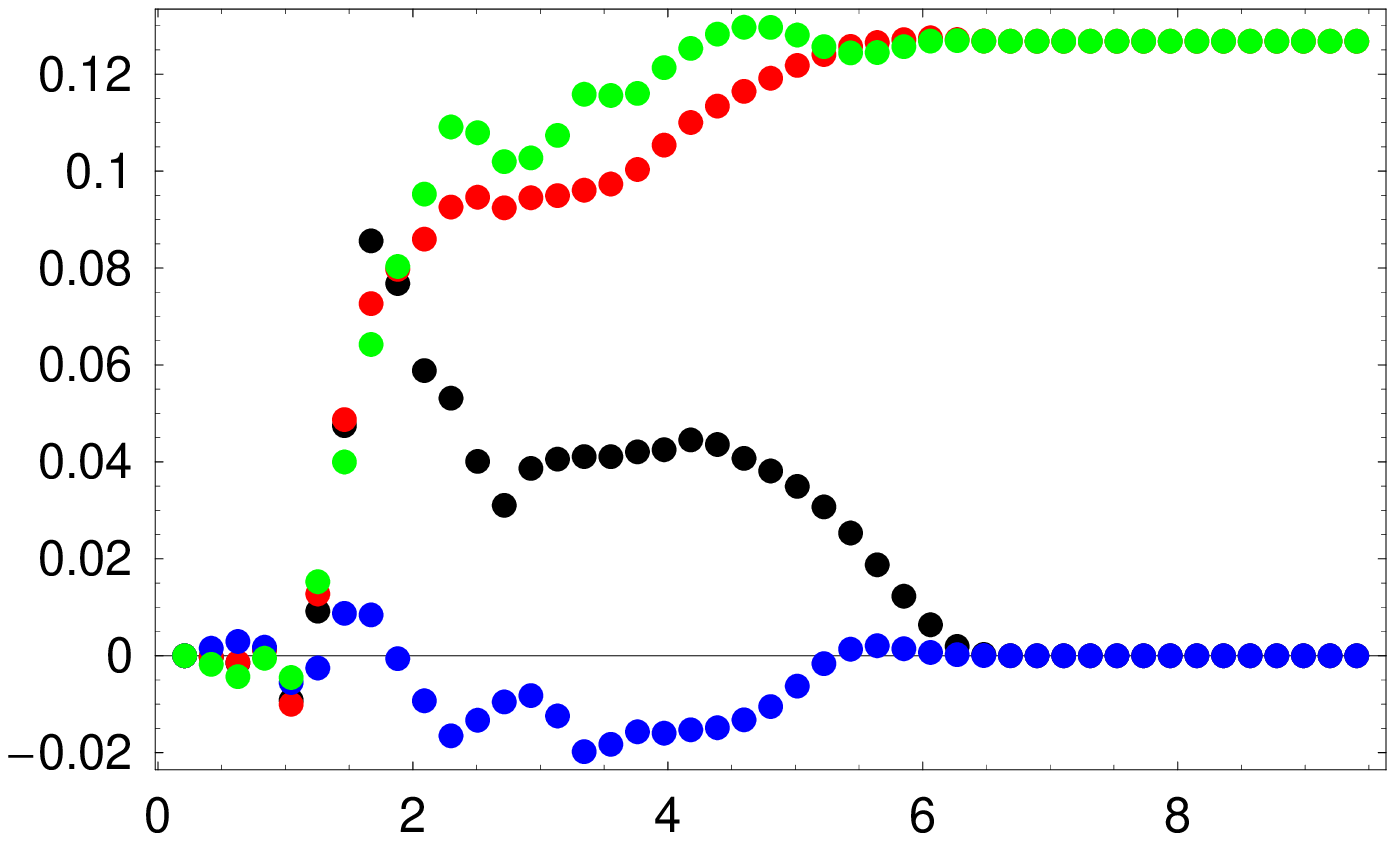}
    &
    \includegraphics[width=0.28\textwidth]{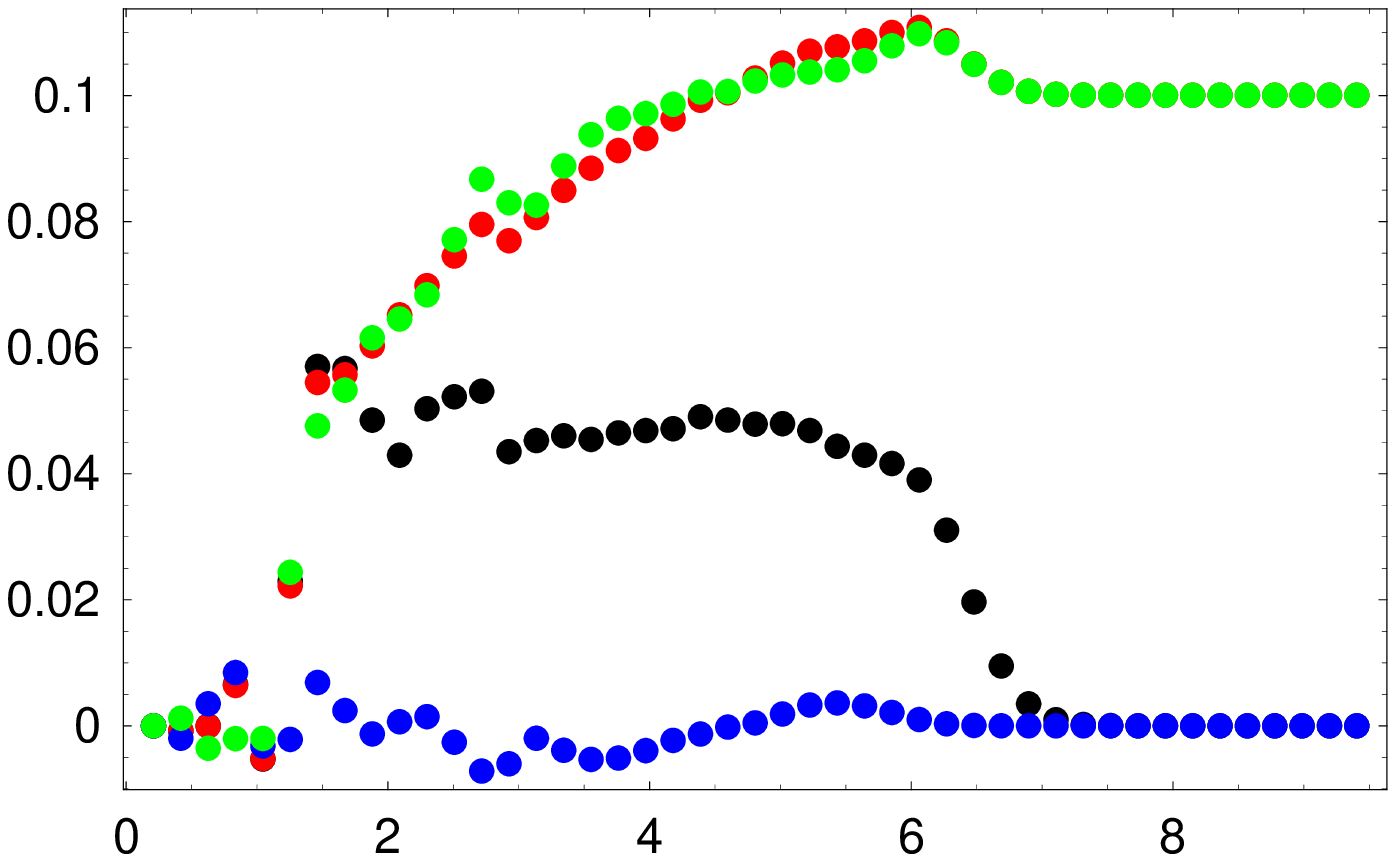}
    & \raisebox{3cm}{\begin{turn}{270} Non local fluxes \end{turn}} 
    \\*[0cm]  
    $\log_{10}k$
    &
    $\log_{10}k$
    &
    $\log_{10}k$
    \end{tabular}
    \end{center}
\caption{Spectra, total and non local fluxes for three values of $P_m$ and $\alpha=-5/2$. The kinetic (magnetic) spectra corresponds to black (red) dots. 
The fluxes $\Pi_{UU}(n)$ and $\Pi_{BU}(n)$ are represented by blue and green dots.
The red dots correspond to $\Pi_{UU}(n) + \Pi_{BU}(n)$,
the black dots to $\Pi_{UU}(n) + \Pi_{BU}(n) + \Pi_{UB}(n) + \Pi_{BB}(n)$.
The  blue, red and black full lines correspond respectively to $\sum_{j=0}^n {\cal D}_U(j)$, $\sum_{j=0}^n {\cal D}_B(j)$ and $\sum_{j=0}^n \left({\cal D}_U(j)+{\cal D}_B(j)\right)$.
} 
\label{fluxes}
\end{figure}

In this section we set $\alpha=-5/2$ and $\nu=10^{-8}$ and consider the dynamo saturated regime for three values of $P_m=10^{-3}, 1$ and $10^{4}$. The kinetic and magnetic spectra 
are plotted in the top row of figure \ref{fluxes}. Both spectra have inertial ranges of Kolmogorov types, scaling in $k_n^{-2/3}$ (scaling in $k^{-5/3}$ in the spectral space).
For $P_m=10^{-3}$ we identify clearly that the magnetic dissipation scale is much smaller than 
the viscous dissipation scale. However the distinction between them is not so clear  for  $P_m=10^{4}$.

The total and non local part of the fluxes are plotted in middle and bottom rows of figure \ref{fluxes}. The non local part of $\Pi_{UU}(n)$ is found to be always much smaller than $\Pi_{UU}(n)$, implying that the energy transfers are mainly local. In the other hand, the importance of the non local part of $\Pi_{BU}(n)$ versus the local one depends on $P_m$. 

\begin{figure}[ht]
\begin{center}
   \includegraphics[width=0.5\textwidth]{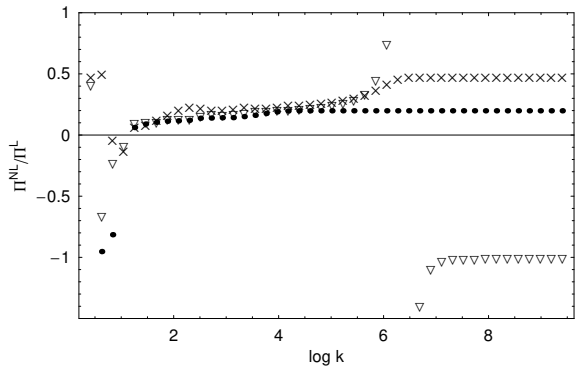}
\end{center}
\caption{Ratio $\Pi_{BU}^{Non Local}(n) / \Pi_{BU}^{Local}(n)$ for $P_m$ = $10^{-3}$ (dots), $1$ (cross) and $10^4$ (triangle).} 
\label{fluxratio}
\end{figure}

In figure \ref{fluxratio} the ratio $\Pi_{BU}^{Non Local}(n) / \Pi_{BU}^{Local}(n)$ is plotted for the three values of $P_m$. 
For $P_m = 10^{-3}$ this ratio is about 20 \%. For $P_m=1$ and
for scales smaller than the viscous scale $k_{\nu}\approx 10^6$, this ratio increases up to 50\%. Finally, for $P_m = 10^4$ there is a discontinuity at $k_{\nu}$, the ratio being then equal to -100\% at smaller scales. These are first evidences of non local interactions.
\newpage
\subsection{Energy transfers}
The four energy transfers are plotted in Appendix \ref{Transfersresults} in figures \ref{transfer-3},
\ref{transfer0} and \ref{transfer5} for respectively $P_m=10^{-3}, 1$ and $10^{5}$.
In each figure the column from left to right corresponds to $\alpha =
-\infty, -5/2, -3/2, -1$ and $-1/2$. The row from second to bottom corresponds to
the transfer ${\cal T}_{UU}(q,n)$, ${\cal T}_{BU}(q,n)$, ${\cal T}_{UB}(q,n)$ and ${\cal T}_{BB}(q,n)$.
The transfers are plotted versus $\log_{10} q$, for three values of $n$ which are indicated
by the dashed vertical lines on the spectra plots on top row and by the red, green or blue dots. The transfers being time-dependent we plot their time-average with error bars corresponding to the standard deviation of the mean. This gives some estimation of the robustness of the results. Some quantities are much more noisy than the others and then less reliable.
The local transfers seem to be always dominant whatever the values of $P_m$ or $\alpha$. 
However there are also some evidence of non local transfers which are discussed below.\\

\begin{figure}
\begin{center}
  \begin{tabular}{@{}c@{\hspace{0em}}c@{\hspace{0em}}c@{\hspace{0em}}c@{\hspace{0em}}}
  &$P_m = 10^{-3}$ & $P_m = 1$ & $P_m = 10^{5}$ \\
  \raisebox{2cm}{${\cal T}_{UU}(q)$} 
  &  
  \includegraphics[width=0.3\textwidth]{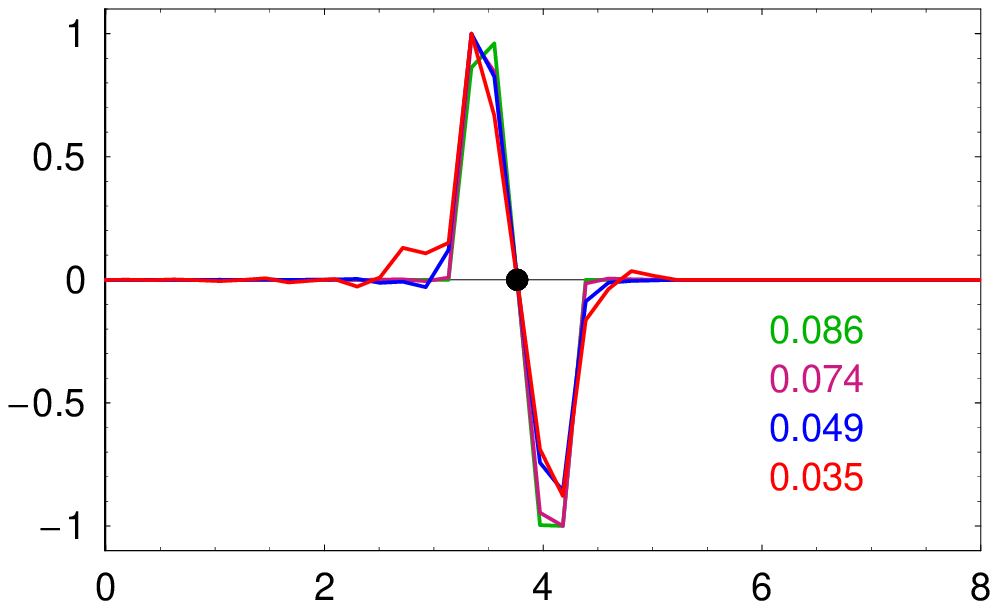}
    &
    \includegraphics[width=0.3\textwidth]{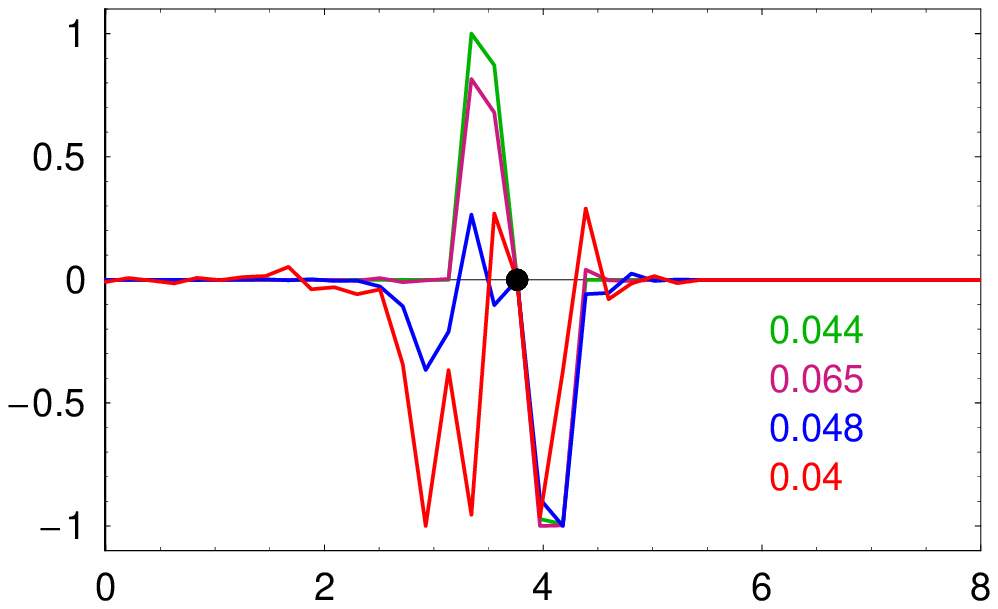}
   &
    \includegraphics[width=0.3\textwidth]{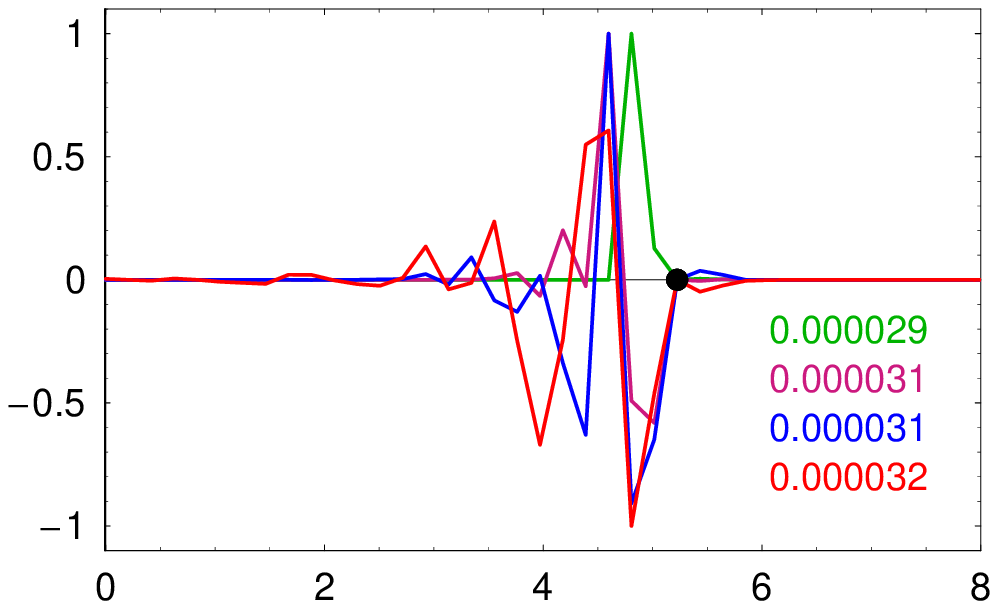}
    \\
    \raisebox{2cm}{${\cal T}_{BU}(q)$} 
    &
    \includegraphics[width=0.3\textwidth]{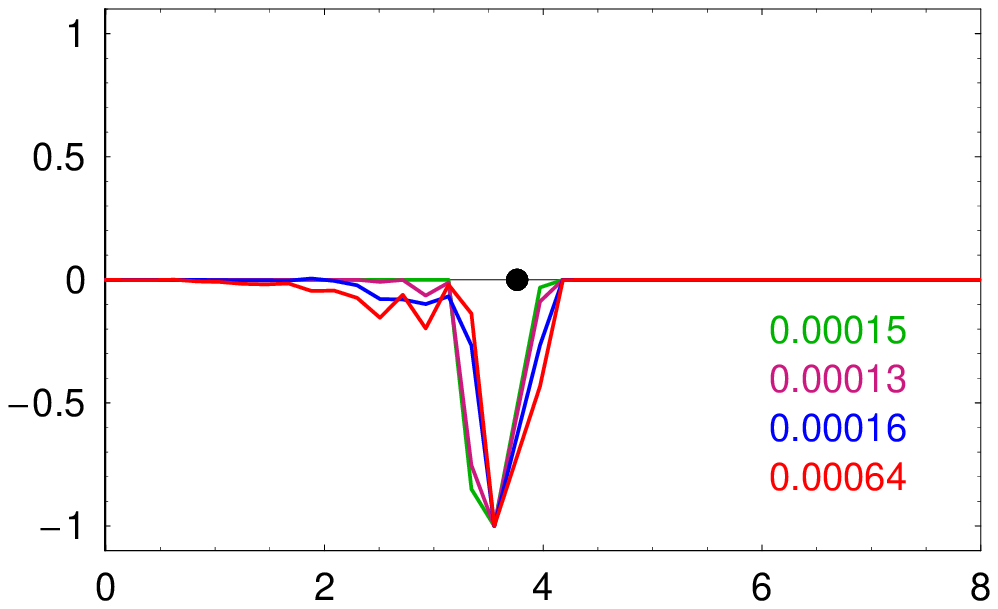}
    &
    \includegraphics[width=0.3\textwidth]{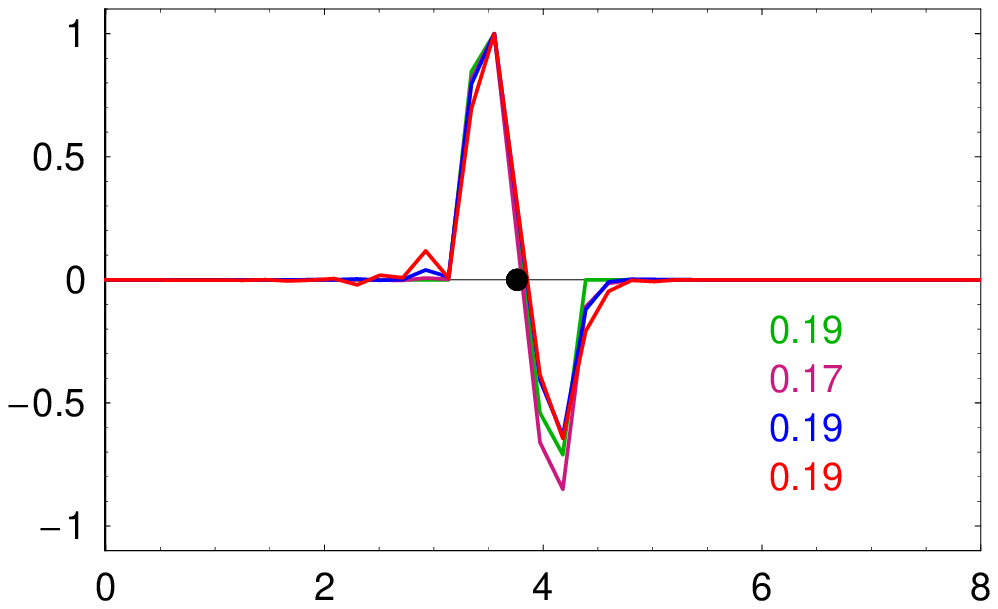}
   &
    \includegraphics[width=0.3\textwidth]{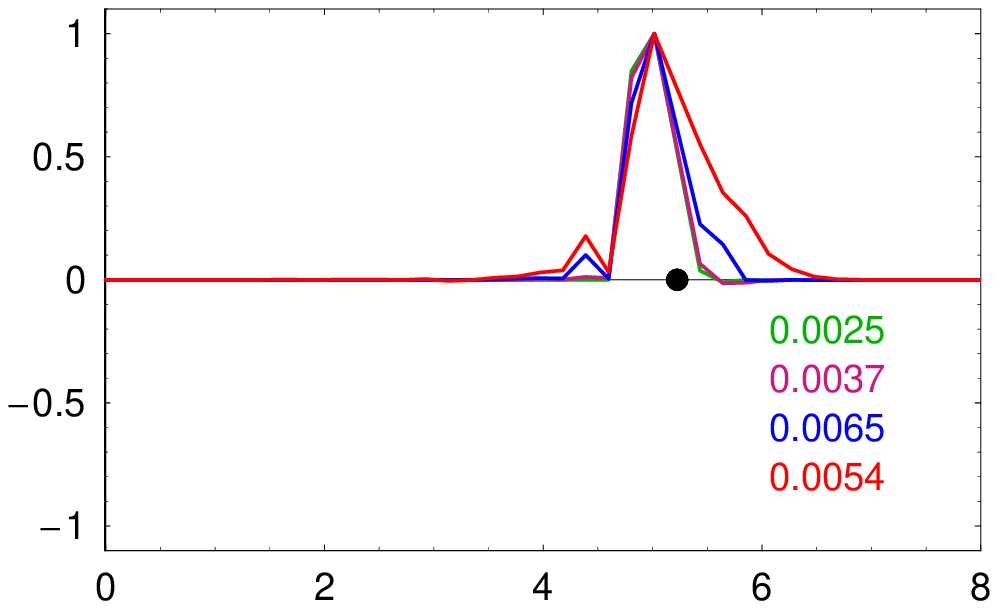}
    \\
    \raisebox{2cm}{${\cal T}_{UB}(q)$}
    &
    \includegraphics[width=0.3\textwidth]{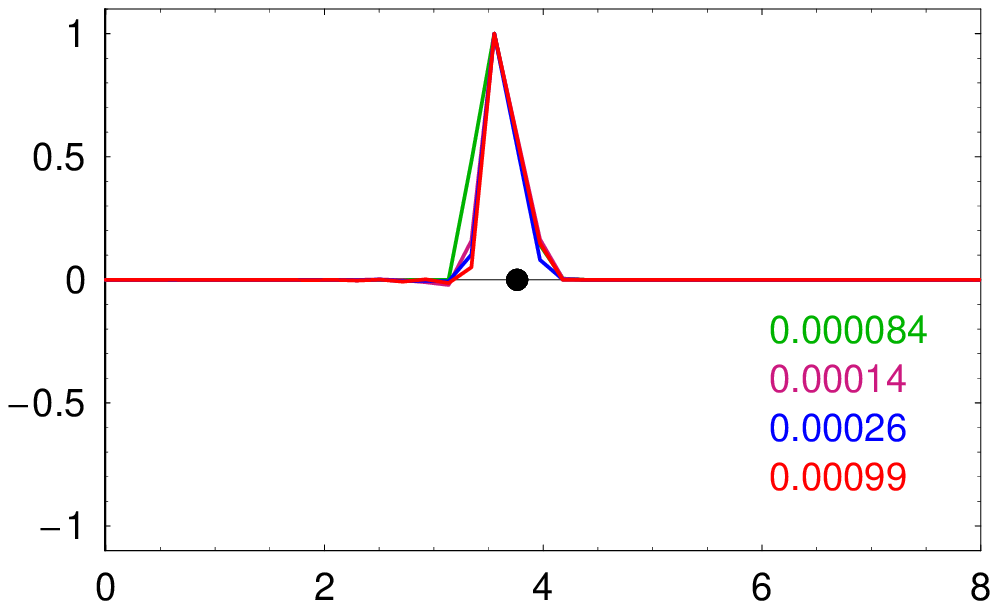}
    &
    \includegraphics[width=0.3\textwidth]{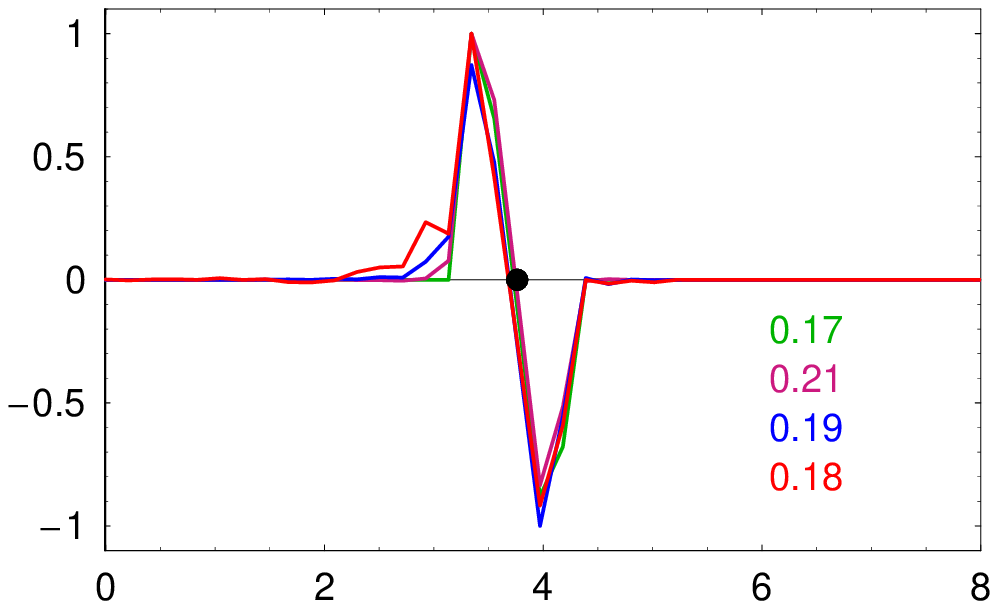}
   &
    \includegraphics[width=0.3\textwidth]{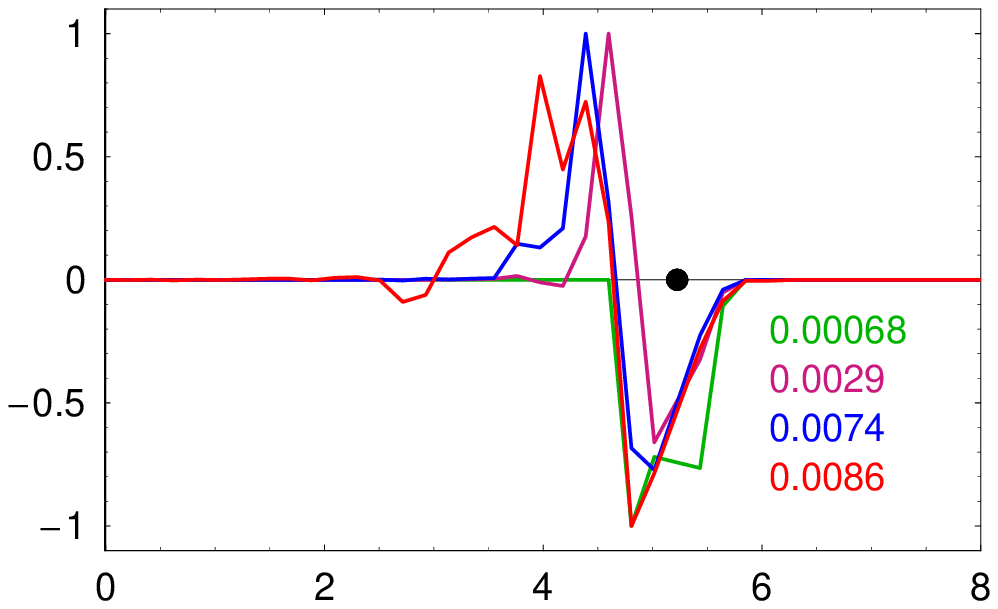}
    \\
    \raisebox{2cm}{${\cal T}_{BB}(q)$} 
    &
    \includegraphics[width=0.3\textwidth]{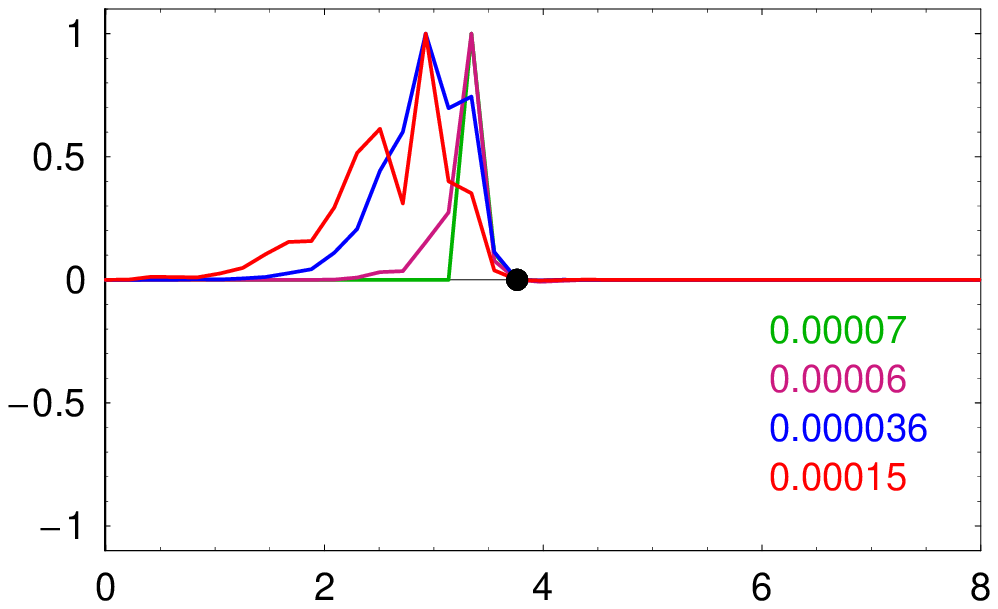}
    &
    \includegraphics[width=0.3\textwidth]{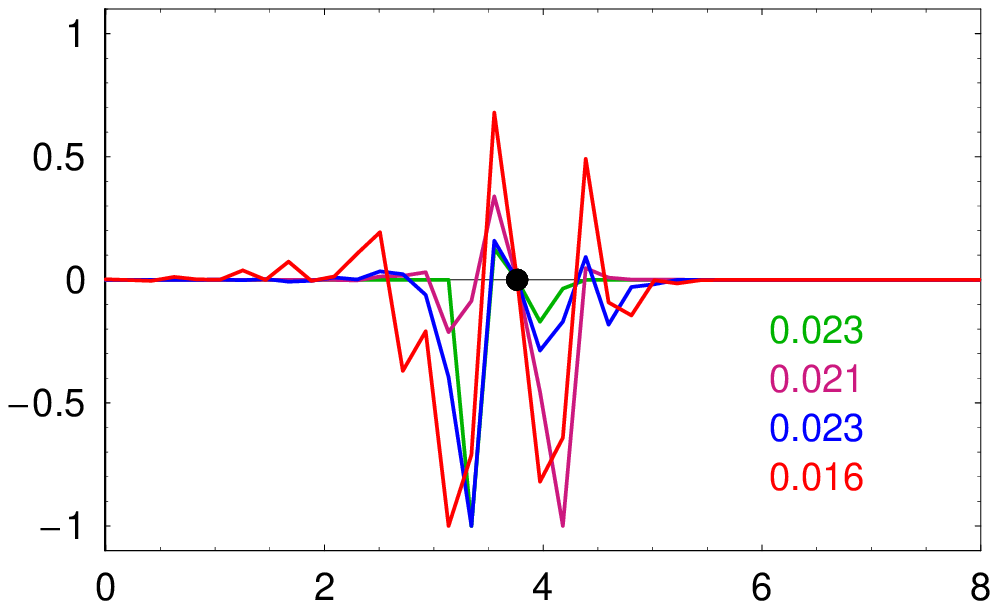}
   &
    \includegraphics[width=0.3\textwidth]{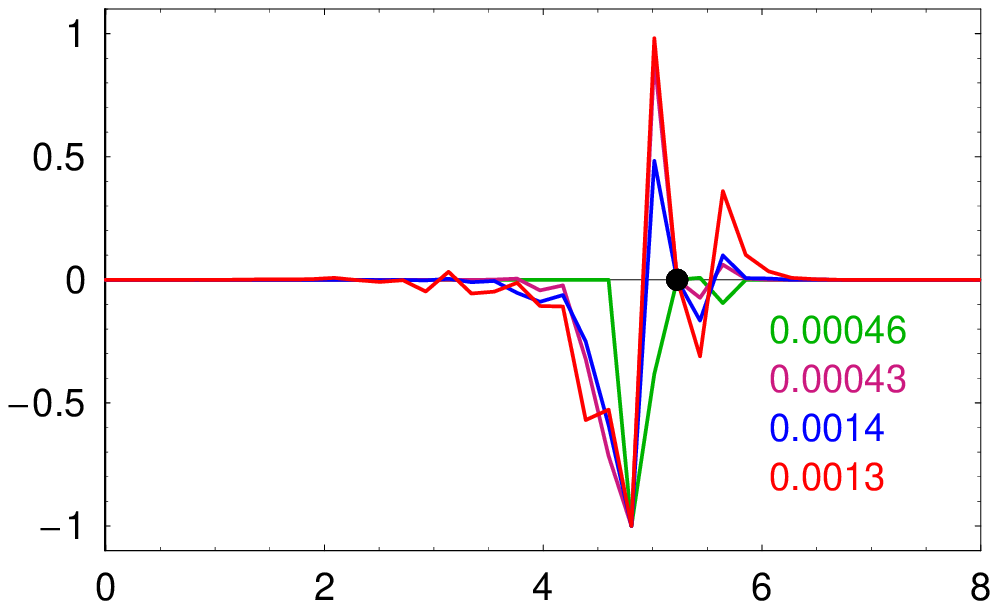} 
    \\
    &
    $\log_{10} q$
    &
    $\log_{10} q$
    &
    $\log_{10} q$
  \end{tabular}
  \end{center}
\caption{From left to right column, $P_m = 10^{-3}, 1, 10^{5}$.
From top to bottom, the plots correspond to ${\cal T}_{UU}(q,n)$; ${\cal T}_{BU}(q,n)$; ${\cal T}_{UB}(q,n)$; ${\cal T}_{BB}(q,n)$ for $n$ indicated by the black dot.
Each curve correspond to $\alpha = - \infty$ (green),
$-5/2$ (magenta), $-3/2$ (blue) and $-1$ (red).}
\label{NLtransfers}
\end{figure}

In figure \ref{NLtransfers} some typical results are presented for the three
values of $P_m=10^{-3}, 1$ and $10^{5}$ (from left to right column), each row
from top to bottom corresponding to
the transfers ${\cal T}_{UU}(q,n)$, ${\cal T}_{BU}(q,n)$, ${\cal T}_{UB}(q,n)$ and ${\cal T}_{BB}(q,n)$ for one given value of $n$ denoted by the black dot. The curves correspond to $\alpha = - \infty$ (green),
$-5/2$ (magenta), $-3/2$ (blue) and $-1$ (red).
\begin{itemize}
	\item 
For $P_m=10^{-3}$, $|{\cal T}_{UU}| \gg |{\cal T}_{BU}|$, implying that
the dominant energy transfer which feeds the kinetic energy
is a local direct cascade of kinetic energy.\\
	\item 
	For $P_m=1$ and $P_m=10^5$, $|{\cal T}_{UU}| \ll |{\cal T}_{BU}|$ implying that
the kinetic energy is mainly obtained from magnetic field. This transfer 
${\cal T}_{BU}$ is found to be mainly local. \\
	\item 
	For $P_m=10^{-3}$ we find that ${\cal T}_{BU}$
is mainly local and always negative. It means that energy is transferred locally from
$U_n$ to $B_n$ (which is consistent with the curve ${\cal T}_{UB}$ just below).
In addition, we see that the curve of ${\cal T}_{BU}$ extends towards larger and larger scales when $\alpha$ goes to zero. Though it is small (up to 20\%), it is a clear evidence of non local transfer from small scale kinetic to large scale magnetic energy. We interpret it
as an alpha-effect in the sense of mean field theory.\\
	\item 
	The curves of ${\cal T}_{UB}$ for $P_m=10^5$ show also clear evidence of non local transfers. In this case, $B_n$ is fed by scales of $U$  much larger than $n$. There is also some local transfer
back from  $B_n$ to $U_n$ shown by the curves of ${\cal T}_{BU}$.\\
	\item 
	For $P_m=10^{-3}$ and though much smaller than ${\cal T}_{UB}$, there is clear evidence of non local direct cascade of magnetic energy as shown by ${\cal T}_{BB}$.\\
	\item 
	Finally, for $P_m=1$ and $P_m=10^5$, ${\cal T}_{BB}$ shows evidence of some kind of non local inverse cascade though much smaller than ${\cal T}_{UB}$.
\end{itemize}
In figures \ref{interpretationlowPm}, \ref{interpretationPm1} and \ref{interpretationlargePm}
of Appendix \ref{sec:interp}
we give some qualitative illustration of the previous interpretation of the results.
\section{Discussion}
The main originality of the shell model presented in (\ref{eq_b})(\ref{Qn}) is that it
takes into account all possible non local interactions between different shells.
In essence it is a non local MHD version of the Sabra model.
Deriving the model we find that there is one free parameter that we call $\alpha$
(and which is different from the alpha-effect of the mean-field dynamo theory).
In some sense this parameter controls the strength of non locality of the model.
An estimation of $\alpha$ has been done on the basis of simple probabilistic arguments of 
possible triad interactions, leading to $\alpha=-7/2$. 
However, in free decaying hydrodynamic turbulence, we find numerically that the value $\alpha=-5/2$ gives the right slope of the kinetic spectrum at large scales ($k^{4}$ as predicted by the turbulence theory in spectral space). Therefore $\alpha=-5/2$ seems the most appropriate value for hydrodynamic turbulence. 
In comparison we note that the local Sabra model (corresponding here to $\alpha=-\infty$) leads to a slope 
about $k^9$.\\
In the other hand, in MHD turbulence it is not obvious that $\alpha$ has to be the same as in hydrodynamic turbulence. Actually, it is even not obvious that $\alpha$ should not depend from the scale $n$, which is not possible in our model. Therefore we investigated several values of $\alpha$ ranging several values from $- \infty$ to zero. Several values of $P_m$ have been investigated depending if it is lower than, equal to or larger than 1.
In order to characterize the non local energy transfers from $X_q$ to $Y_n$ within any possible triad, the quantities ${\cal T}_{XY}(q,n)$ have been derived an plotted versus $q$ for a few values of $n$. Though most of them are local, several energy transfers have been found to be partially non local, depending on $P_m$.
\ack
This work was supported by the Econet Program 10257QL. Multiprocessor computational resources were supported by special program of the Russian Academy of Science. RS thanks  BRHE post-doctoral fellowship program (CRDF grant Y2-MP-09-02) for financial support. The authors thank A. Alexakis, P. Frick, H. Politano and A. Schekochihin for fruitful discussions.
\appendix
\section{Possible triads in logarithmic shell models}
\label{appendixpossible}
For a given shell $n$, we first take benefit from the fact that the triads ($n,p,q$) and ($n,q,p$)
are identical and then the representation in the plane $(p,q)$ is symmetric with respect to the diagonal.
We then limit our demonstration to half of (\ref{possible}):
\begin{itemize}
	\item Assuming $p \le n-1$, we have $\lambda^n + \lambda^p \le \lambda^n + \lambda^{n-1} $.
	Then from (\ref{lambda}) and (\ref{kp}) we have $|\textbf{k}_3| \le k_0 \lambda^{n+1}$, implying $q \le n+1$.
	\item For $p=n$, we have $\lambda^n + \lambda^p = 2 \lambda^n $. From (\ref{lambda}) we have $2 \le \lambda^2$.
	Therefore from (\ref{kp}) we have $|\textbf{k}_3| \le k_0 \lambda^{n+2}$ and then $q \le n+2$.
	\item From (\ref{lambda}) we have $\lambda^{p+1} - \lambda^p \ge \lambda^{p-1}$. For $p \ge n+1$ 
	we then have $\lambda^{p+1} - \lambda^p \ge \lambda^{n}$, implying from (\ref{kp}) that $|\textbf{k}_3| \le k_0 \lambda^{p+1}$
	and then $q \le p+1$.
	\item Assuming that $p \le n-3$, from (\ref{kp}) we have $|\textbf{k}_3| \ge k_0|\lambda^{n-1} - \lambda^{n-4}|$.
	Then (\ref{lambda}) implies $\lambda^{n-1} \ge \lambda^{n-2} + \lambda^{n-3} > \lambda^{n-2} + \lambda^{n-4}$.
	Then we have $|\textbf{k}_3| > k_0 \lambda^{n-2} \ge \lambda^{n-1}$
	and then $q \ge n-1$.
	\item For $p=n-2$, from (\ref{kp}) we have $|\textbf{k}_3| \ge k_0|\lambda^{n-1} - \lambda^{n-3}|$.
	From (\ref{lambda}) we have $\lambda^{n-1} - \lambda^{n-3} \ge \lambda^{n-2}$, implying 
	$|\textbf{k}_3| \ge \lambda^{n-2}$
	and then $q \ge n-2$.
\end{itemize}

\section{Expressions of the energy transfers}
\label{Apptransfers}
The energy transfer (\ref{transferT}) can be written with the help of (\ref{S}) in the form
\begin{equation}
	T(X_q|Z_p|Y_n) = \frac{1}{3}\Re\left\{i k_n Y_n^* M_n(X_q,Z_p) - i k_q X_q^* M_q(Y_n,Z_p)\right\}.
	\label{transferT2}
\end{equation}
The symmetric bilinear form  $M_n(X_q,Z_p)$ is defined as follows
\begin{eqnarray}
	M_n(U_q,U_p) &=& L_n(U_q,U_p,+a)+L_n(U_p,U_q,+a)\\
	M_n(B_q,B_p) &=& L_n(B_q,B_p,-a)+L_n(B_p,B_q,-a)\\
	M_n(U_q,B_p) &=& L_n(U_q,B_p,+b)+L_n(B_p,U_q,-b)\\
	M_n(B_q,U_p) &=& L_n(B_q,U_p,-b)+L_n(U_p,B_q,+b)
	\end{eqnarray}
with $L_n(X_p,Y_q,c)$ given by
\begin{equation}
	L_n(X_q, Z_p, c) = \left\{ \begin{tabular}{@{}l@{\hspace{1em}}l@{\hspace{1em}}l@{\hspace{3em}}l@{}}
	                                   $T_{p-q} c^3_{p-q} X_q\; Z_{p}$ & for $q \le n-2$ &and $p =n-1$\\
	                                   $T_{n-q} c^2_{n-q} X_q^* Z_{p}$ & for $q \le n-1$ &and $p =n+1$\\
	                                   $T_{q-n} c^1_{q-n} X_q^* Z_{p}$ & for $q \ge n+1$ &and $p =q+1$
	                                 \end{tabular}
	                                 \right..
\end{equation}
We note that $L_n(X_p,Y_q,c)$ is related to $Q_n(X,Y,c)$, defined in (\ref{Qn}), by
\begin{equation}
	Q_n(X,Y,c) = \sum_{p,q}L_n(X_p,Y_q,c).
\end{equation}
\section{Energy transfers results}
\label{Transfersresults}
\begin{figure}[ht] \begin{tabular}{@{}c@{\hspace{0em}}c@{\hspace{0em}}c@{\hspace{0em}}c@{\hspace{0em}}c@{\hspace{0em}}c@{\hspace{0em}}c@{}}
          &
    \begin{turn}{0} Spectra \end{turn}
        &
    \begin{turn}{0} ${\cal T}_{UU}(q,n)$ \end{turn}
        &
    \begin{turn}{0}${\cal T}_{BU}(q,n)$\end{turn}
        &
   \begin{turn}{0} ${\cal T}_{UB}(q,n)$\end{turn}
        &
       \begin{turn}{0} ${\cal T}_{BB}(q,n)$\end{turn}
       &
       \\*[0cm]
    \begin{turn}{90}$\quad \quad \quad \quad\alpha=-0.5$ \end{turn}
        &
    \includegraphics[width=0.28\textwidth,angle=90]{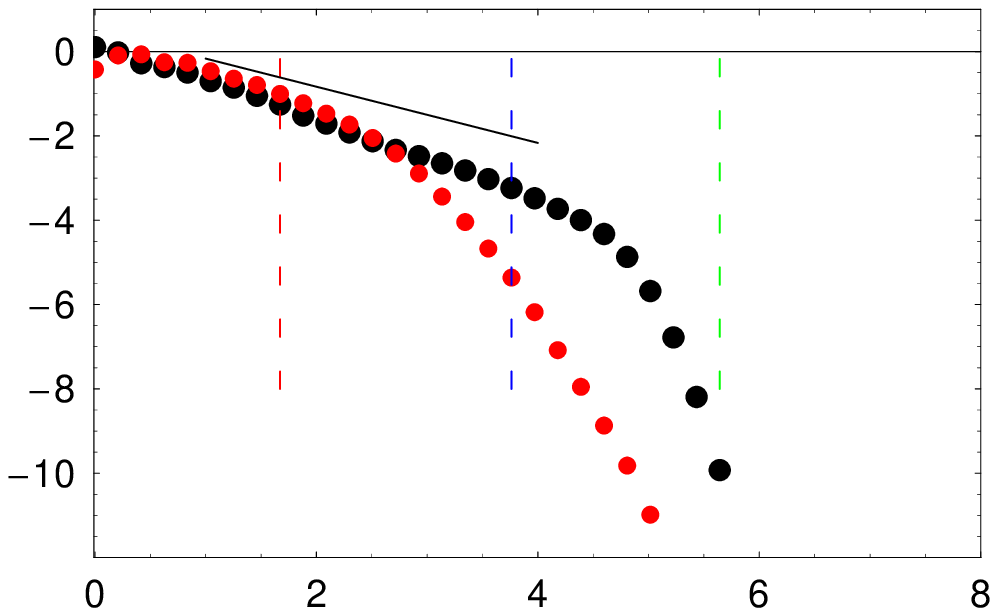}
        &
    \includegraphics[width=0.28\textwidth,angle=90]{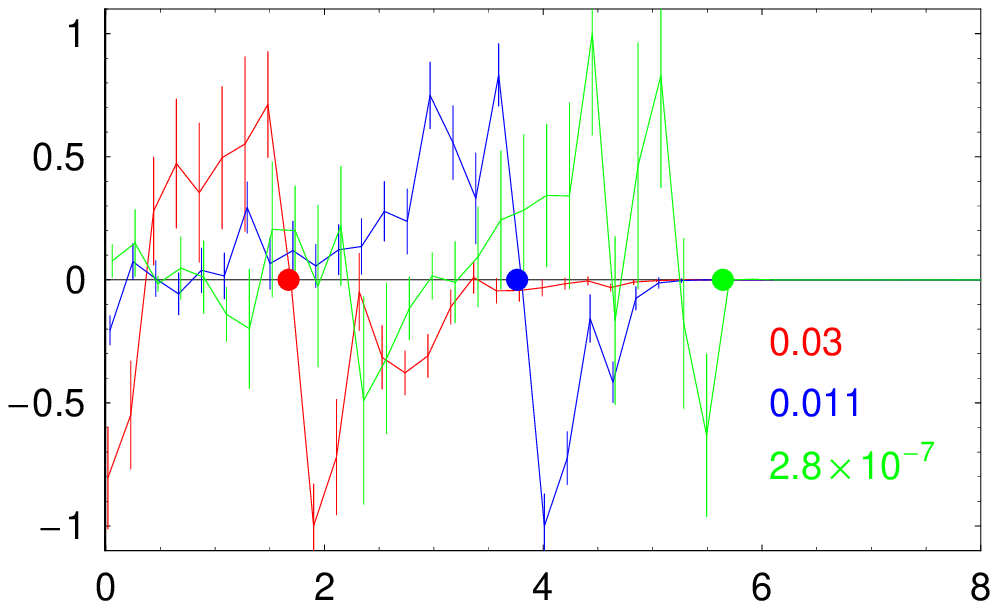}
        &
    \includegraphics[width=0.28\textwidth,angle=90]{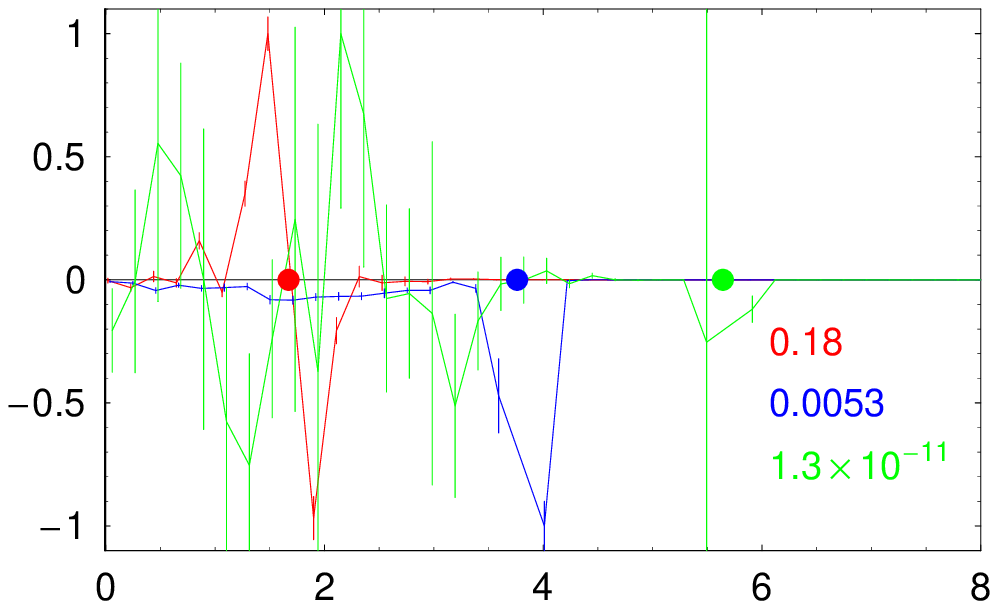}
        &
    \includegraphics[width=0.28\textwidth,angle=90]{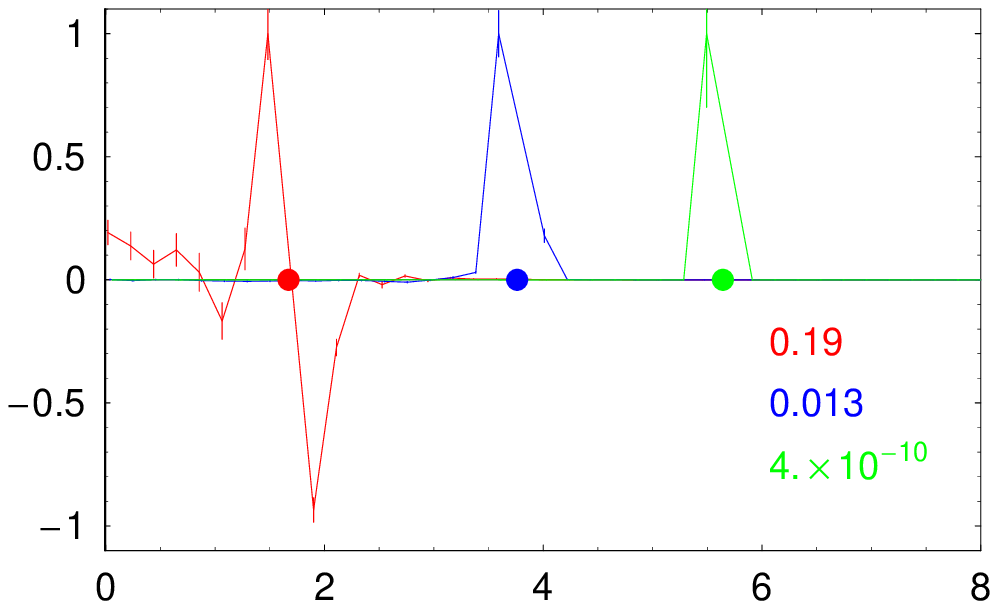}
        &
    \includegraphics[width=0.28\textwidth,angle=90]{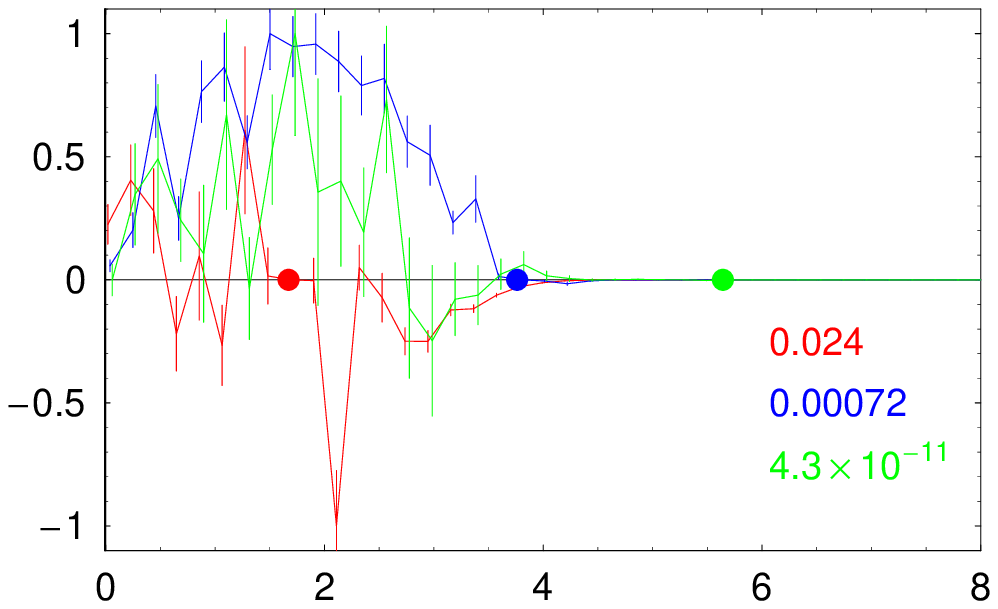}
           &
       \begin{turn}{90} $\quad \quad \quad \quad \quad \quad \quad  \log_{10}q$ \end{turn}\\*[-0.2cm]
    \begin{turn}{90}$\quad \quad \quad \quad  \alpha=-1$ \end{turn}
        &
    \includegraphics[width=0.28\textwidth,angle=90]{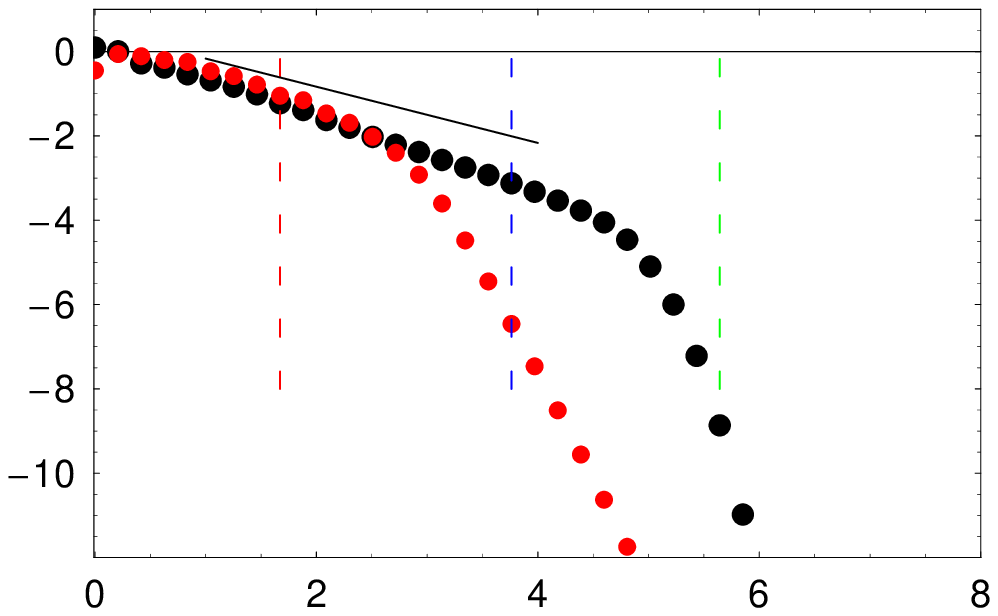}
        &
    \includegraphics[width=0.28\textwidth,angle=90]{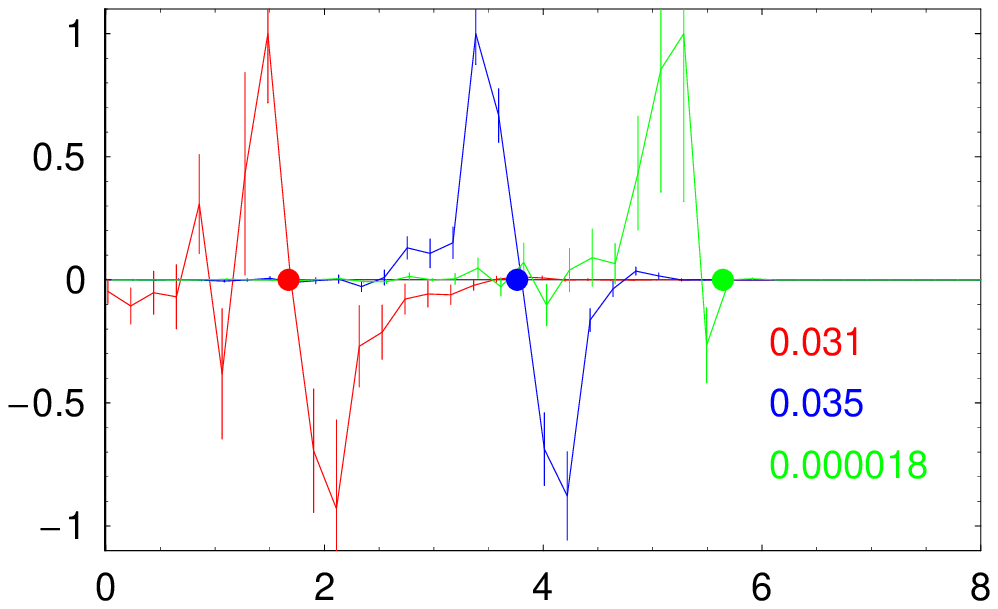}
        &
    \includegraphics[width=0.28\textwidth,angle=90]{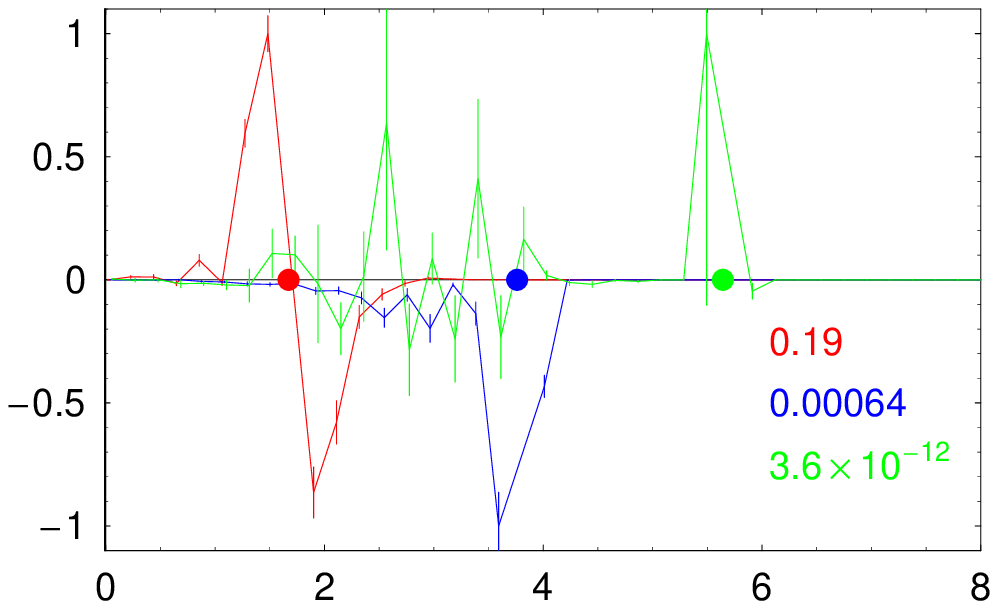}
        &
    \includegraphics[width=0.28\textwidth,angle=90]{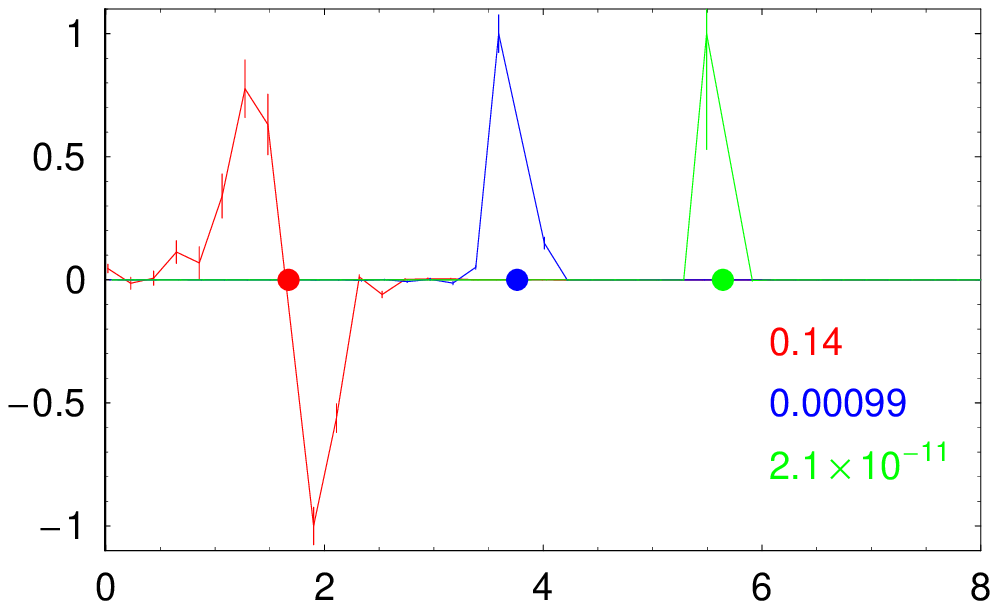}
        &
    \includegraphics[width=0.28\textwidth,angle=90]{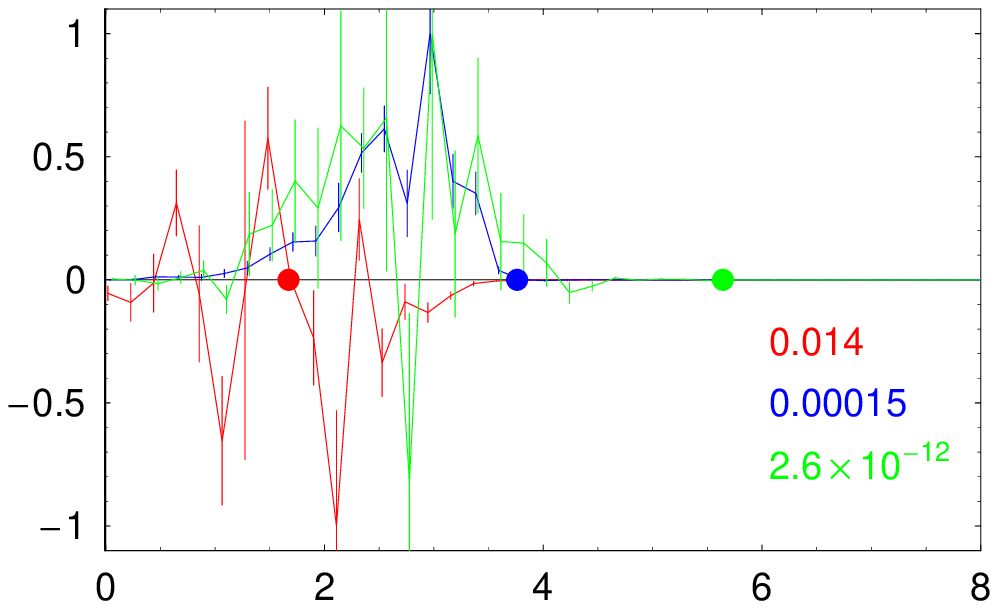}
               &
       \begin{turn}{90} $\quad \quad \quad \quad \quad \quad \quad \log_{10}q$ \end{turn}\\*[-0.2cm]
    \begin{turn}{90}$\quad \quad \quad \quad  \alpha=-1.5$ \end{turn}
        &
    \includegraphics[width=0.28\textwidth,angle=90]{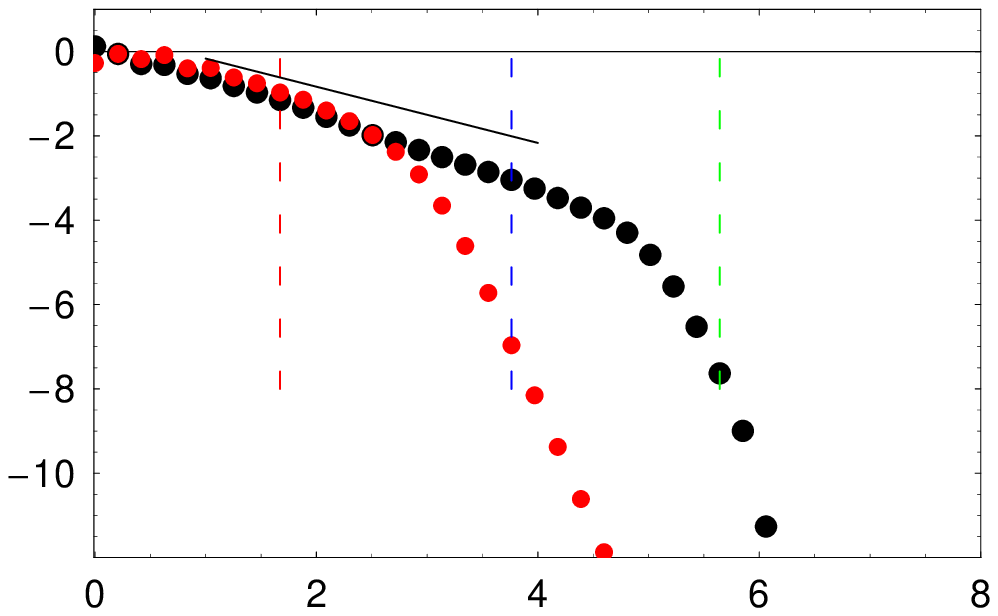}
        &
    \includegraphics[width=0.28\textwidth,angle=90]{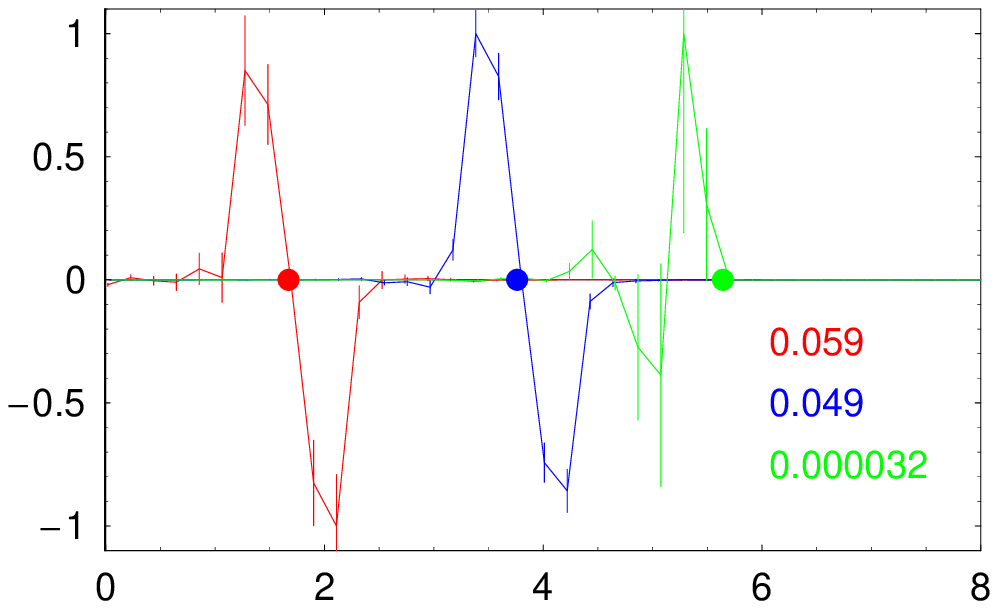}
        &
    \includegraphics[width=0.28\textwidth,angle=90]{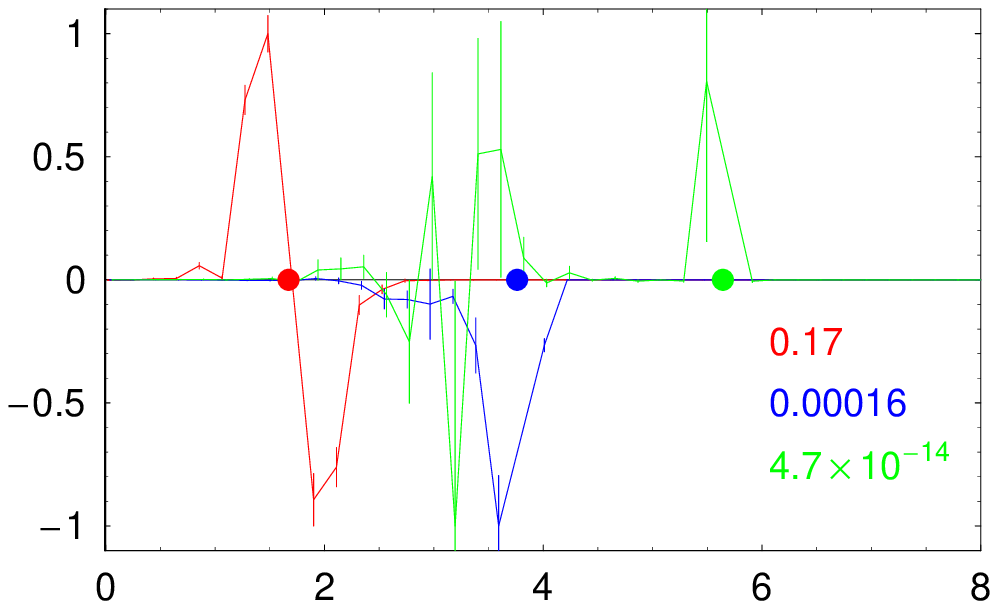}
        &
    \includegraphics[width=0.28\textwidth,angle=90]{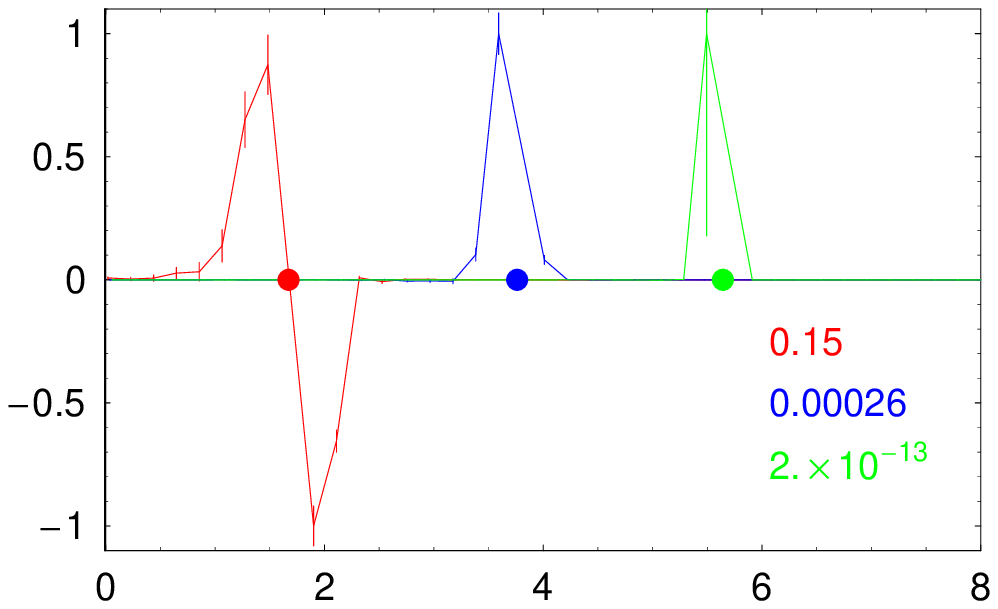}
        &
    \includegraphics[width=0.28\textwidth,angle=90]{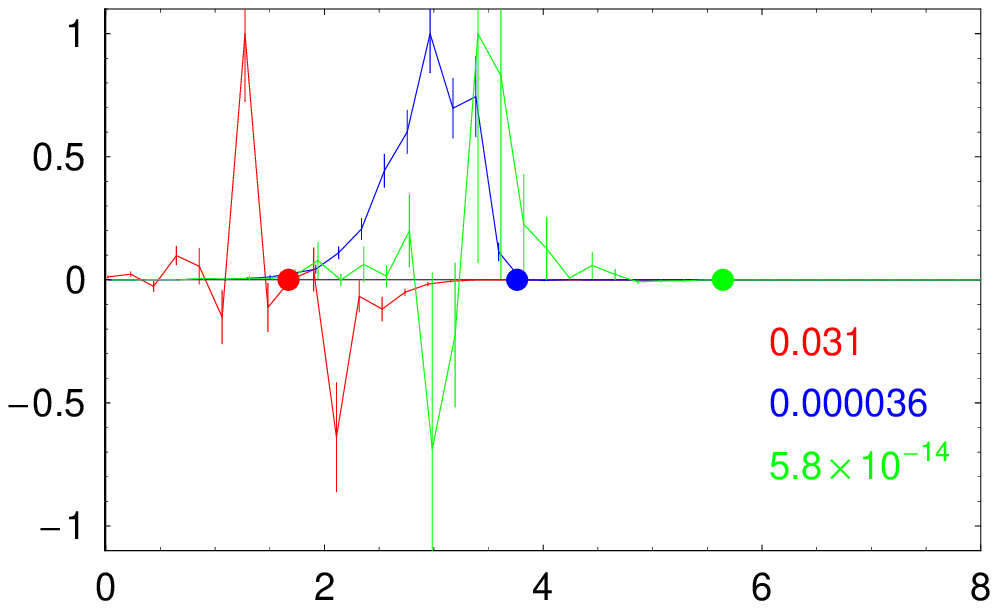}
               &
       \begin{turn}{90} $\quad \quad \quad \quad \quad \quad \quad  \log_{10}q$ \end{turn}\\*[-0.2cm]
    \begin{turn}{90}$\quad \quad \quad \quad  \alpha=-2.5$ \end{turn}
    &
    \includegraphics[width=0.28\textwidth,angle=90]{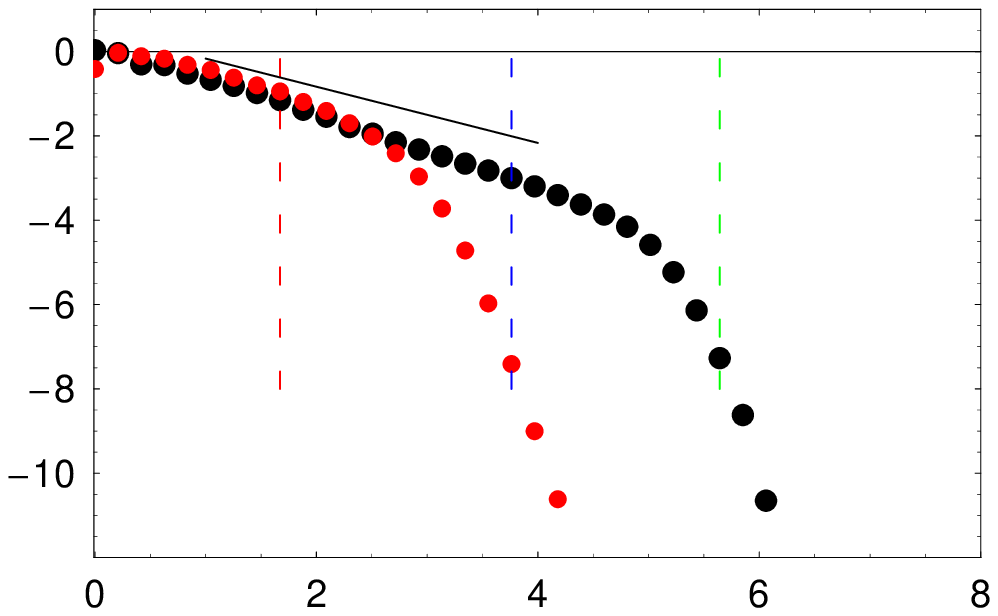}
        &
    \includegraphics[width=0.28\textwidth,angle=90]{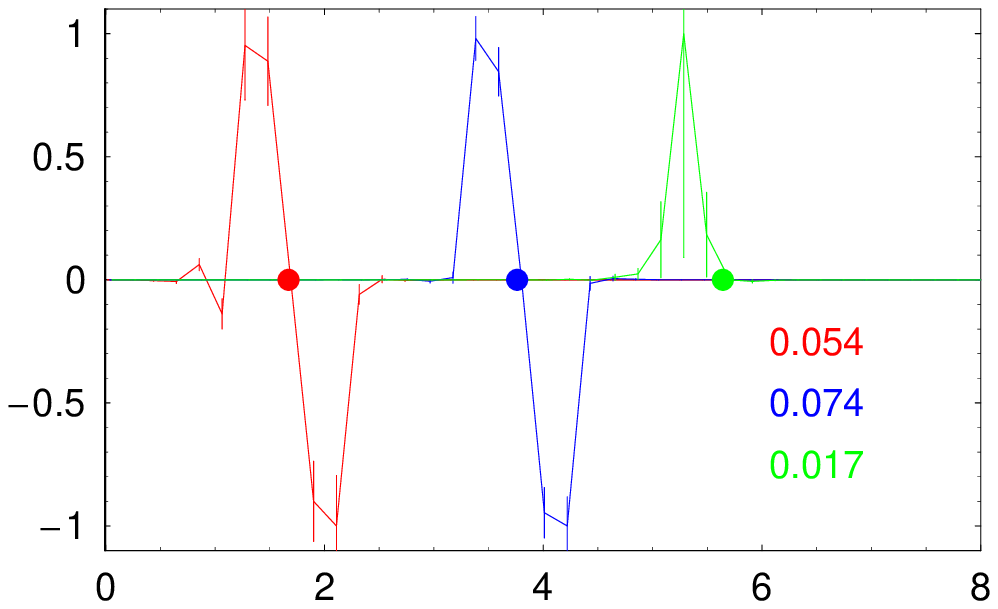}
        &
    \includegraphics[width=0.28\textwidth,angle=90]{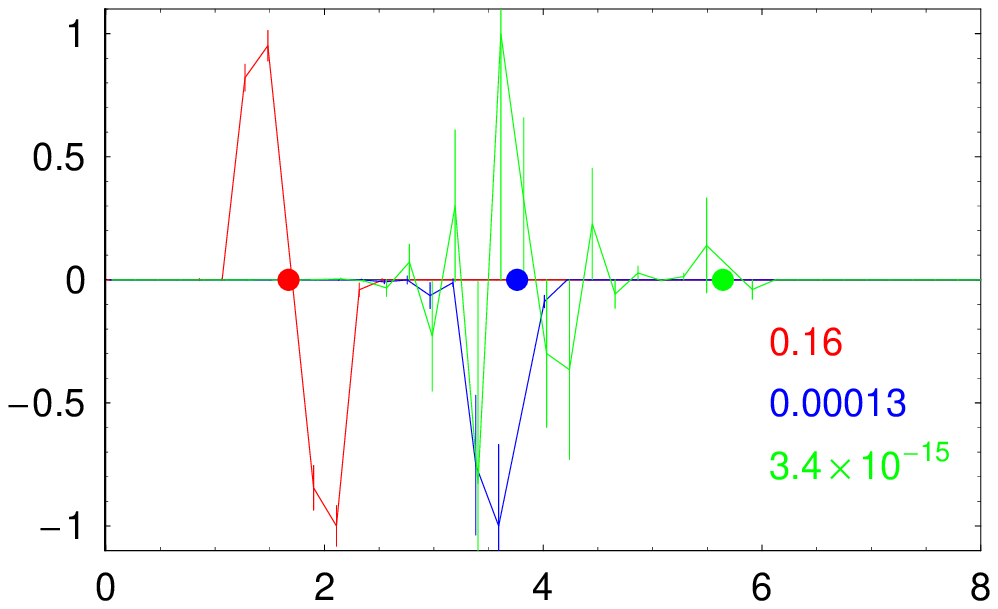}
        &
    \includegraphics[width=0.28\textwidth,angle=90]{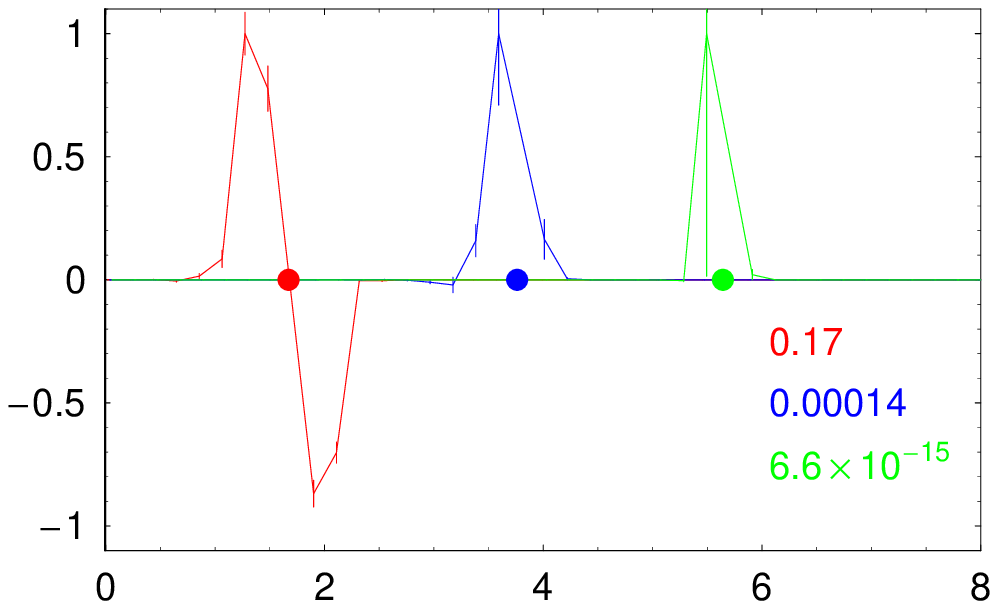}
        &
    \includegraphics[width=0.28\textwidth,angle=90]{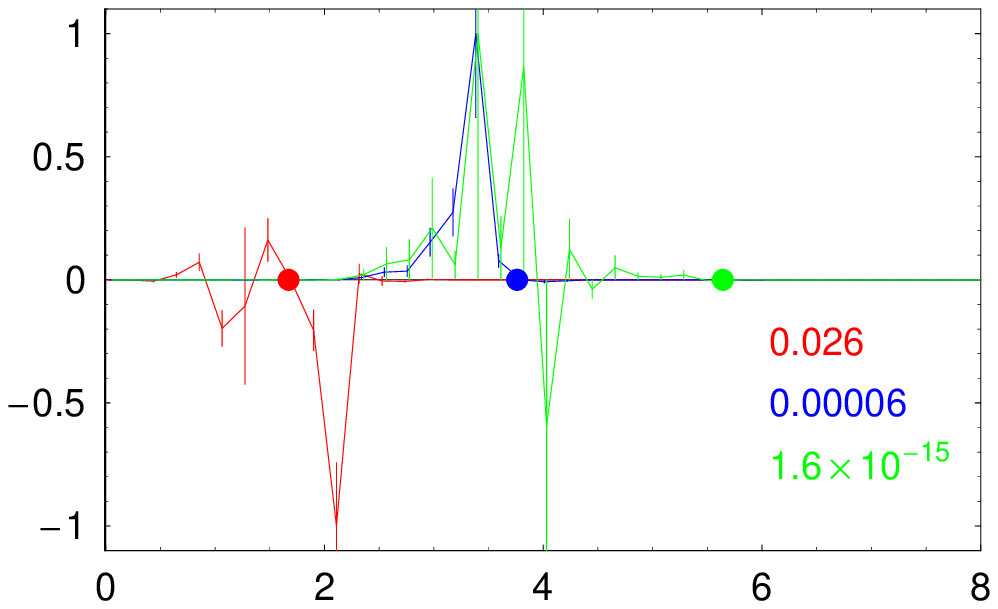}
              &
       \begin{turn}{90} $\quad \quad \quad  \quad \quad \quad \quad \log_{10}q$ \end{turn} \\*[-0.2cm]
    \begin{turn}{90}$\quad \quad \quad  \quad \alpha=-\infty$ \end{turn}
        &
    \includegraphics[width=0.28\textwidth,angle=90]{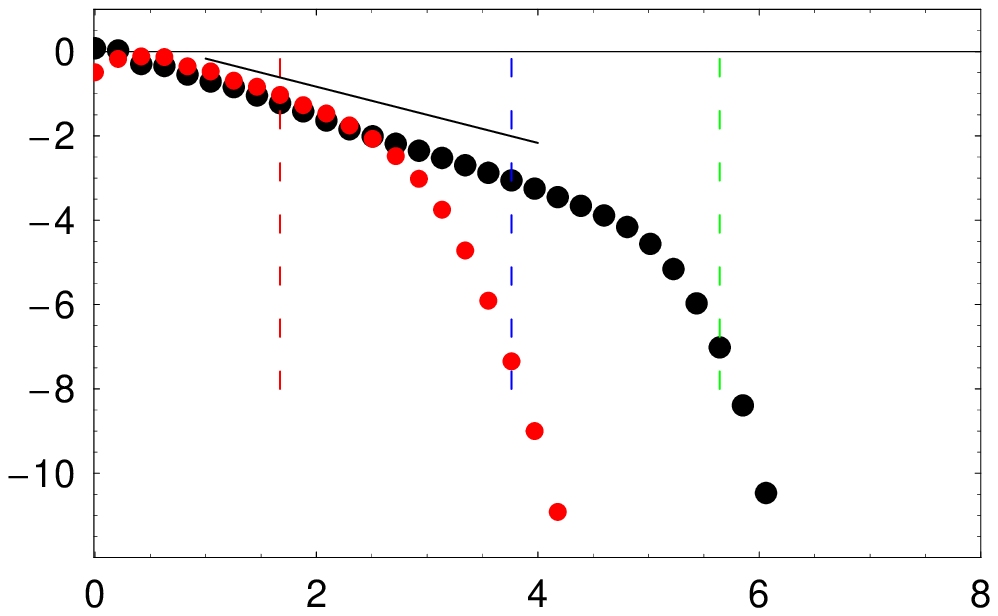}
        &
    \includegraphics[width=0.28\textwidth,angle=90]{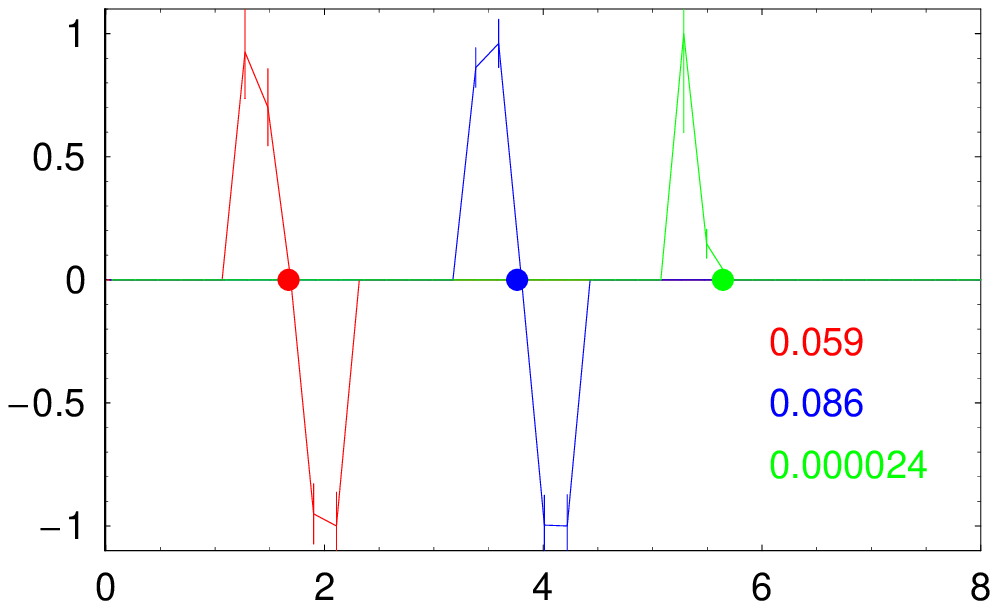}
        &
    \includegraphics[width=0.28\textwidth,angle=90]{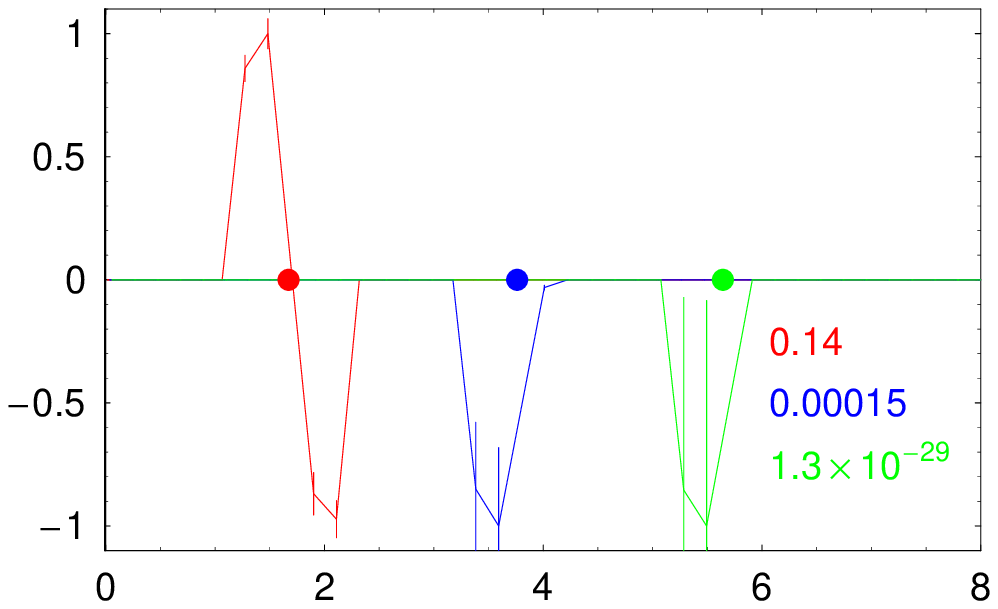}
        &
    \includegraphics[width=0.28\textwidth,angle=90]{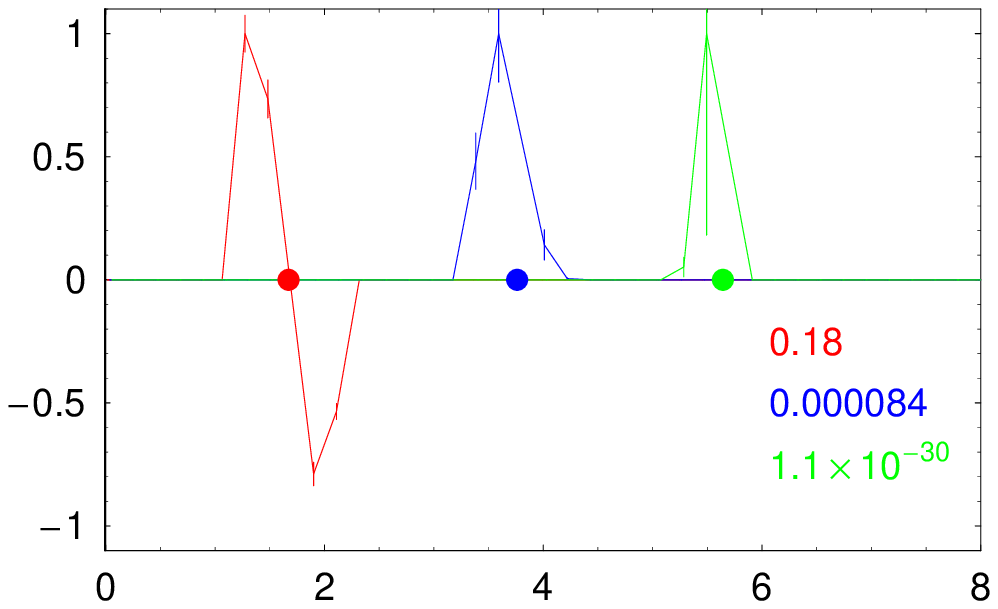}
        &
    \includegraphics[width=0.28\textwidth,angle=90]{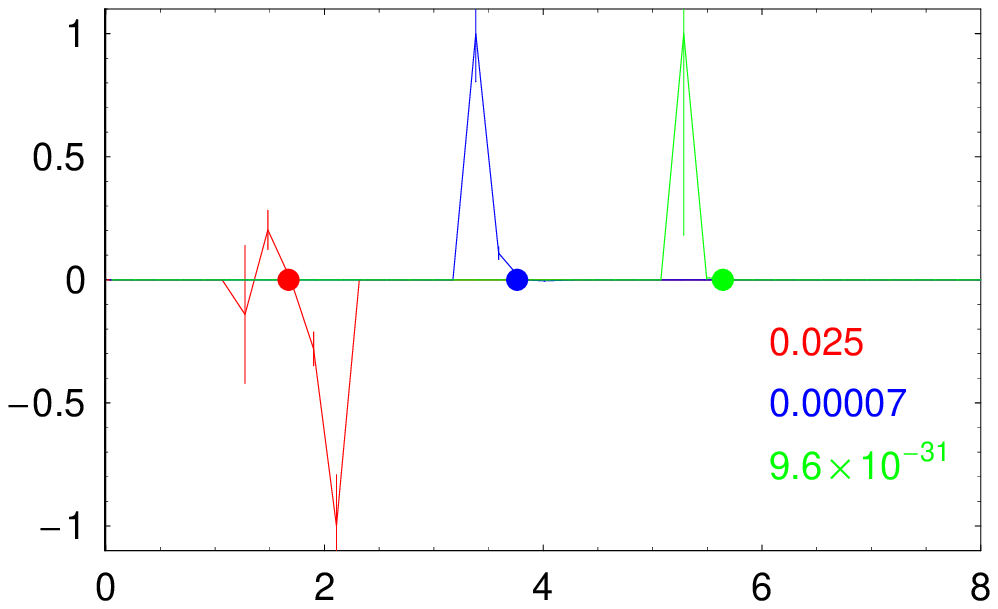}
           &
       \begin{turn}{90} $\quad \quad \quad  \quad \quad \quad \quad \log_{10}q$ \end{turn}
  \end{tabular}
\caption{Spectra and transfer functions for $P_m=10^{-3} (\nu=10^{-7}, \eta=10^{-4})$ and several values of $\alpha$.
From left to right column, $\alpha = -\infty; -2.5; -1.5; -1; -0.5$.
From top to bottom, the plots correspond to the spectra (red dots for kinetic and black for magnetic);
${\cal T}_{UU}(q,n)$; ${\cal T}_{BU}(q,n)$; ${\cal T}_{UB}(q,n)$; ${\cal T}_{BB}(q,n)$.
For a given $\alpha$, the transfer functions are plotted versus $q$ for three values of $n$ indicated by the dashed lines.} 
\label{transfer-3}
\end{figure}
\begin{figure}
  \begin{tabular}{@{}c@{\hspace{0em}}c@{\hspace{0em}}c@{\hspace{0em}}c@{\hspace{0em}}c@{\hspace{0em}}c@{\hspace{0em}}c@{}}
        &
    \begin{turn}{0} Spectra \end{turn}
        &
    \begin{turn}{0} ${\cal T}_{UU}(q,n)$ \end{turn}
        &
    \begin{turn}{0}${\cal T}_{BU}(q,n)$\end{turn}
        &
   \begin{turn}{0} ${\cal T}_{UB}(q,n)$\end{turn}
        &
       \begin{turn}{0} ${\cal T}_{BB}(q,n)$\end{turn}
        & \\*[0cm]
        \begin{turn}{90}$\quad \quad \quad \quad \alpha=-0.5$ \end{turn}
        &
    \includegraphics[width=0.28\textwidth,angle=90]{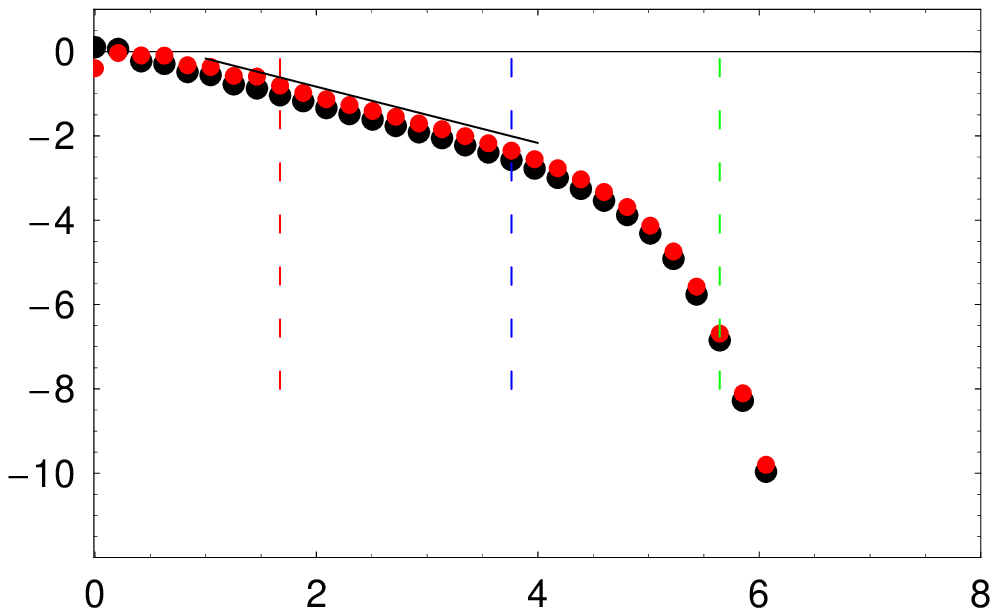}
        &
    \includegraphics[width=0.28\textwidth,angle=90]{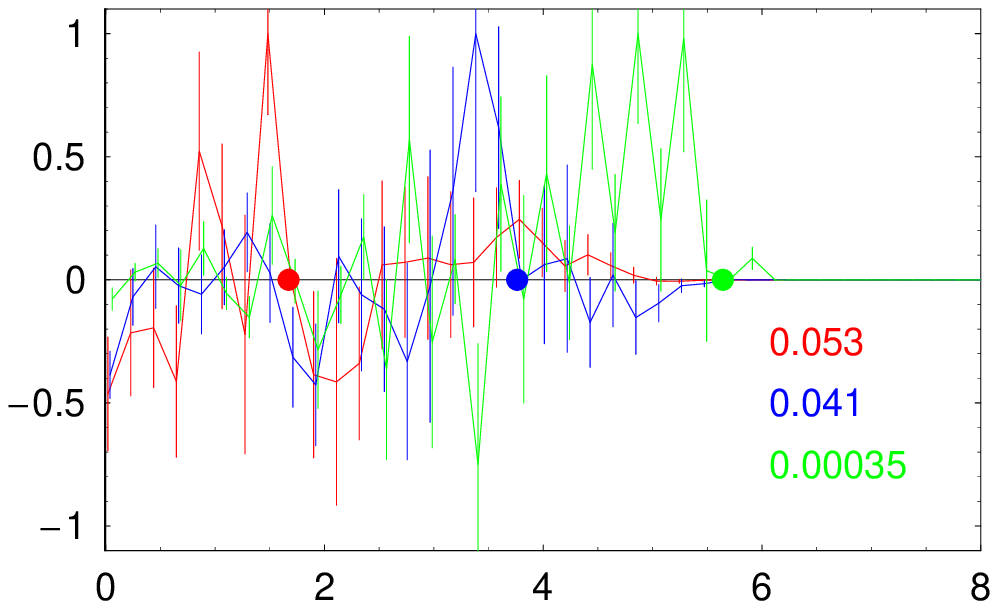}
        &
    \includegraphics[width=0.28\textwidth,angle=90]{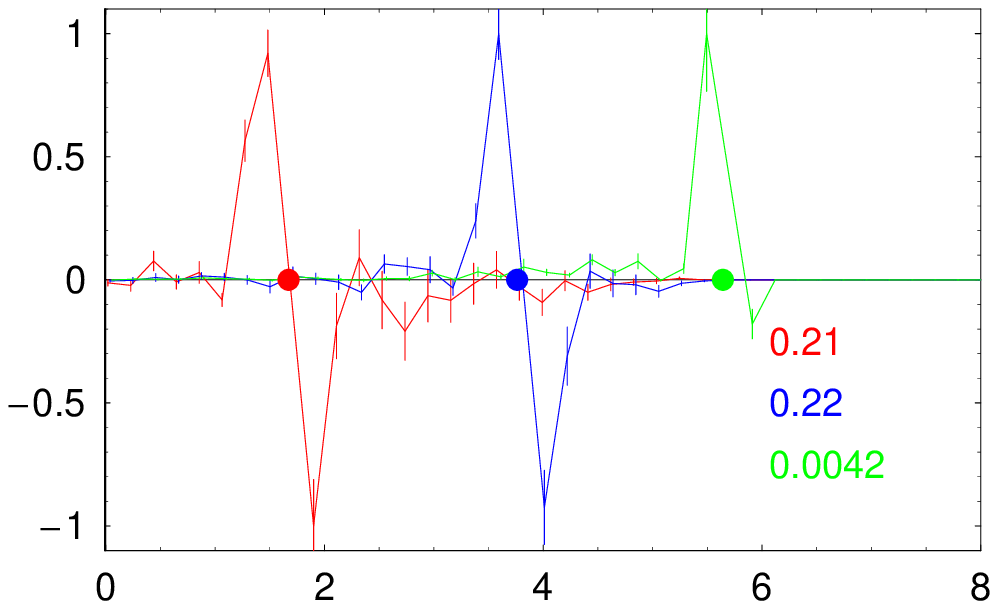}
        &
    \includegraphics[width=0.28\textwidth,angle=90]{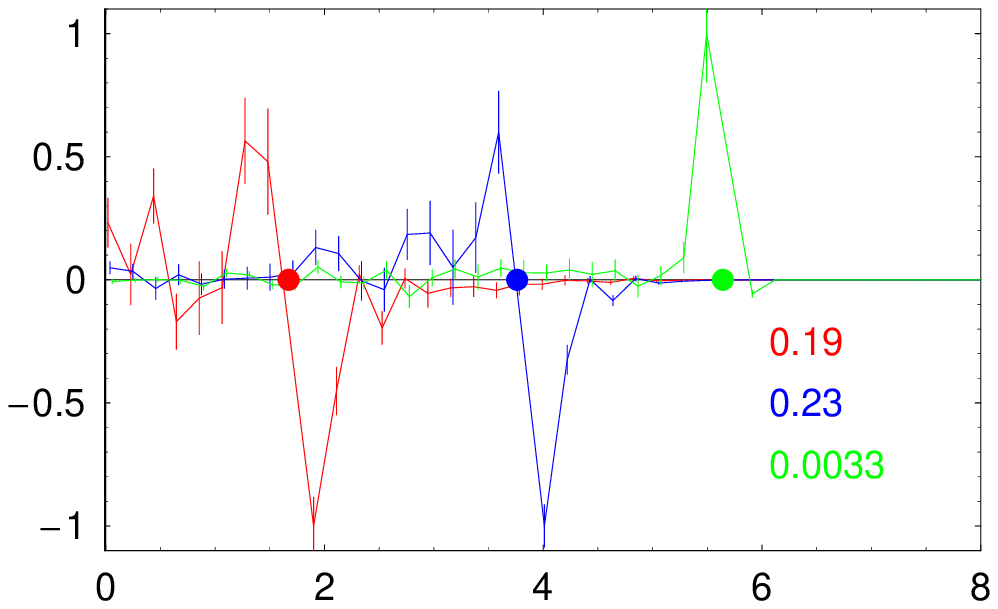}
        &
    \includegraphics[width=0.28\textwidth,angle=90]{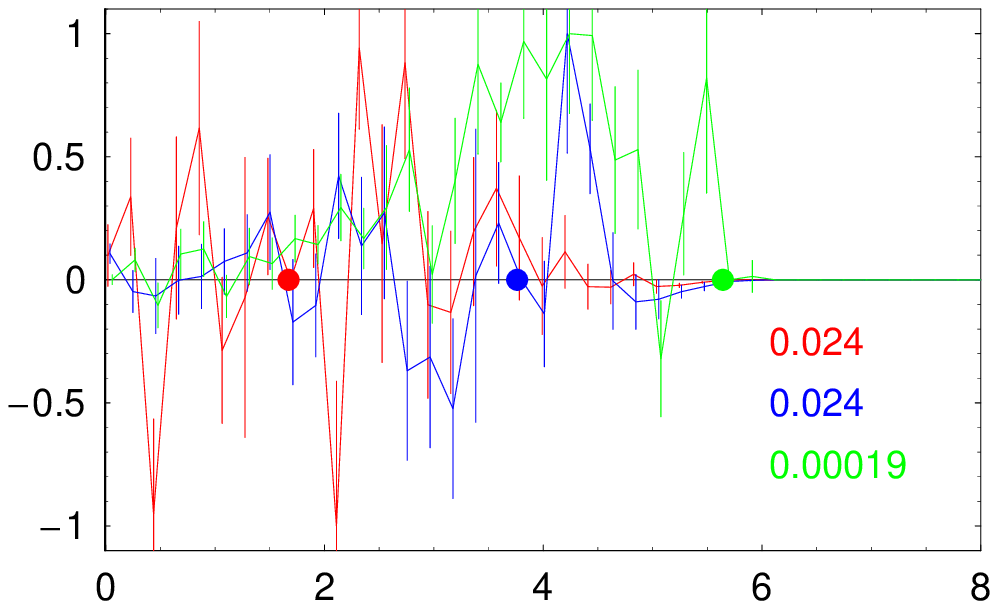}
               &
       \begin{turn}{90} $\quad \quad \quad \quad \quad \quad \quad \log_{10}q$\end{turn}
    \\*[-0.2cm]
        \begin{turn}{90}$\quad \quad \quad \quad \alpha=-1$ \end{turn}
        &
    \includegraphics[width=0.28\textwidth,angle=90]{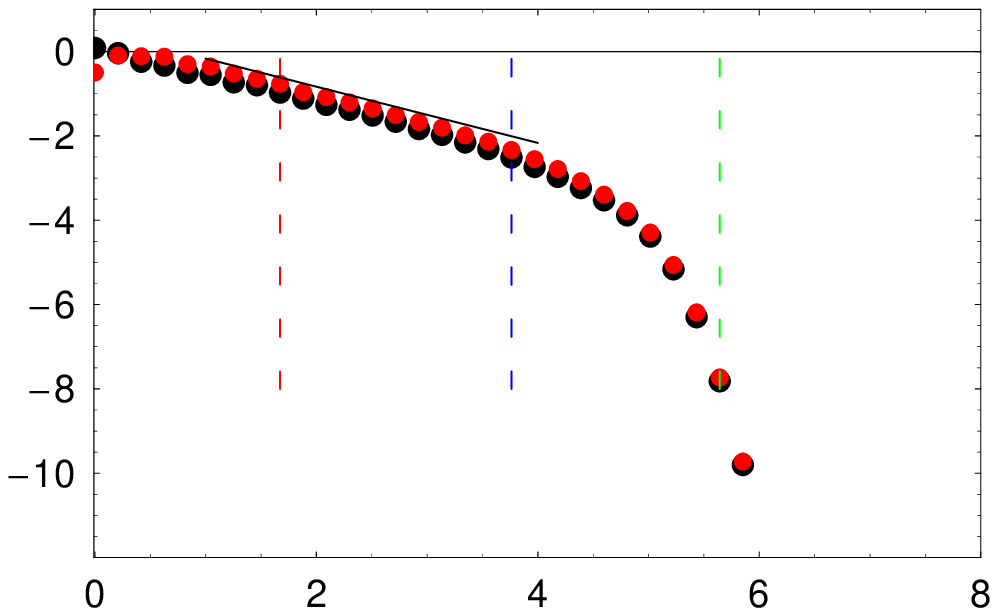}
        &
    \includegraphics[width=0.28\textwidth,angle=90]{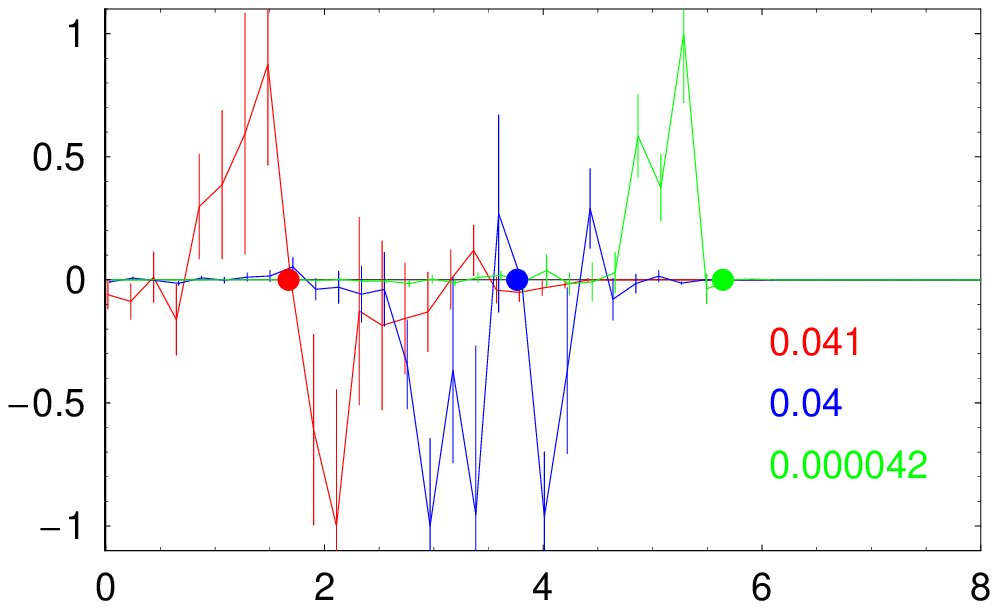}
        &
    \includegraphics[width=0.28\textwidth,angle=90]{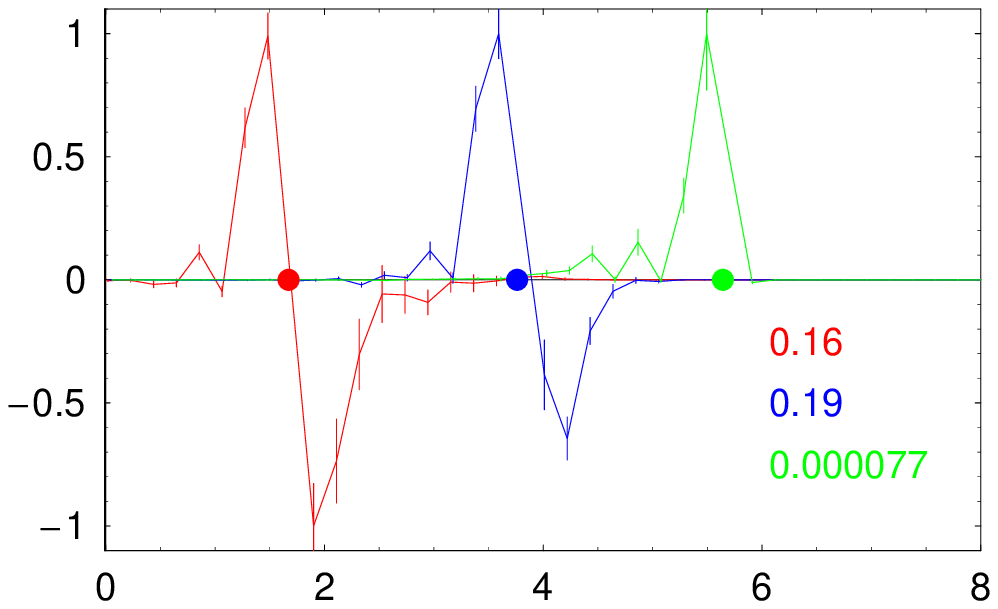}
        &
    \includegraphics[width=0.28\textwidth,angle=90]{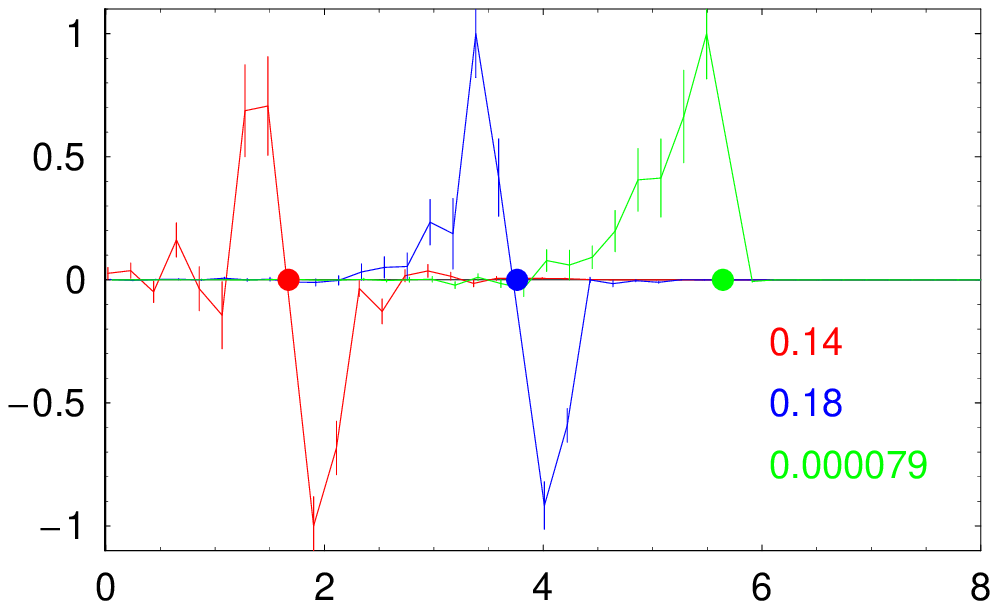}
        &
    \includegraphics[width=0.28\textwidth,angle=90]{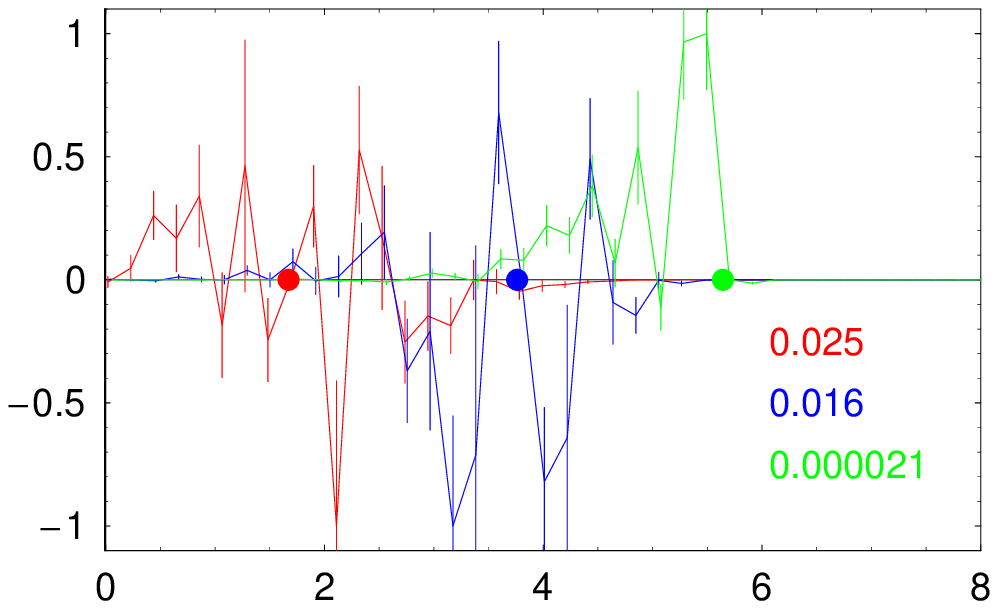}
               &
       \begin{turn}{90} $\quad \quad \quad \quad \quad \quad \quad \log_{10}q$\end{turn}
        \\*[-0.2cm]
            \begin{turn}{90}$\quad \quad \quad \quad \alpha=-1.5$ \end{turn}
        &
    \includegraphics[width=0.28\textwidth,angle=90]{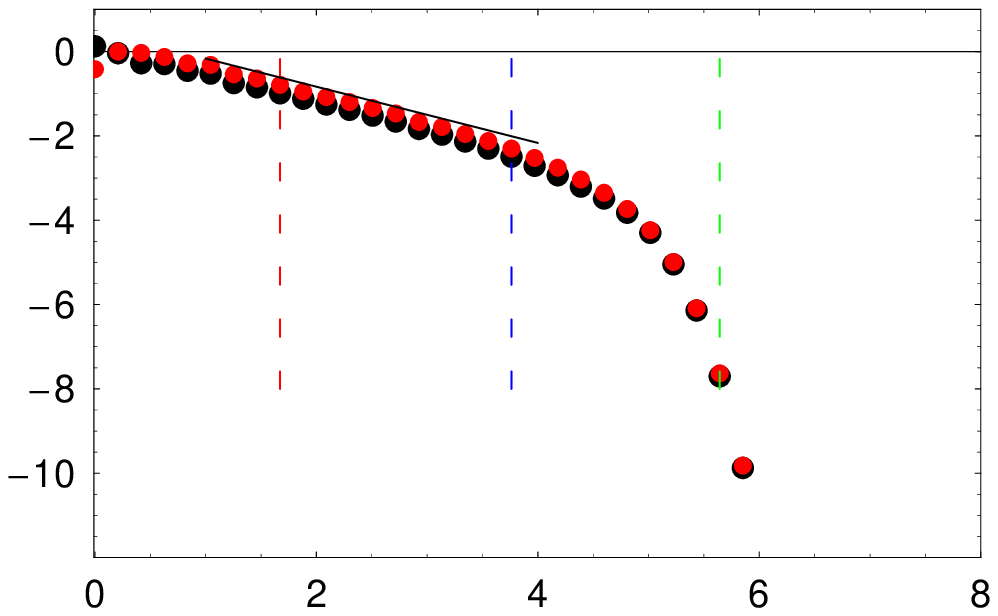}
        &
    \includegraphics[width=0.28\textwidth,angle=90]{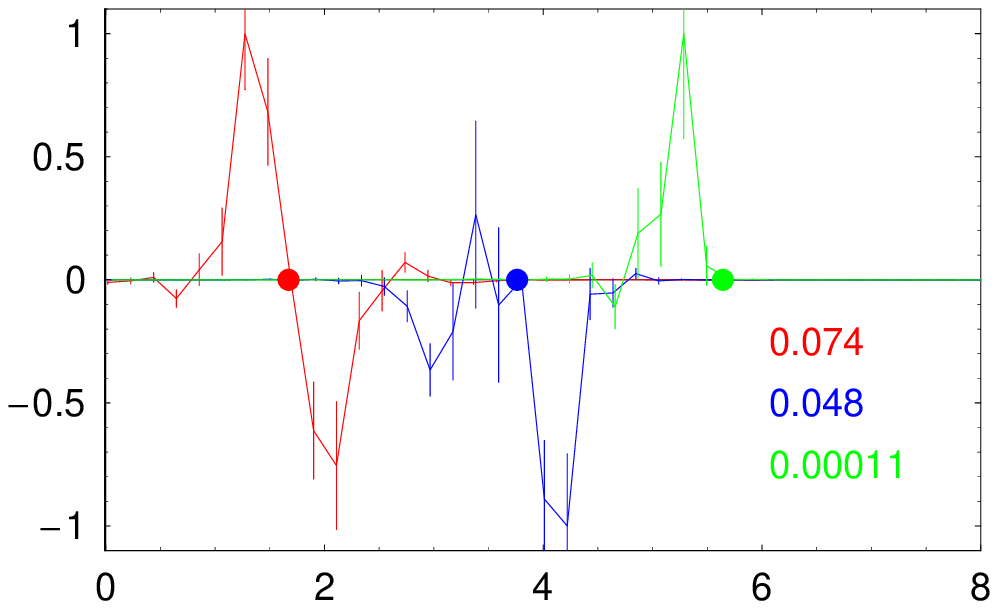}
        &
    \includegraphics[width=0.28\textwidth,angle=90]{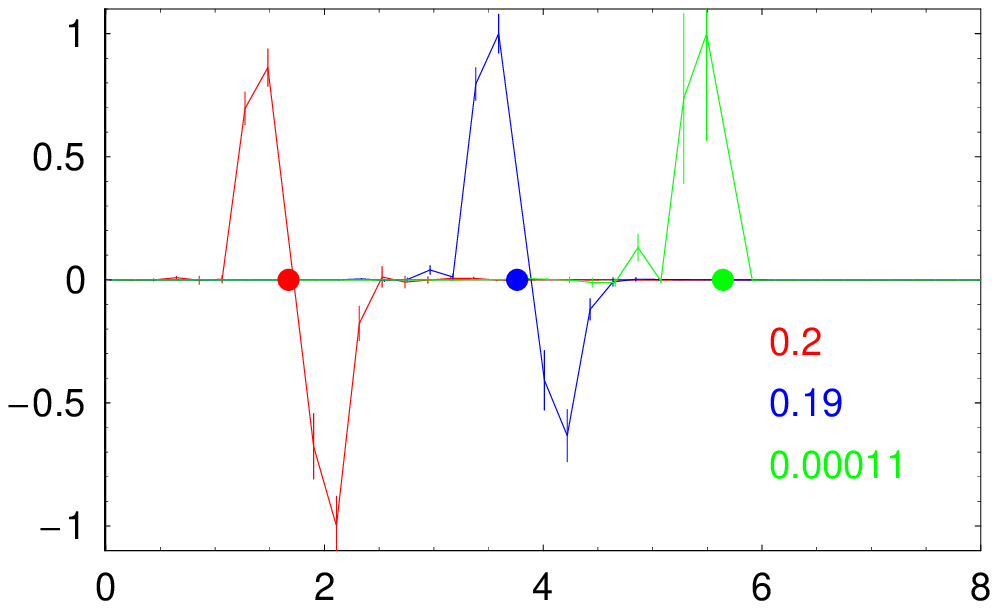}
        &
    \includegraphics[width=0.28\textwidth,angle=90]{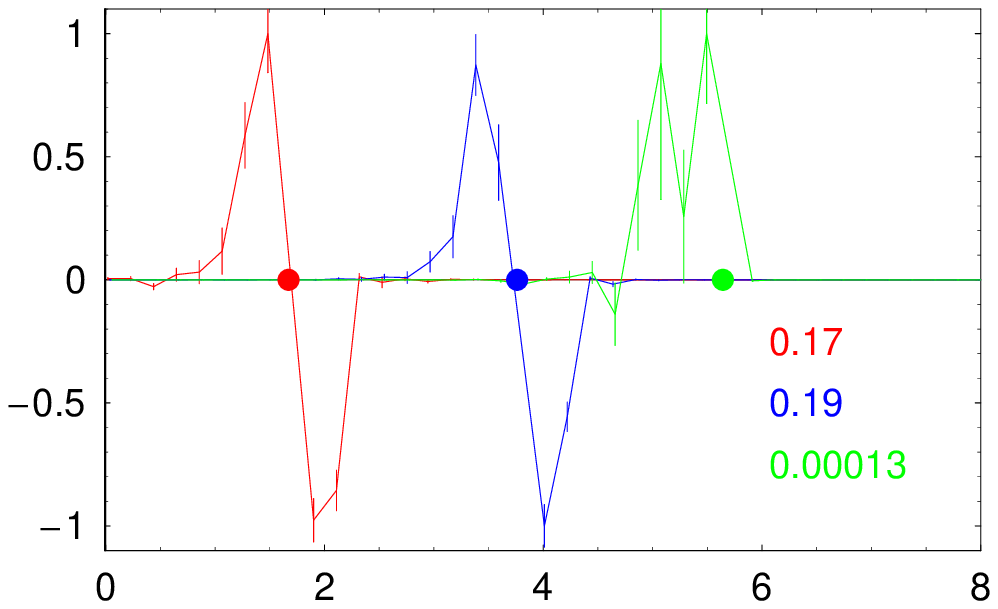}
        &
    \includegraphics[width=0.28\textwidth,angle=90]{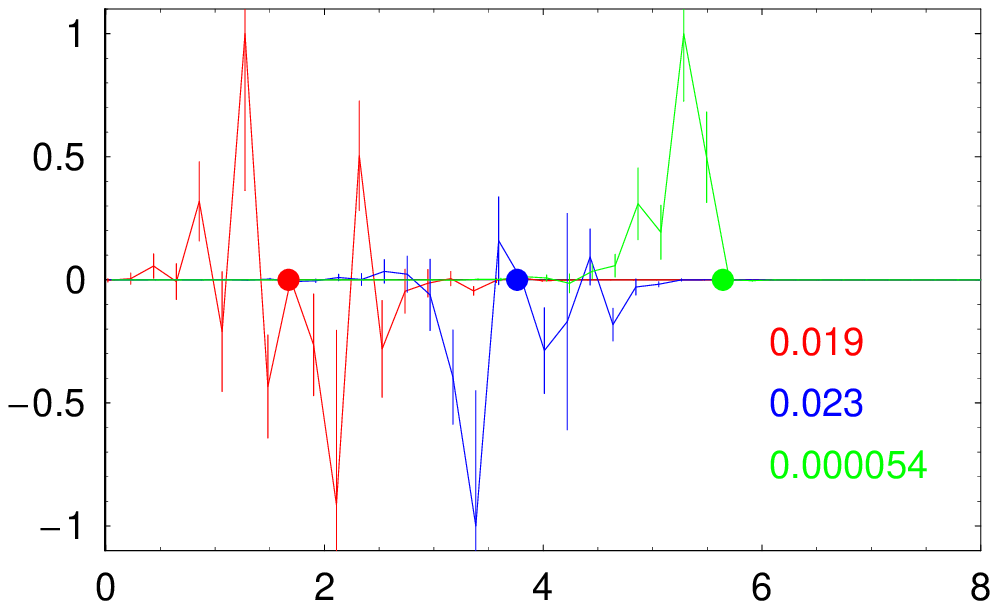}
               &
       \begin{turn}{90} $\quad \quad \quad \quad \quad \quad \quad \log_{10}q$ \end{turn}
        \\*[-0.2cm]
         \begin{turn}{90}$\quad \quad \quad \quad \alpha=-2.5$ \end{turn}
    &
    \includegraphics[width=0.28\textwidth,angle=90]{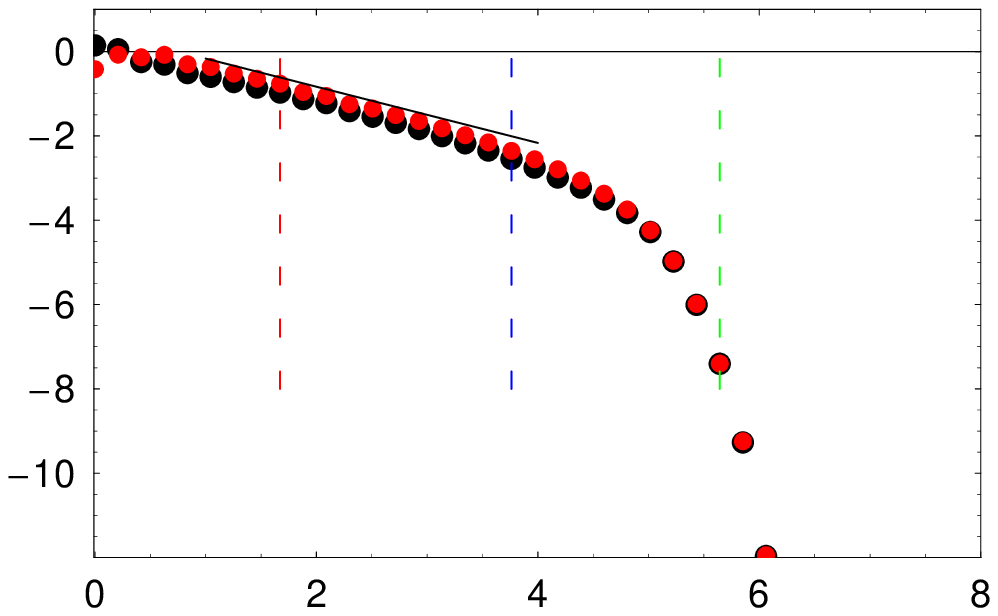}
        &
    \includegraphics[width=0.28\textwidth,angle=90]{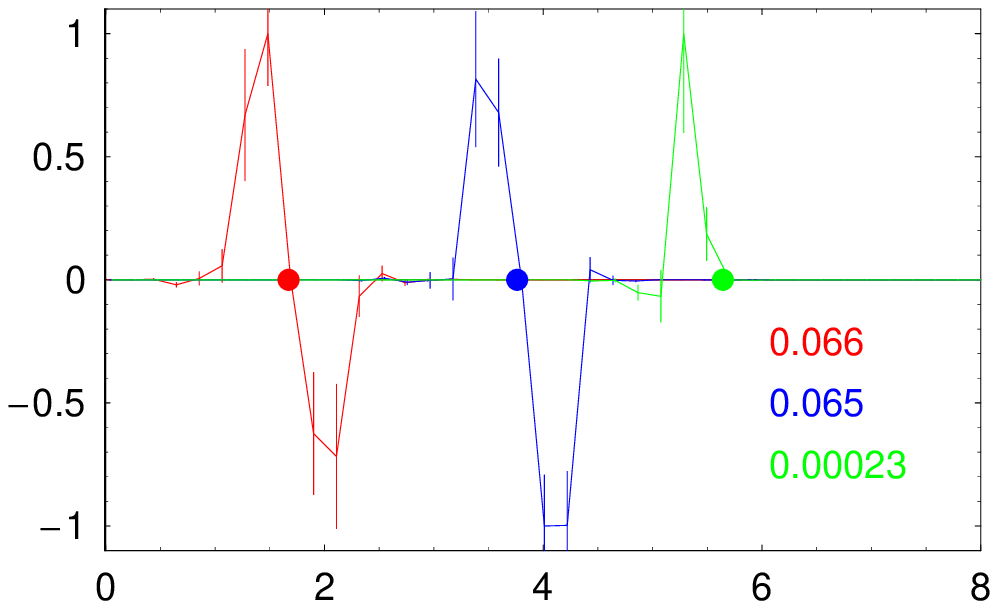}
        &
    \includegraphics[width=0.28\textwidth,angle=90]{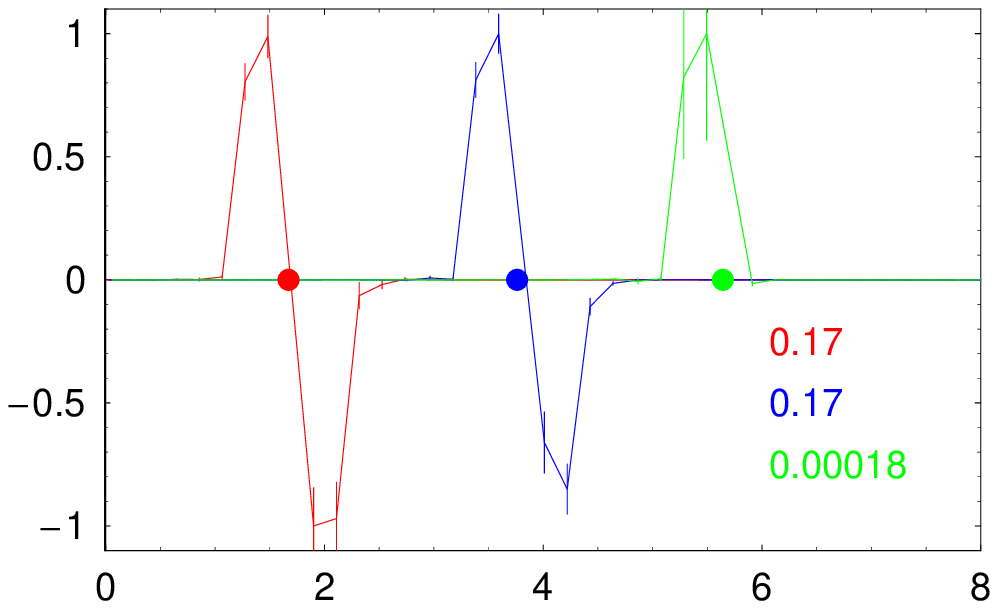}
        &
    \includegraphics[width=0.28\textwidth,angle=90]{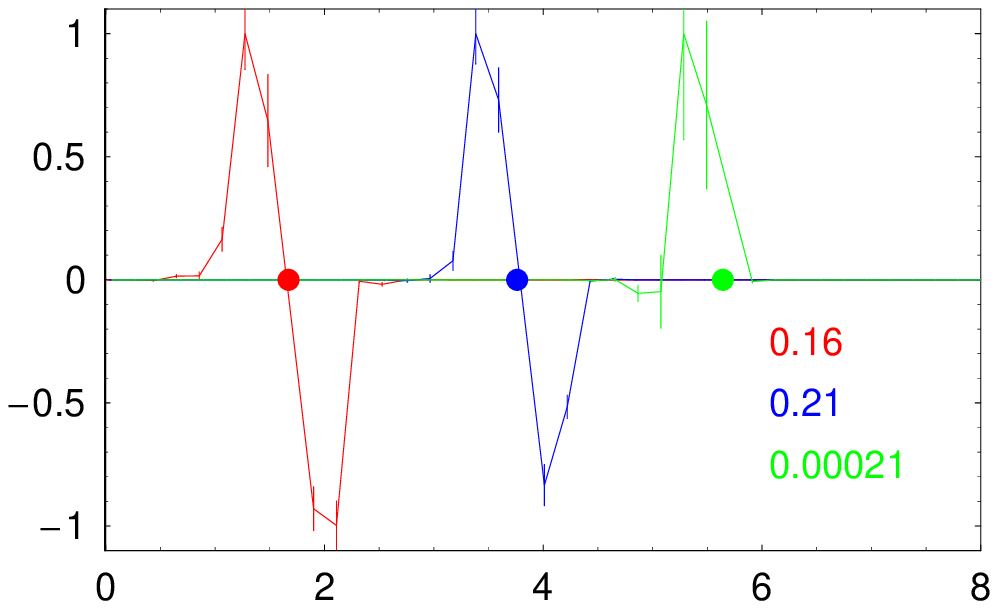}
        &
    \includegraphics[width=0.28\textwidth,angle=90]{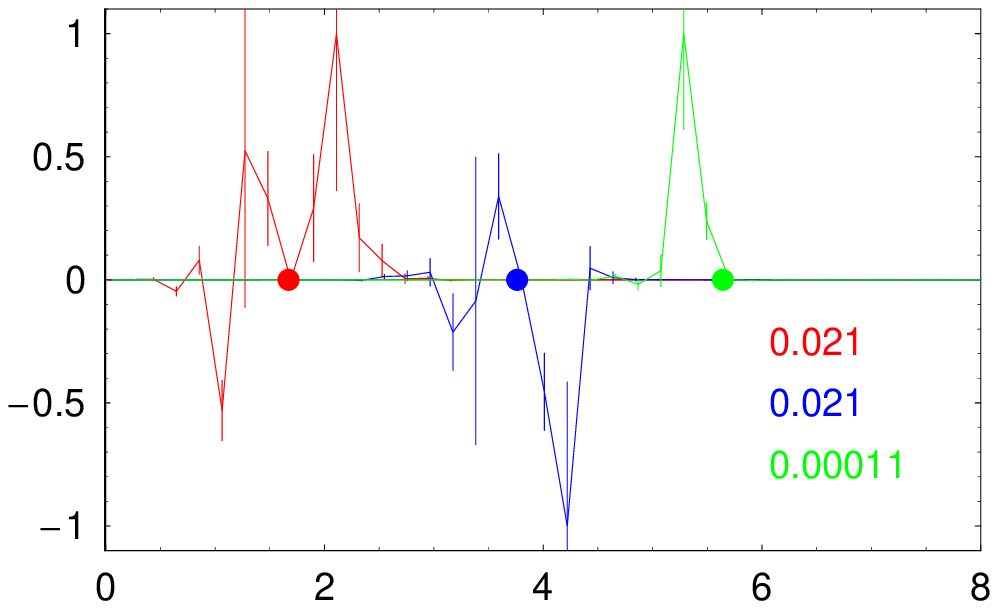}
               &
       \begin{turn}{90} $\quad \quad \quad \quad \quad \quad \quad \log_{10}q$ \end{turn}
        \\*[-0.2cm]
        \begin{turn}{90}$\quad \quad \quad \quad \alpha=-\infty$ \end{turn}
        &
    \includegraphics[width=0.28\textwidth,angle=90]{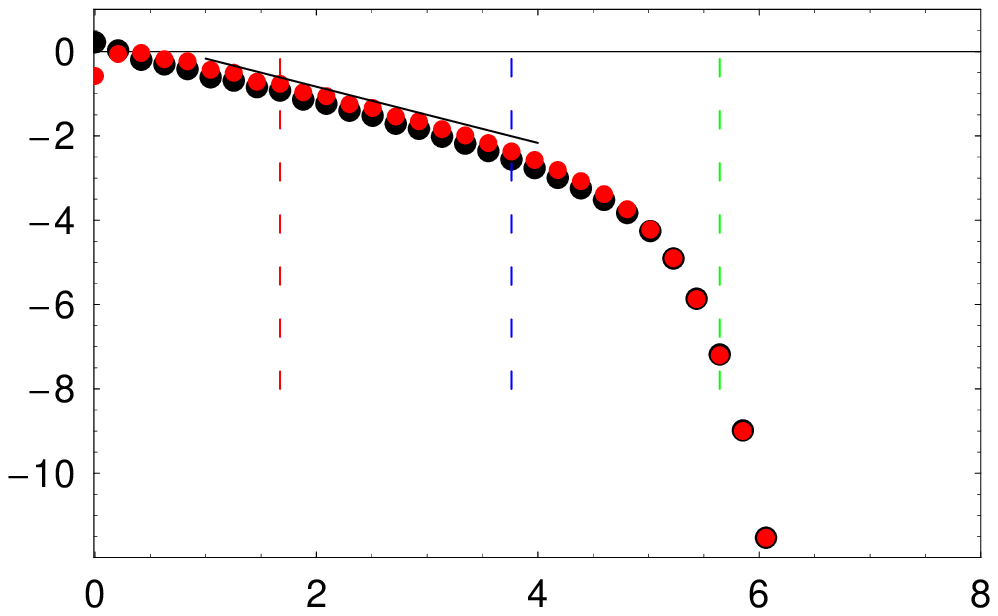}
        &
    \includegraphics[width=0.28\textwidth,angle=90]{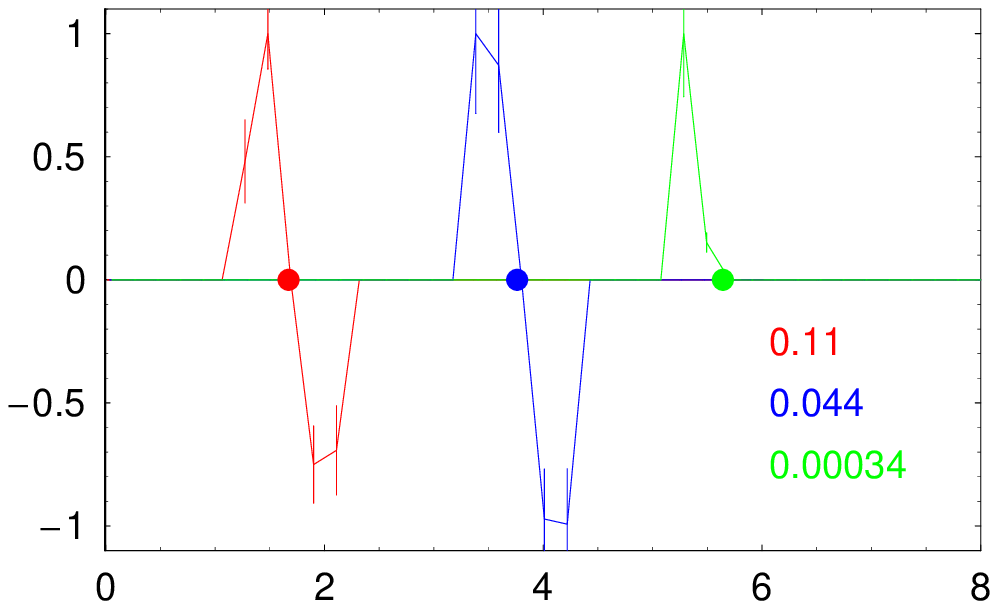}
        &
    \includegraphics[width=0.28\textwidth,angle=90]{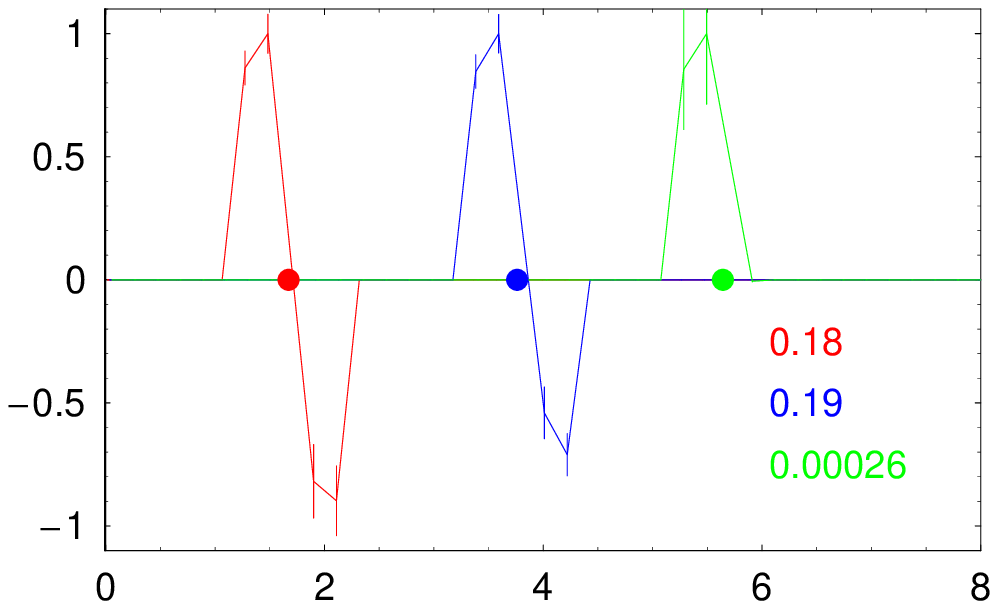}
        &
    \includegraphics[width=0.28\textwidth,angle=90]{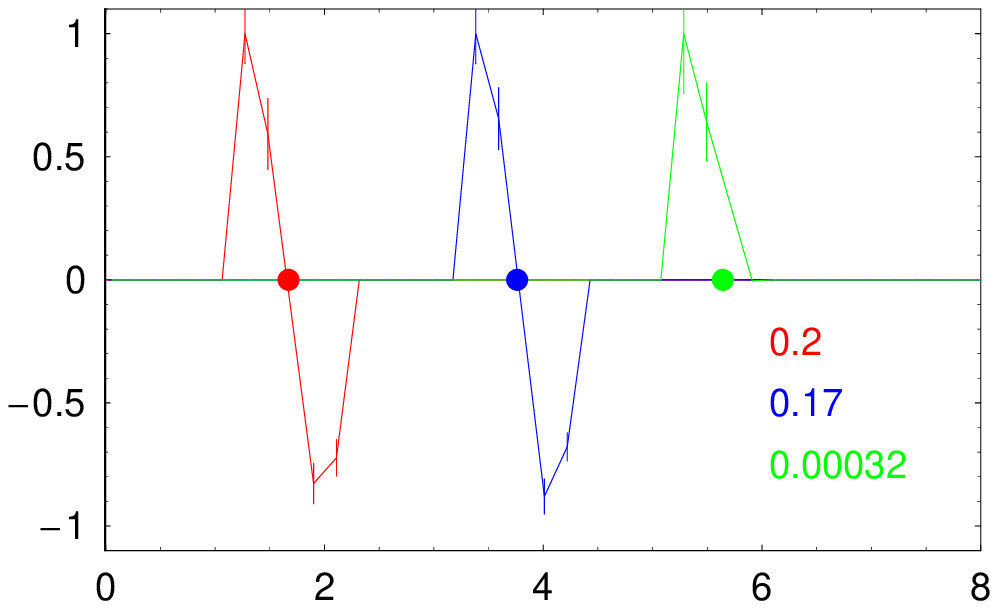}
        &
    \includegraphics[width=0.28\textwidth,angle=90]{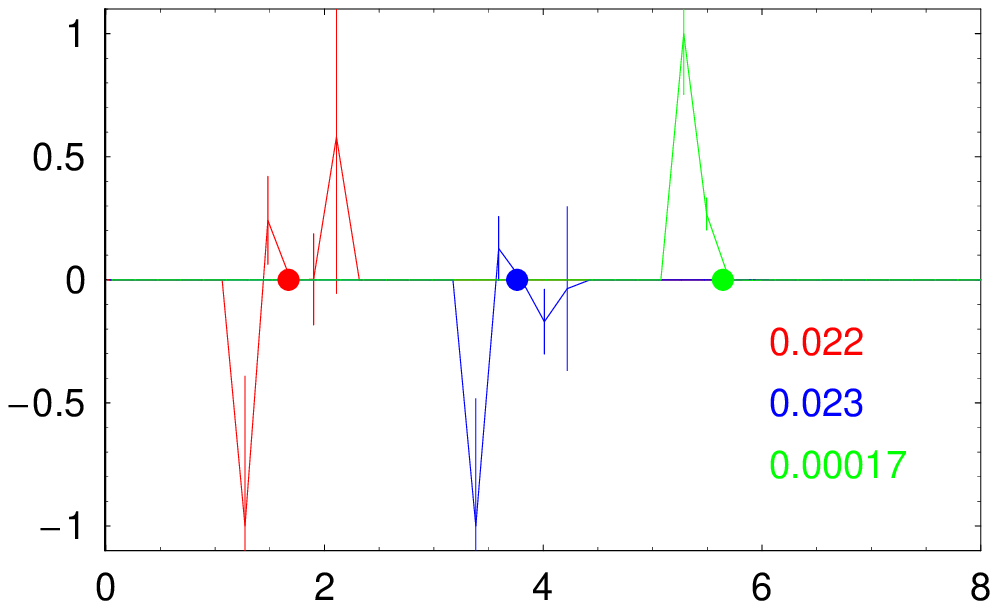}
               &
       \begin{turn}{90} $\quad \quad \quad \quad \quad \quad \quad \log_{10}q$ \end{turn}
  \end{tabular}
\caption{Same caption as figure \ref{transfer-3} but for $P_m=1 (\nu=\eta=10^{-7})$.} 
\label{transfer0}
\end{figure}
\begin{figure} \begin{tabular}{@{}c@{\hspace{0em}}c@{\hspace{0em}}c@{\hspace{0em}}c@{\hspace{0em}}c@{\hspace{0em}}c@{\hspace{0em}}c@{}}
         &
    \begin{turn}{0} Spectra \end{turn}
        &
    \begin{turn}{0} ${\cal T}_{UU}(q,n)$ \end{turn}
        &
    \begin{turn}{0}${\cal T}_{BU}(q,n)$\end{turn}
        &
   \begin{turn}{0} ${\cal T}_{UB}(q,n)$\end{turn}
        &
   \begin{turn}{0} ${\cal T}_{BB}(q,n)$\end{turn}
        & \\*[0cm]
   \begin{turn}{90}$\quad \quad \quad \quad \alpha=-0.5$ \end{turn}
        &
   \includegraphics[width=0.28\textwidth,angle=90]{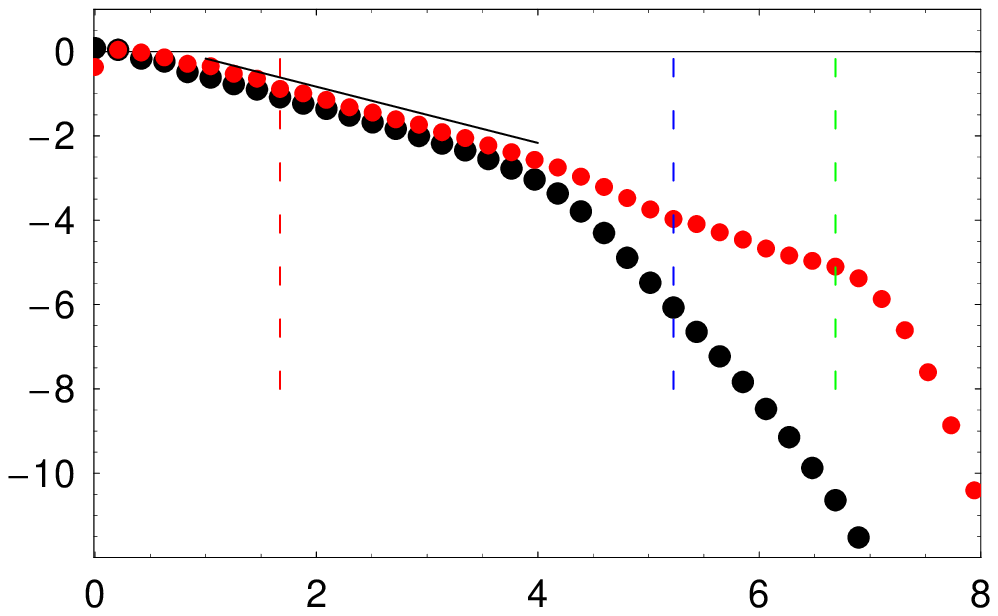}
        &
    \includegraphics[width=0.28\textwidth,angle=90]{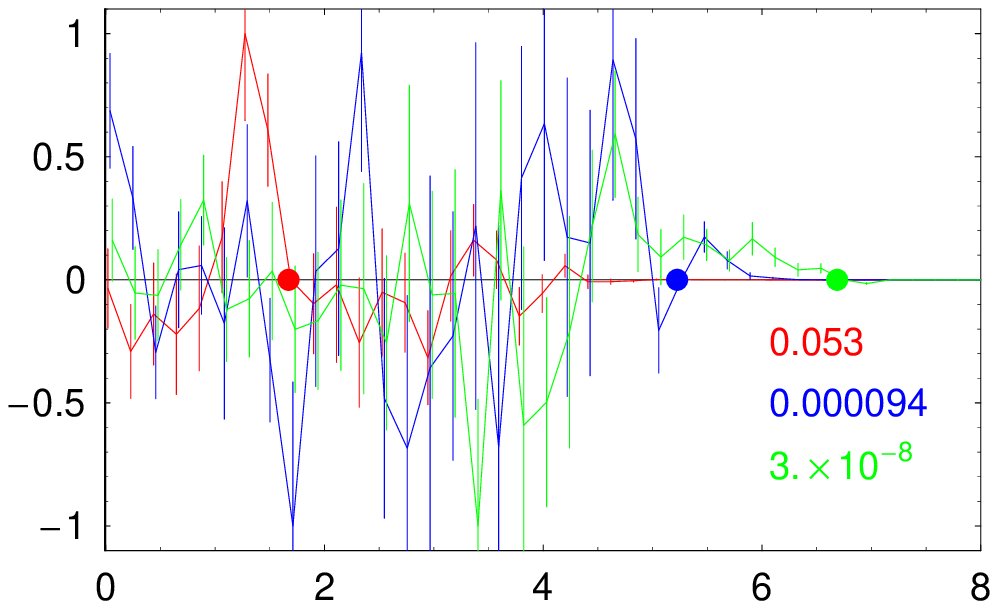}
        &
    \includegraphics[width=0.28\textwidth,angle=90]{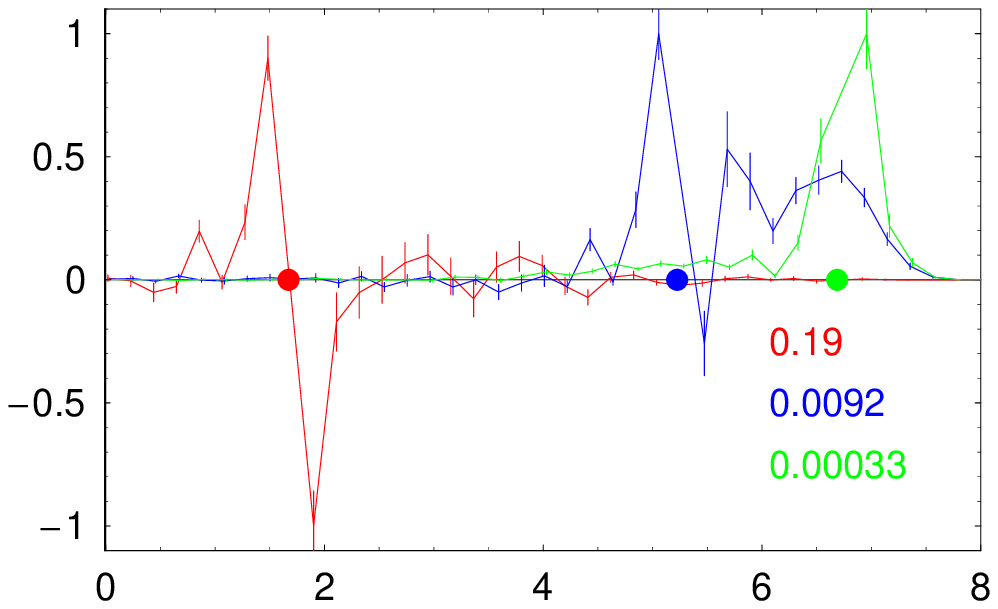}
        &
    \includegraphics[width=0.28\textwidth,angle=90]{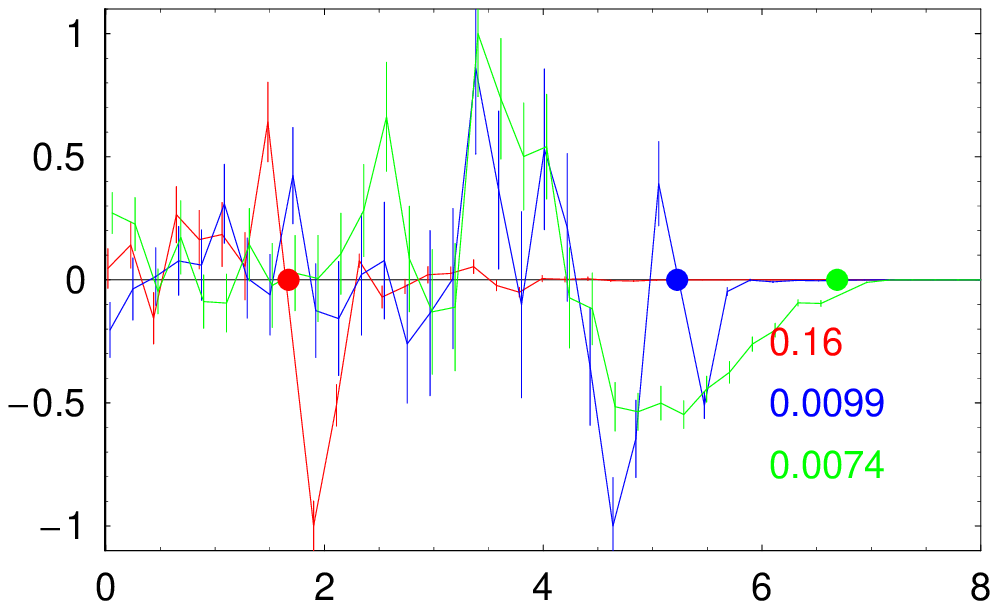}
        &
    \includegraphics[width=0.28\textwidth,angle=90]{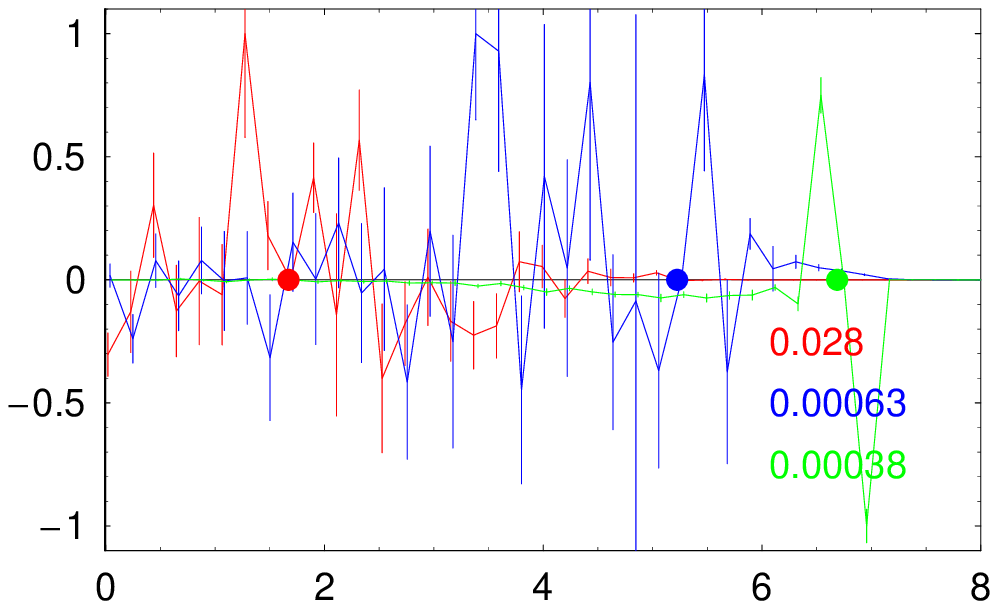}
               &
       \begin{turn}{90} $\quad \quad \quad \quad \quad \quad \quad \log_{10}q$ \end{turn}
    \\*[-0.2cm]
    \begin{turn}{90}$\quad \quad \quad \quad \alpha=-1$ \end{turn}
        &
    \includegraphics[width=0.28\textwidth,angle=90]{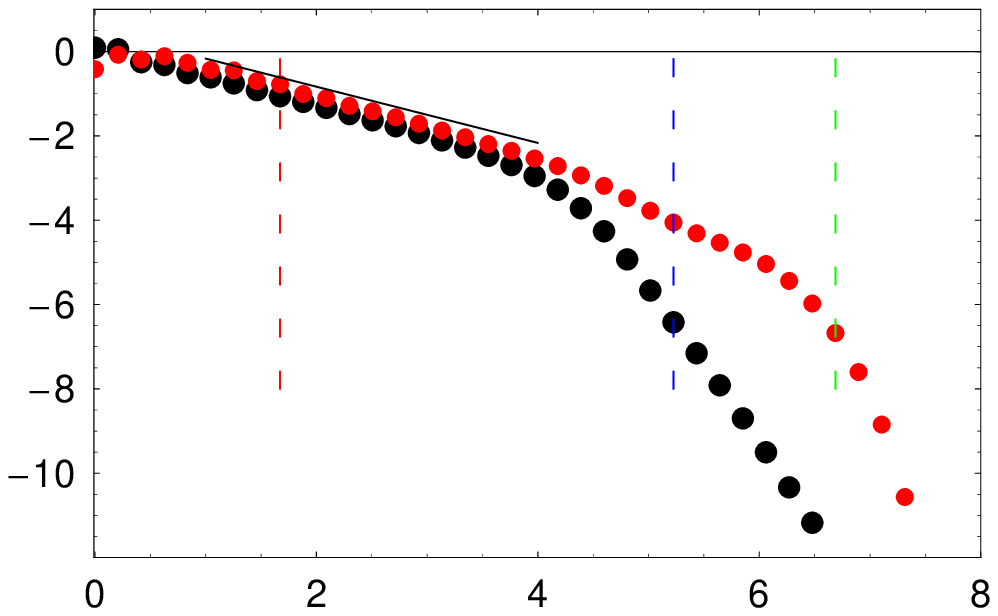}
        &
    \includegraphics[width=0.28\textwidth,angle=90]{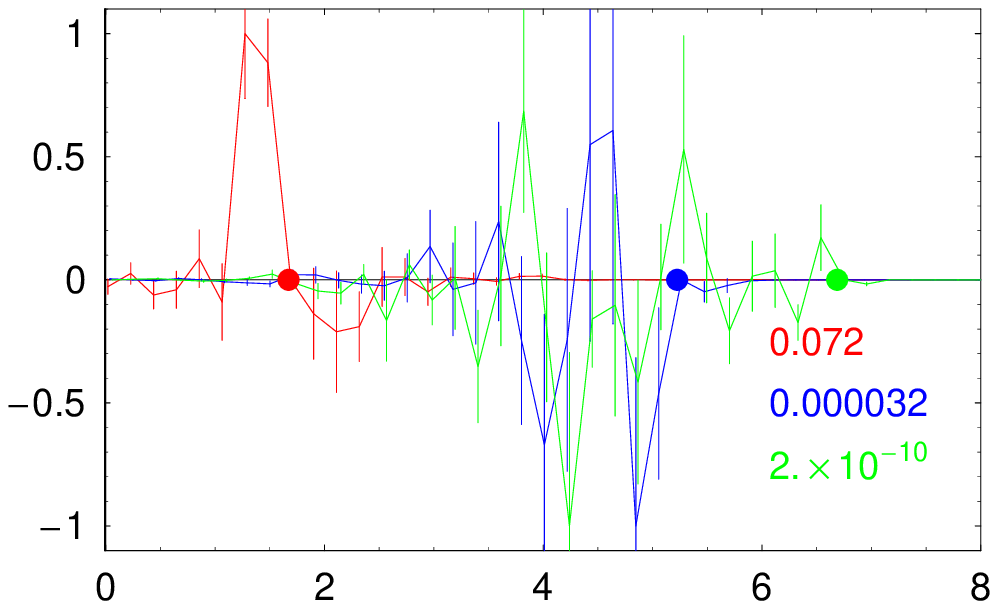}
        &
    \includegraphics[width=0.28\textwidth,angle=90]{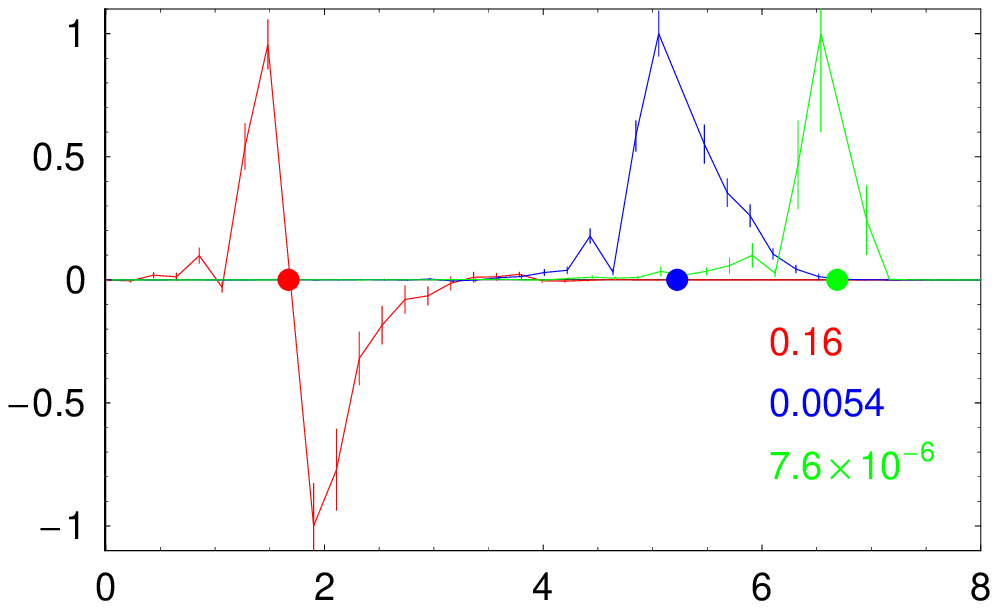}
        &
    \includegraphics[width=0.28\textwidth,angle=90]{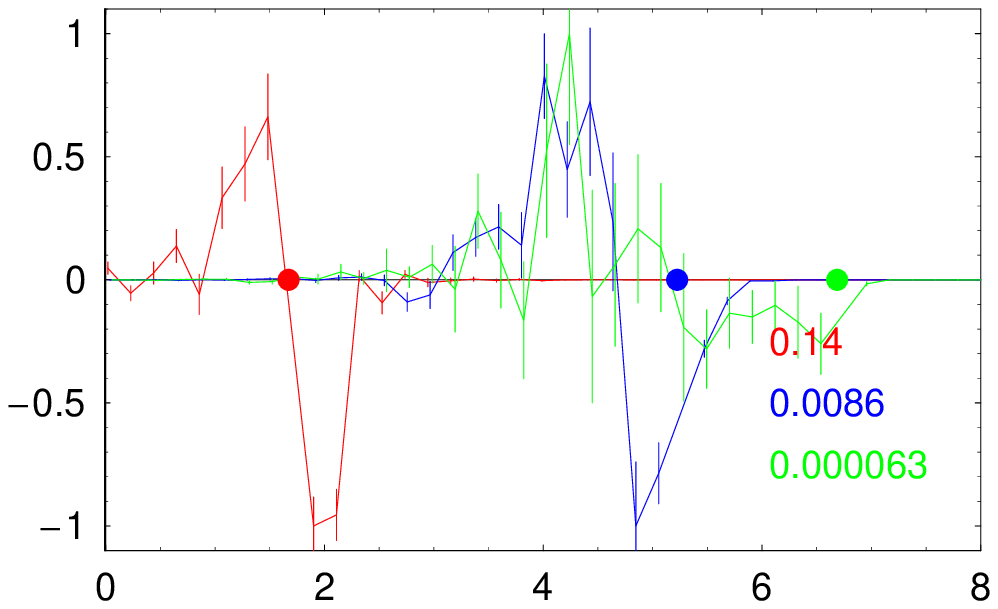}
        &
    \includegraphics[width=0.28\textwidth,angle=90]{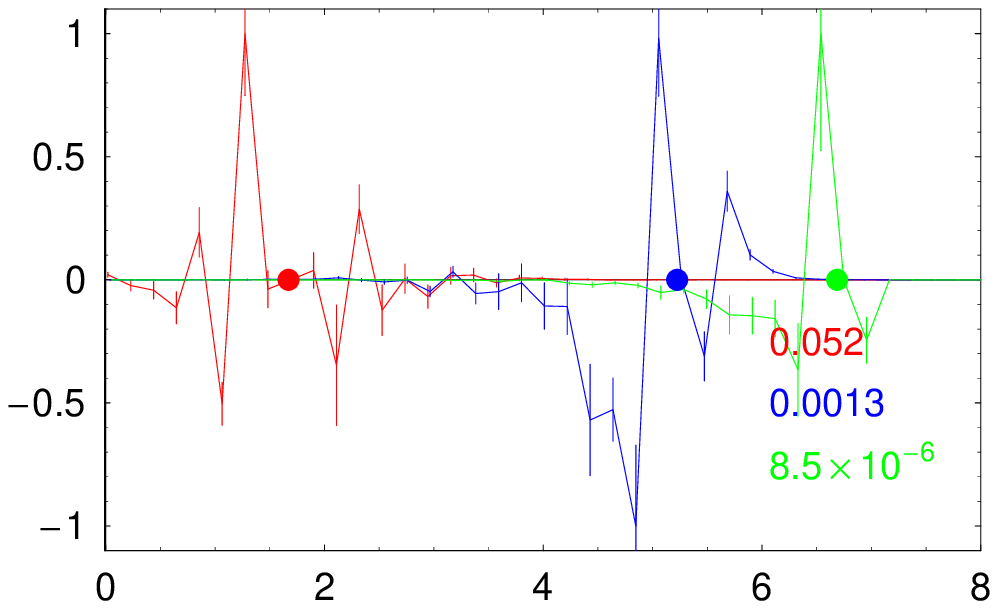}
               &
       \begin{turn}{90} $\quad \quad \quad \quad \quad \quad \quad \log_{10}q$ \end{turn}
        \\*[-0.2cm]
    \begin{turn}{90}$\quad \quad \quad \quad \alpha=-1.5$ \end{turn}
        &
    \includegraphics[width=0.28\textwidth,angle=90]{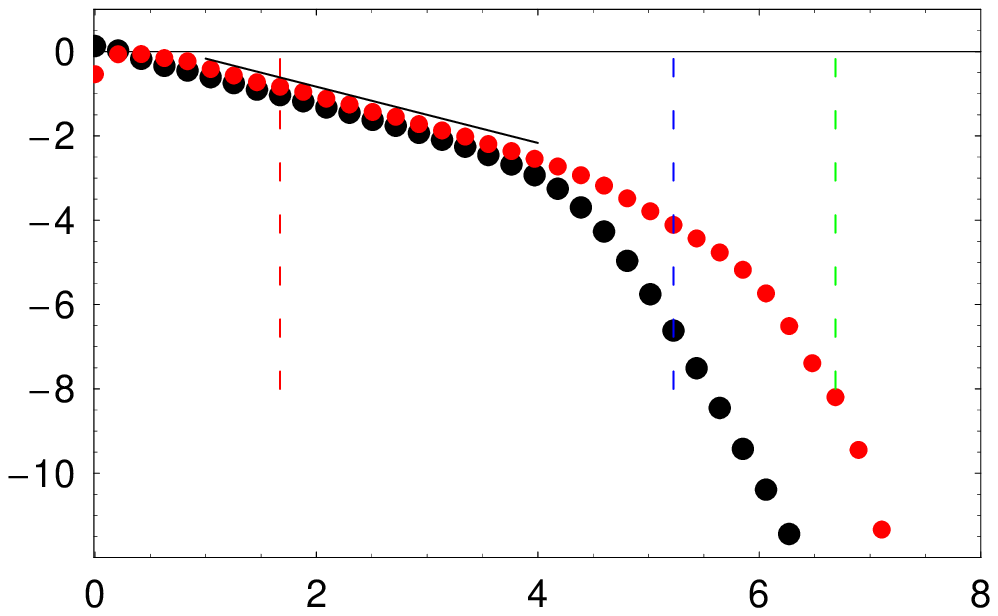}
        &
    \includegraphics[width=0.28\textwidth,angle=90]{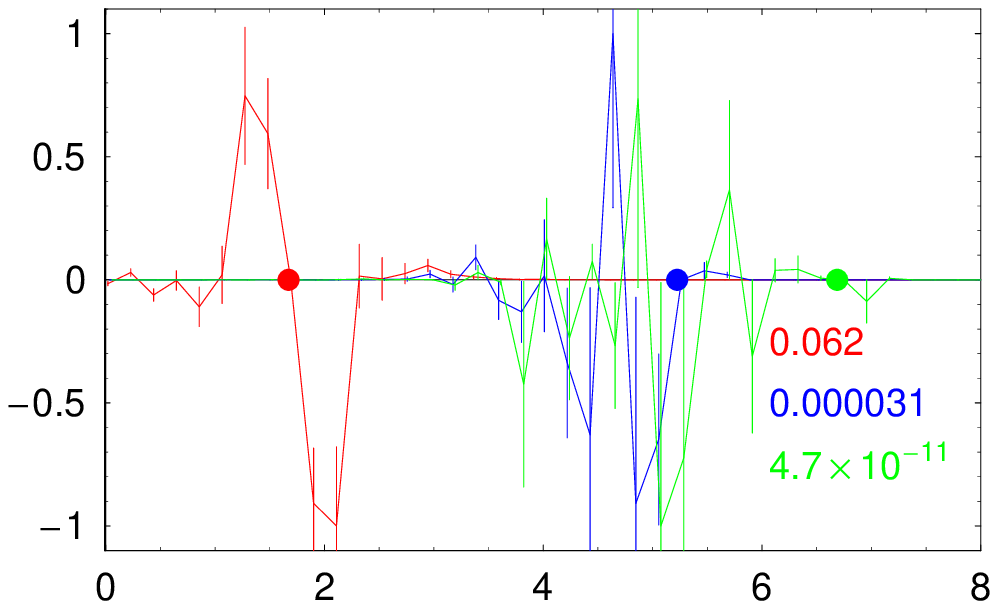}
        &
    \includegraphics[width=0.28\textwidth,angle=90]{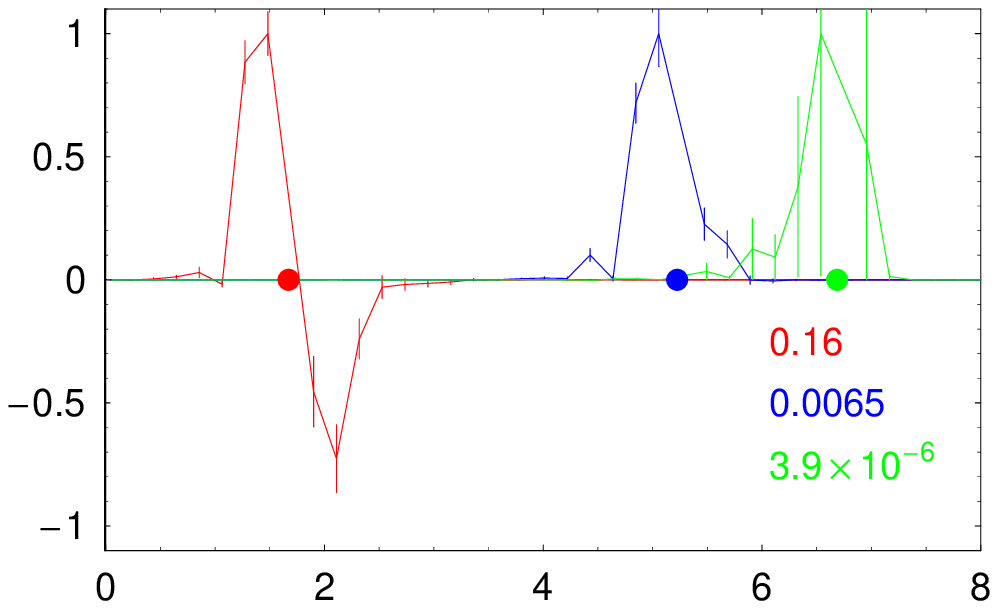}
        &
    \includegraphics[width=0.28\textwidth,angle=90]{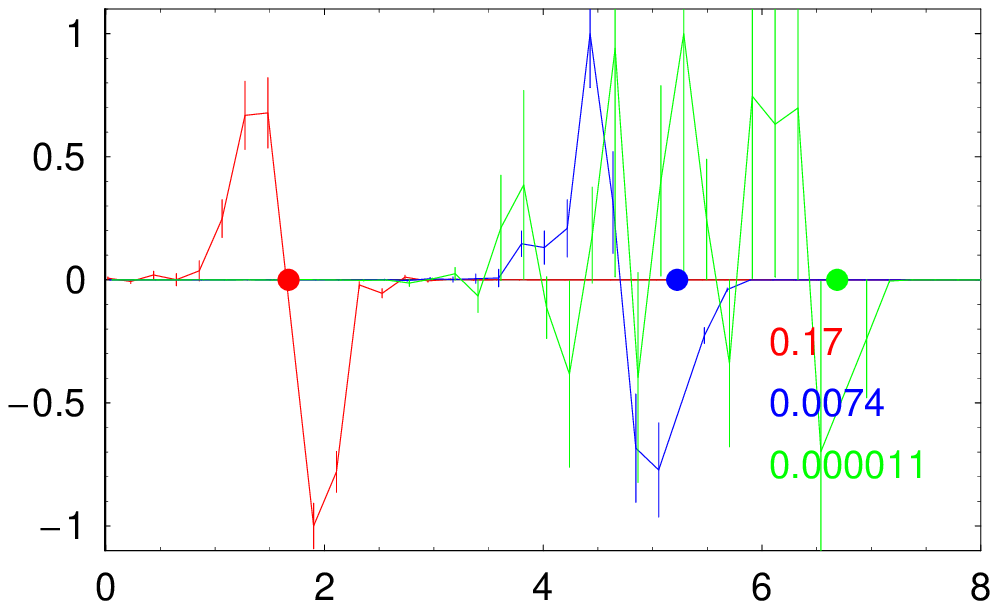}
        &
    \includegraphics[width=0.28\textwidth,angle=90]{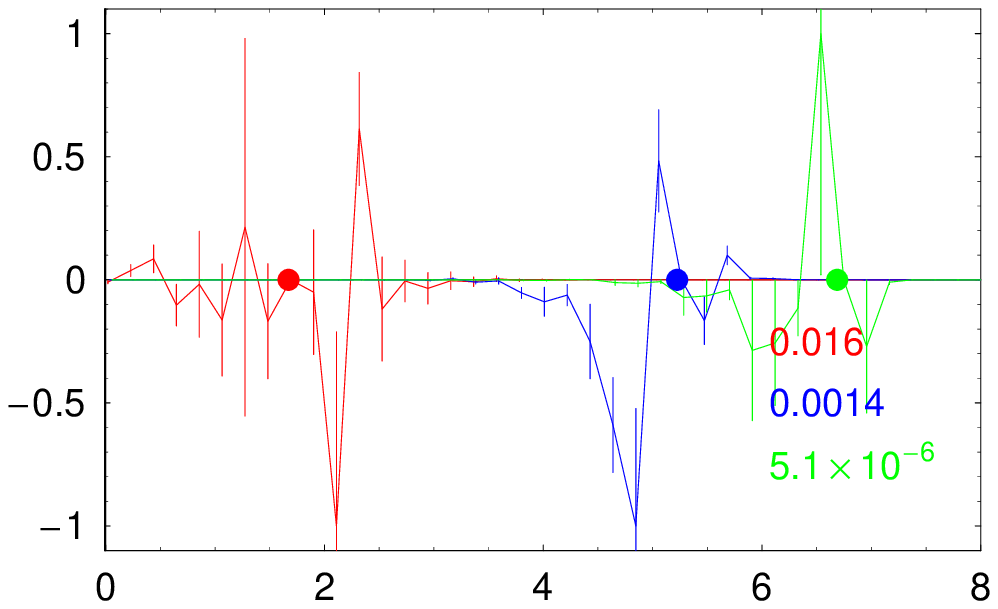}
               &
       \begin{turn}{90} $\quad \quad \quad \quad \quad \quad \quad \log_{10}q$ \end{turn}
        \\*[-0.2cm]
    \begin{turn}{90}$\quad \quad \quad \quad \alpha=-2.5$ \end{turn}
        &
    \includegraphics[width=0.28\textwidth,angle=90]{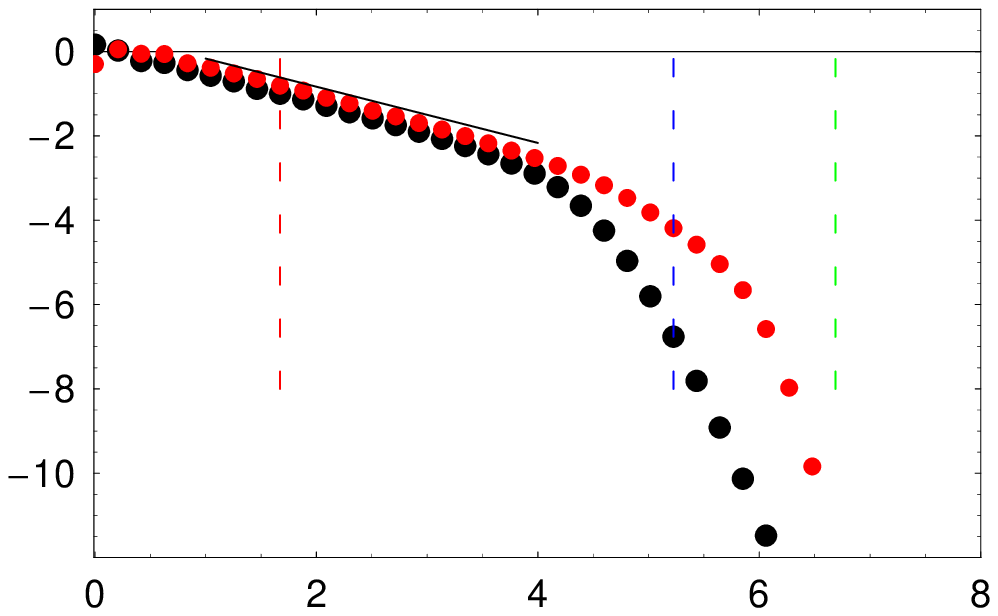}
        &
    \includegraphics[width=0.28\textwidth,angle=90]{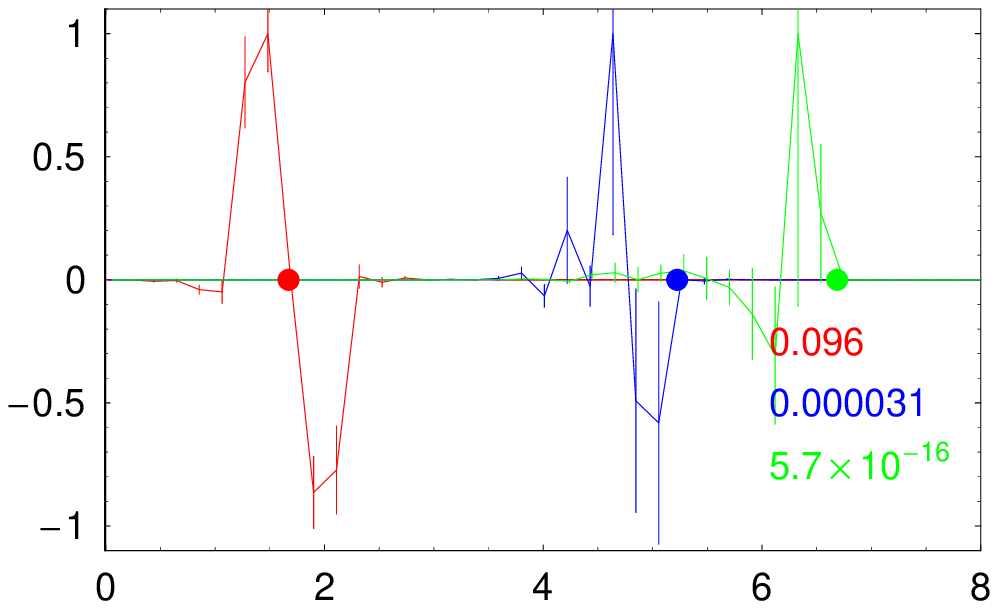}
        &
    \includegraphics[width=0.28\textwidth,angle=90]{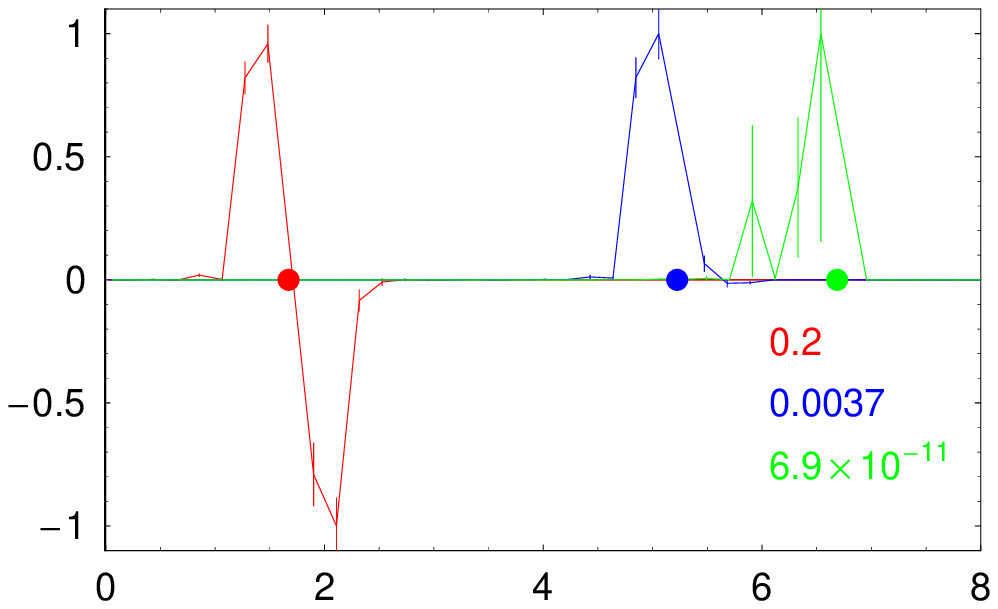}
        &
    \includegraphics[width=0.28\textwidth,angle=90]{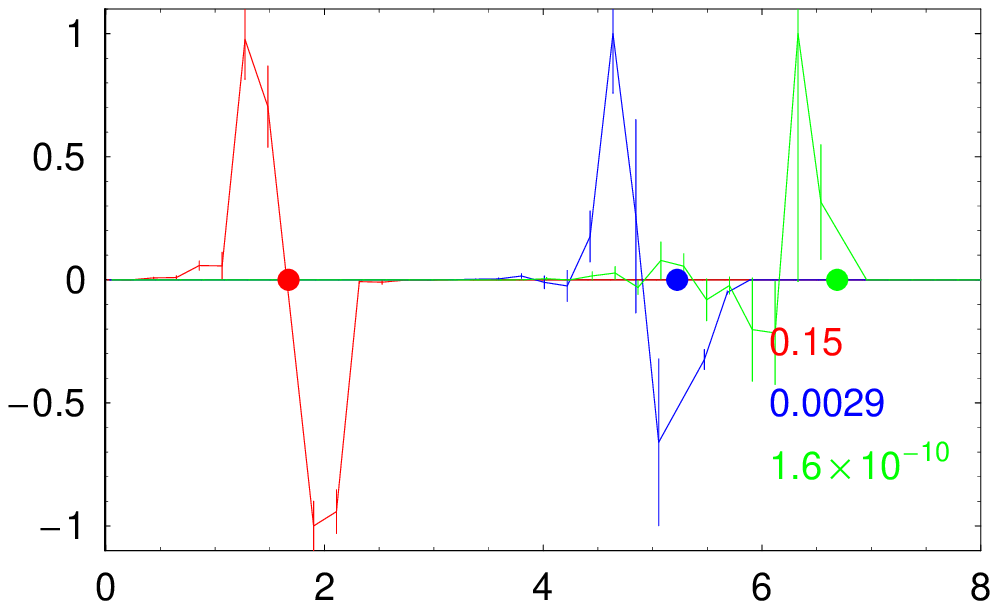}
        &
    \includegraphics[width=0.28\textwidth,angle=90]{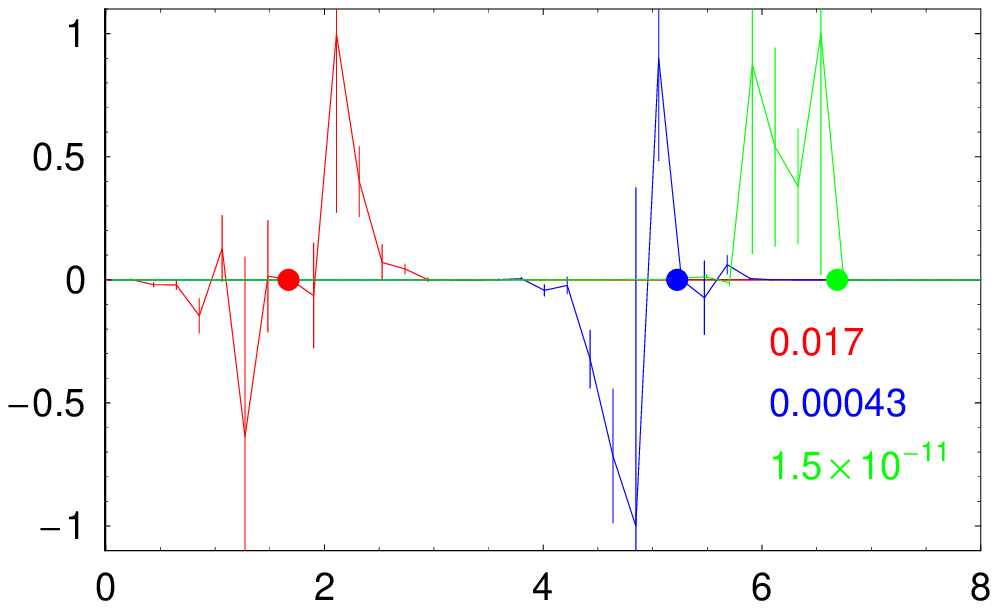}
               &
       \begin{turn}{90} $\quad \quad \quad \quad \quad \quad \quad \log_{10}q$ \end{turn}
        \\*[-0.2cm]
    \begin{turn}{90}$\quad \quad \quad \quad \alpha=-\infty$ \end{turn}
        &
    \includegraphics[width=0.28\textwidth,angle=90]{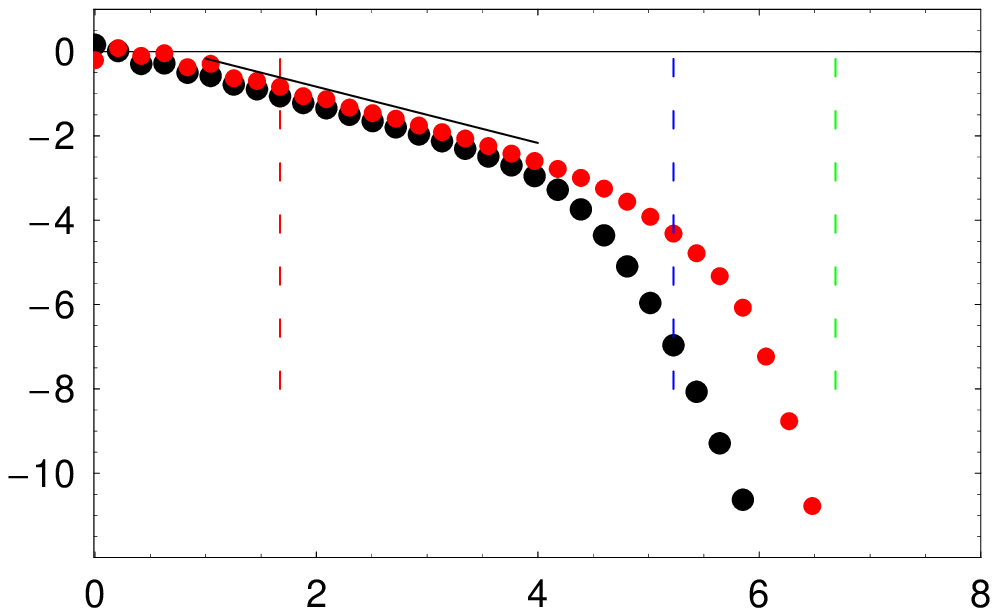}
        &
    \includegraphics[width=0.28\textwidth,angle=90]{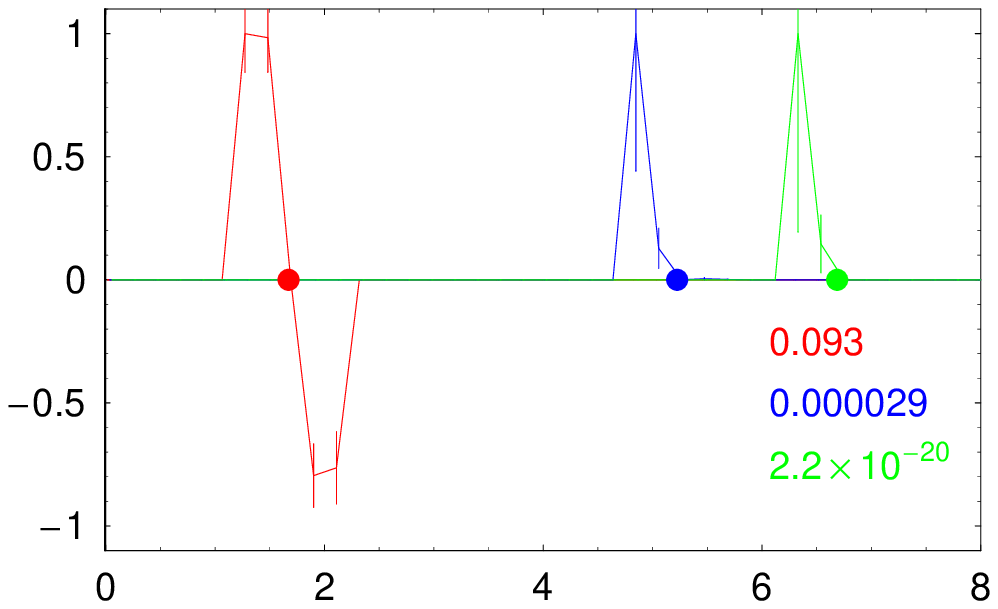}
        &
    \includegraphics[width=0.28\textwidth,angle=90]{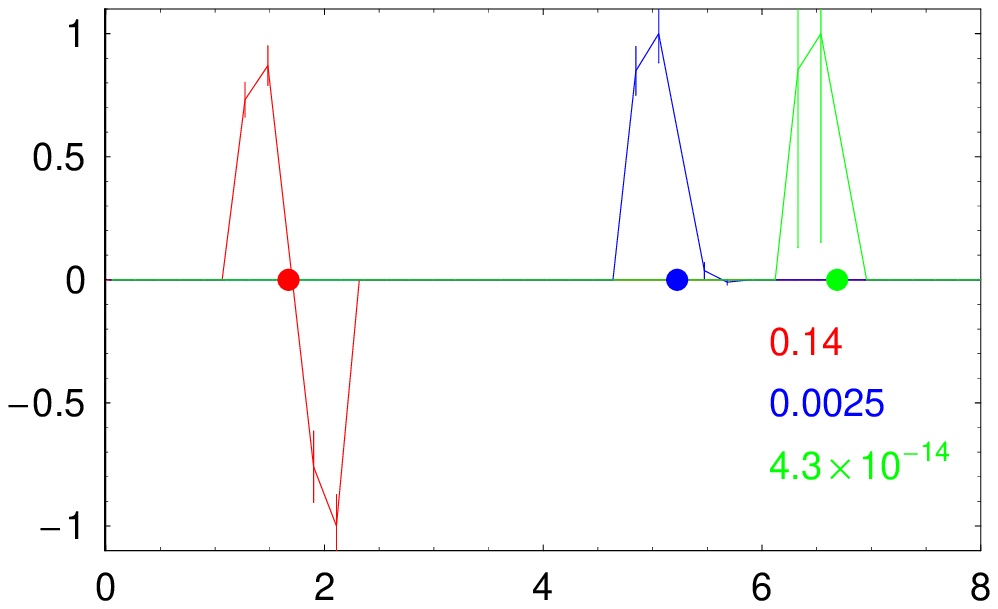}
        &
    \includegraphics[width=0.28\textwidth,angle=90]{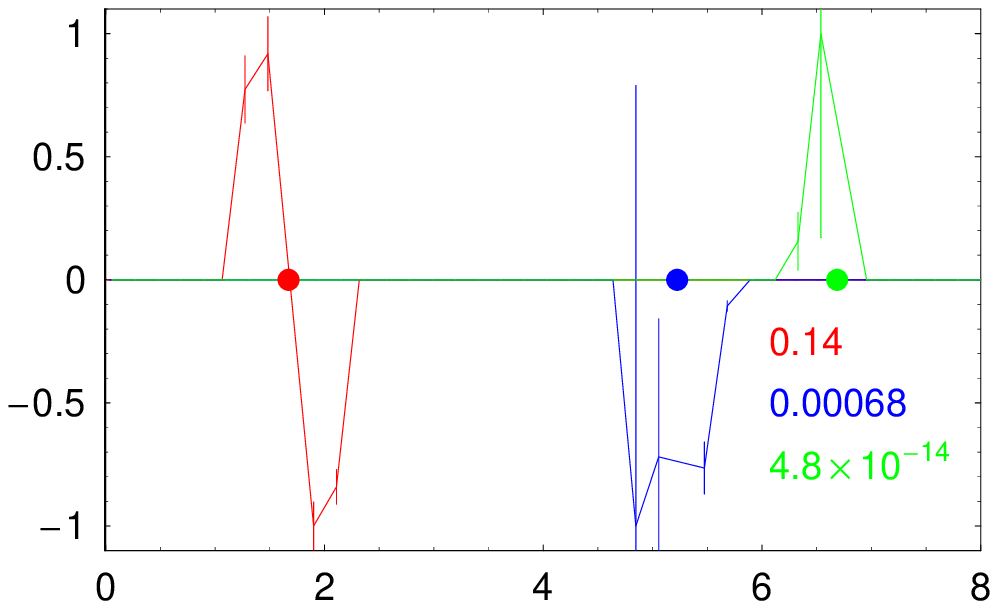}
        &
    \includegraphics[width=0.28\textwidth,angle=90]{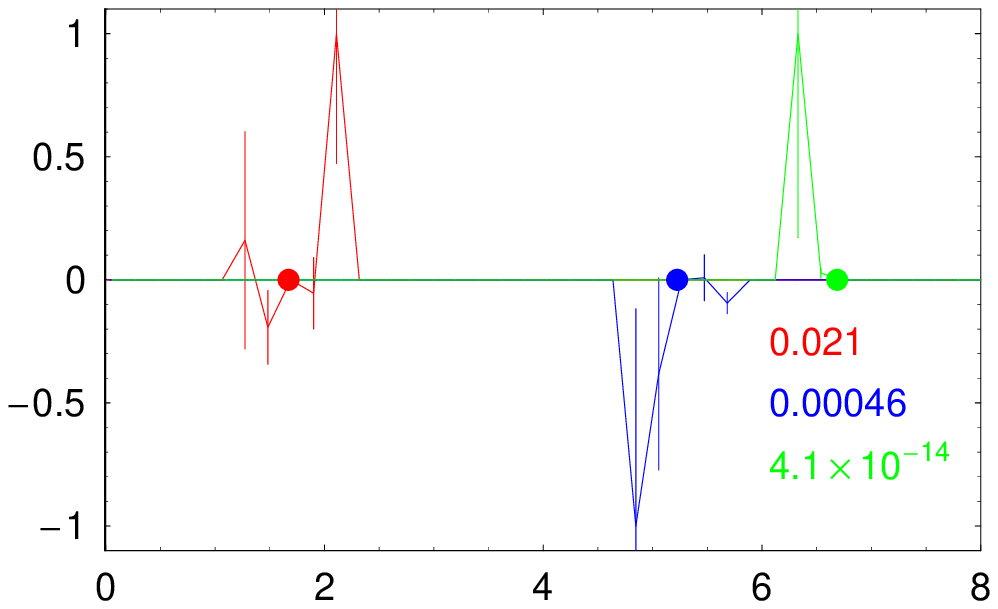}
               &
       \begin{turn}{90} $\quad \quad \quad \quad \quad \quad \quad \log_{10}q$ \end{turn}
  \end{tabular}
\caption{Same caption as figure \ref{transfer-3} but for $P_m=10^5 (\nu=10^{-6}, \eta=10^{-11})$.} 
\label{transfer5}
\end{figure}
\section{Illustration of the energy transfers}
\label{sec:interp}
\begin{figure}
\begin{center}
  \begin{tabular}{@{}c@{\hspace{0em}}}
    \includegraphics[width=0.7\textwidth]{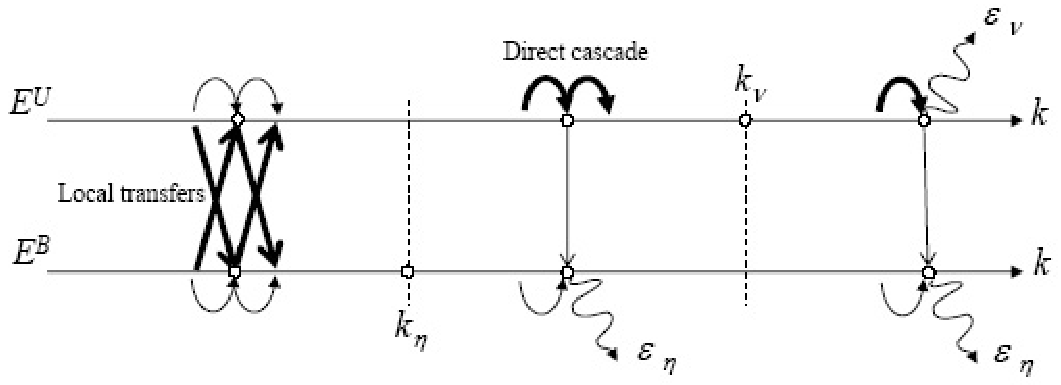}
   \\
    \includegraphics[width=0.7\textwidth]{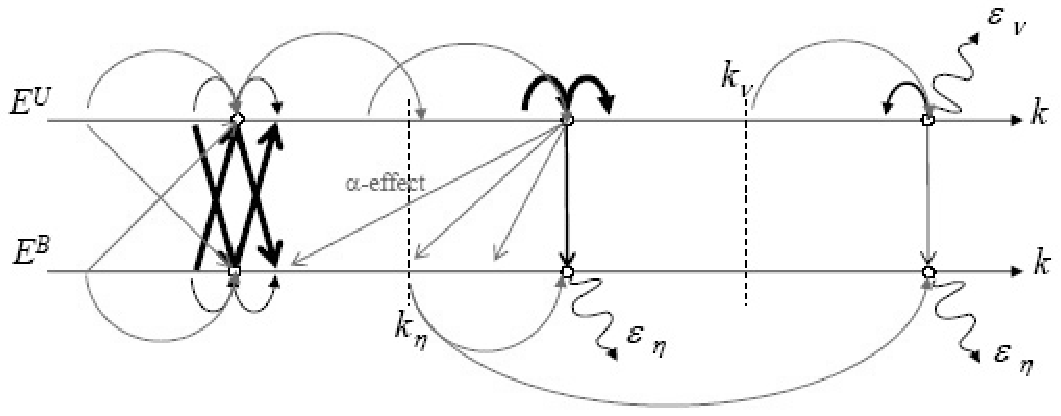}
  \end{tabular}
  \end{center}
\caption{Illustration of the energy transfers for $P_m \ll 1$ for local (top) and non local (bottom) model.
The thickness of the arrow gives some qualitative estimate of the strength of the transfer.} 
\label{interpretationlowPm}
\end{figure}
\begin{figure}
\begin{center}
  \begin{tabular}{@{}c@{\hspace{0em}}}
    \includegraphics[width=0.6\textwidth]{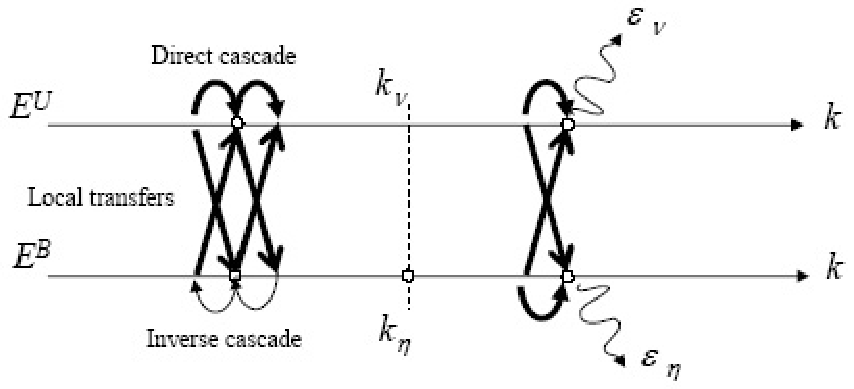}
    \\*[0cm]
    \includegraphics[width=0.6\textwidth]{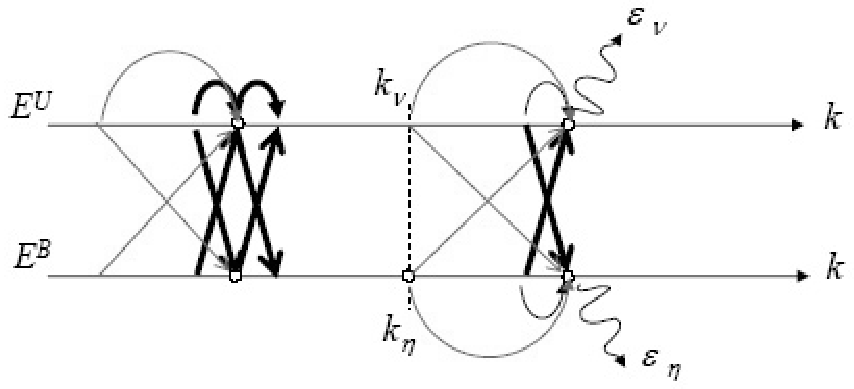}
  \end{tabular}
   \end{center}
\caption{Same as figure \ref{interpretationlowPm} but for $P_m = 1$.} 
\label{interpretationPm1}
\end{figure}
\begin{figure}
\begin{center}
  \begin{tabular}{@{}c@{\hspace{0em}}}
    \includegraphics[width=0.7\textwidth]{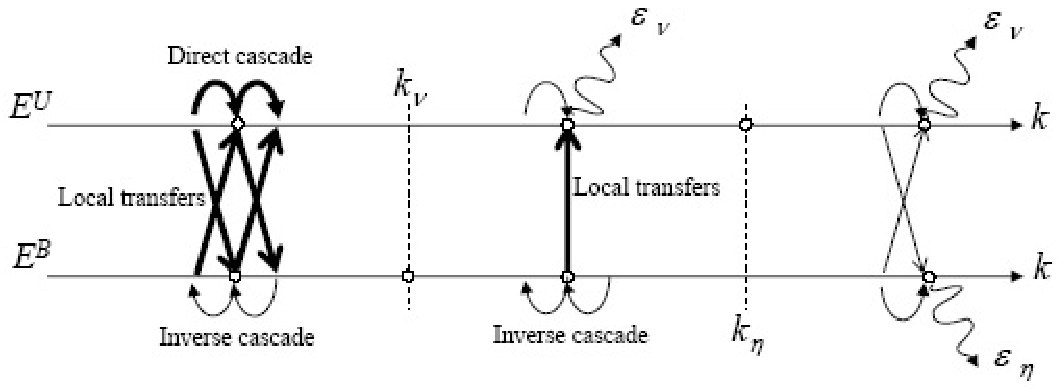}
    \\
    \includegraphics[width=0.7\textwidth]{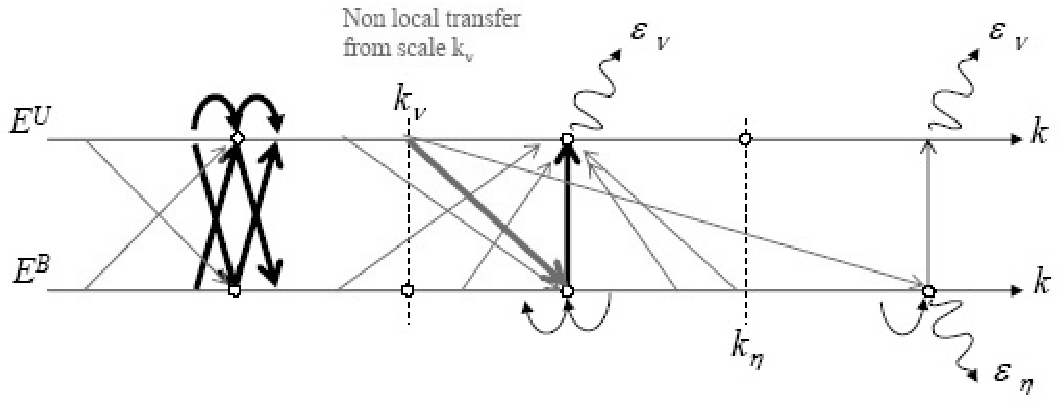}
  \end{tabular}
   \end{center}
\caption{Same as figure \ref{interpretationlowPm} but for $P_m \gg 1$.} 
\label{interpretationlargePm}
\end{figure}
\section*{References}
\bibliography{nonlocal}
\end{document}